\documentclass[11pt]{article}
\usepackage{amscd,amssymb,amsmath,latexsym,enumerate}
\usepackage[mathscr]{euscript}
\usepackage{epsfig}
\usepackage{fancybox}
\usepackage{verbatim}

\usepackage{mathdots}

\usepackage{color}

\usepackage{xcolor}
\definecolor{MyBlue}{cmyk}{1,0.13,0,0.63}
\definecolor{MyGreen}{cmyk}{0.91,0,0.88,0.52}
\newcommand{\mylinkcolor}{MyBlue}
\newcommand{\mycitecolor}{MyGreen}
\newcommand{\myurlcolor}{black}

\usepackage{hyperref}
\hypersetup{%
  bookmarksnumbered=true,bookmarksopen=false,%
  plainpages=false,
  linktocpage=true,%
  colorlinks=true,breaklinks=true,%
  linkcolor=\mylinkcolor,citecolor=\mycitecolor,urlcolor=\myurlcolor,%
  pdfpagelayout=OneColumn,%
  pageanchor=true,%
}

\title{On $\mathbb{Z}_2$-indices for ground states of fermionic chains}

\author{Chris Bourne$^1$, Hermann Schulz-Baldes$^2$ 
\thanks{email: 
\texttt{chris.bourne@tohoku.ac.jp, schuba@mi.uni-erlangen.de}}
\\
{\small $^1$ AIMR, Tohoku University, Sendai, and RIKEN iTHEMS, Wako, Japan}
\\
{\small $^2$ Department Mathematik, FAU Erlangen-N\"urnberg, Germany}
}

\date{\today}

\newtheorem{theo}{Theorem}[section]
\newtheorem{proposi}[theo]{Proposition}
\newtheorem{lemma}[theo]{Lemma}
\newtheorem{coro}[theo]{Corollary}

\newtheorem{defini}[theo]{Definition}
\newtheorem{remark}[theo]{Remark}
\newtheorem{example}[theo]{Example}

\newcommand{\CM}{{\mathbb C}}
\newcommand{\NM}{{\mathbb N}}
\newcommand{\RM}{{\mathbb R}}

\newcommand{\ZM}{{\mathbb Z}}

\newcommand{\Pp}{{\cal P}}
\newcommand{\PP}{{\bf P}}

\newcommand{\HH}{{\bf H}}
\newcommand{\UU}{{\bf U}}

\newcommand{\NN}{{\bf N}}

\newcommand{\Ff}{{\cal F}}
\newcommand{\Gg}{{\cal G}}

\newcommand{\Uu}{{\cal U}}

\newcommand{\Cc}{{\cal C}}

\newcommand{\Hh}{{\cal H}}

\newcommand{\one}{{\bf 1}}

\newcommand{\ev}{{\mbox{\rm\tiny ev}}}
\newcommand{\odd}{{\mbox{\rm\tiny od}}}
\newcommand{\bound}{{\mbox{\rm\tiny bd}}}

\newcommand{\Ker}{{\rm Ker}} 
\newcommand{\Ran}{{\rm Ran}}

\newcommand{\ii}{{i}}

\newcommand{\ph}{{\mbox{\rm\tiny ph}}}

\newcommand{\fraka}{{\mathfrak a}}
\newcommand{\frakb}{{\mathfrak b}}
\newcommand{\frakbtilde}{\tilde{\mathfrak b}}
\newcommand{\frakc}{{\mathfrak c}}
\newcommand{\frakd}{{\mathfrak d}}
\newcommand{\frakn}{{\mathfrak n}}

\def\calL{\mathcal{L}}
\def\calO{\mathcal{O}}

\def\calB{\mathcal{B}}
\def\calH{\mathcal{H}}

\def\calU{\mathcal{U}}

\newcommand{\ol}{\overline}

\begin{document}

\maketitle

\vspace{.5cm}

\begin{abstract}
For parity-conserving fermionic chains, we review how to associate $\mathbb{Z}_2$-indices
to ground states in finite systems with quadratic and higher-order interactions 
as well as to quasifree 
ground states on the infinite CAR algebra.
It is shown that the $\mathbb{Z}_2$-valued 
spectral flow provides a topological obstruction for two systems to have 
the same $\mathbb{Z}_2$-index. 
A rudimentary definition of a $\mathbb{Z}_2$-phase label for 
a class of parity-invariant and pure ground states of the one-dimensional infinite CAR algebra
is also provided.  Ground states with differing phase labels cannot be connected without a closing 
of the spectral gap of the infinite GNS Hamiltonian.
\hfill
\end{abstract}


\tableofcontents

\section{Introduction}

Rigorous analysis of condensed matter systems using topological methods has made substantial progress in 
the past 10--15 years. Topological insulators and superconductors 
have shown that invariants from differential topology (and their extensions in noncommutative geometry) 
give rise to stable and novel physical phenomena, see~\cite{PS} for references.

\vspace{0.2cm}

There have also been significant developments in the analytical understanding of gapped ground states of 
many-body spin systems and their relation to topological order. 
Improved Lieb--Robinson bounds and the area law for the decay of entanglement entropy~\cite{Hastings07} 
are among many 
 non-trivial results concerning properties of uniformly gapped ground states 
of frustration-free spin systems~\cite{BMNS, BN, BO, OgataIII}. See~\cite{NSY18} for a comprehensive review. 
In dimensions greater than one, where braiding may occur, analytic results are much harder to obtain, 
though important examples such as Kitaev's toric code~\cite{KitAnyon} can be treated within the 
framework of frustration-free ground states. Newer methods for higher-dimensional spin systems 
are also in development~\cite{Cha18}. There has also been several results concerning 
stability of topological invariants such as the Hall conductance in interacting 
fermion systems~\cite{BBDRF1,BBDRF2,BDRF18,GMP, HM, MonacoTeufel}.

\vspace{0.2cm}

There have been efforts in the physics community to connect these 
two areas of topological  physics via the study of interacting topological phases. 
While a precise characterisation of interacting phases
 remains in development, following a proposal of Kitaev, it is currently 
expected that symmetry protected topological (SPT) phases of gapped ground states are described 
using a generalised cohomology theory~\cite{FreedHopkins, SSR,Xiong18}. Roughly speaking, 
such theories construct a homotopy group of deformation 
classes of invertible topological field theories or 
short-range entangled states with specified additional input, {\it e.g.} symmetries and dimension.  
For the case of fermions, which we focus on in this manuscript, $\mathbb{Z}_2$-graded tensor networks 
provide a convenient toolset to construct such field theories, see~\cite{BWH, FPEPS, KTY18, TurzilloYou}.

\vspace{0.2cm}

The goals of this paper are much more modest. Our aim is to review the $\mathbb{Z}_2$-index 
associated to one-dimensional fermionic ground states considered by Kitaev~\cite{Kit} 
as an indication of Majorana fermions at the boundary of one-dimensional superconducting wires. 
This $\mathbb{Z}_2$-phase label is now regarded as the one-dimensional SPT phase of 
gapped and parity-symmetric  fermionic systems without additional symmetries. While some  
properties of infinite systems and the thermodynamic limit 
can be obtained by a careful treatment of finite systems, 
rigourous studies of infinite fermionic systems directly are less common.
One reason is that ground states in infinite systems are generally understood via techniques 
from operator algebras and, as such, require a more involved framework.

\vspace{0.2cm}

The $\mathbb{Z}_2$-indices for ground states of finite fermionic chains 
with quadratic and higher-order interactions are first reviewed. We also consider 
$\mathbb{Z}_2$-indices for  quasifree ground states 
of infinite systems, which generalise the finite-dimensional $\mathbb{Z}_2$-index. 
The exposition on quasifree ground states is closely related to 
work by Araki, Evans and Matsui on the $XY$-chain and the phase transition of the 
$2$-dimensional Ising model~\cite{Araki,ArakiEvans, ArakiMatsui85}. 
Many have noted 
that the quadratic finite Kitaev chain is the same as the quantum Ising chain under the Jordan--Wigner 
transform. But a more systematic treatment on the connections between spin chains in quantum 
statistical mechanics and fermionic gapped ground state phases, particularly in infinite systems, 
appears to be absent in the literature. 
As such, these concepts are reviewed in detail.

\vspace{0.2cm}

A key connection is also shown  between the $\mathbb{Z}_2$-ground state 
index and the $\mathbb{Z}_2$-valued spectral flow recently studied 
in~\cite{CPS}. (Let us stress that the $\mathbb{Z}_2$-valued spectral flow is \emph{unrelated} to the spectral 
flow of the quasiadiabatic evolution of ground states \cite{NSY18}, see Section \ref{sec:Z_2_prelims}.) 
Indeed for finite quadratic chains and quasifree ground states of the CAR algebra, 
the $\mathbb{Z}_2$-valued spectral flow is shown to encode the topological obstruction 
for two Hamiltonians to have the same $\mathbb{Z}_2$-ground state index.  
For systems with periodic or anti-periodic boundary conditions, this topological obstruction can be detected 
via the insertion of a flux quanta through a local cell  and the associated $\mathbb{Z}_2$-valued spectral flow. 
Finite chains with twisted boundary conditions as studied in~\cite{TwistedKitaev} 
also provide an example. 
We remark that fermionic interactions with periodic or anti-periodic boundary conditions 
become highly non-local if one takes the  Jordan--Wigner transformation and considers 
the  corresponding bosonic Hamiltonian. Therefore, such Hamiltonians will in general violate the Local Topological Quantum 
Order condition used in~\cite{MZ, MN18}  to show stability of a ground state gap.

\vspace{0.2cm}

One of the motivations to study flux insertions is
to analyse topological properties of Hamiltonians and their ground states. 
 By connecting 
flux insertion to $\mathbb{Z}_2$-spectral flow, an index-theoretic construction, 
the topological nature of the ground states under consideration becomes manifest. 
Flux insertion has also been used in higher-dimensional systems to construct 
a many-body index for charge transport~\cite{BBDRF2} as well as show the stability of the Hall conductance 
under interactions~\cite{BBDRF1}. 
These observations open a potential pathway to study 
topological invariants of higher dimensional interacting systems of fermions
 by inserting (non-abelian) monopoles as in \cite{CSBMonopole,DS}.

\vspace{0.2cm}

While much of the manuscript is review, we do provide a  candidate  for a 
$\mathbb{Z}_2$-index of pure, gapped and parity-invariant ground states on 
the one-dimensional infinite CAR algebra that can be used as a phase label.
To the best of our knowledge, the 
construction is new, though it heavily relies  on the split property of one-dimensional 
ground states~\cite{Matsui01, Matsui13} 
as well as the infinite Jordan--Wigner 
transform~\cite[Chapter 6.5]{EvansKawahigashi}. The use of the split property as 
a tool to  characterise ground state SPT phases was first noted by Ogata~\cite{OgataTRS}.
Results from~\cite{NSY18} give tools to show basic stability properties of this index, including invariance under 
a $C^1$-path of uniformly gapped Hamiltonians satisfying extra compatibility conditions.  
We also show that if two gapped ground states have differing phase labels, 
then the spectral gap of the infinite GNS Hamiltonian must close for paths of 
ground states connecting the two systems. This gives us some confidence that the suggested 
phase label is a useful one.

\subsection*{Outline}

Section \ref{sec:prelim} gives a  brief summary of the operator algebra approach to  fermionic ground states and the 
$\mathbb{Z}_2$-valued spectral flow. 
The paper  is then divided into 2 relatively distinct  parts corresponding to 
finite and infinite chains, where  the characterisation of the ground state changes from the 
lowest-energy eigenvector to the operator algebraic definition.

\vspace{0.2cm}

Section \ref{sec:finite_quadratic} considers finite chains with Hamiltonians quadratic 
in the creation and annihilation operators. In this setting, the $\mathbb{Z}_2$-index is 
defined as the homotopy type of a Bogoliubov transformation that diagonalises the 
Hamiltonian. The example of the Kitaev Hamiltonian is studied in detail. While the ground state $\mathbb{Z}_2$-index 
can in principle be defined for any positive quadratic Hamiltonian, it is in general much 
easier to compute for closed chains  with periodic or anti-periodic boundary 
conditions. For chains with open boundary conditions, different phases can be 
differentiated by 
the existence or non-existence of Majorana boundary states. 
We also show that the $\mathbb{Z}_2$-valued spectral flow gives a topological 
obstruction for two Hamiltonians to have the same $\mathbb{Z}_2$-index.  
The Martingale method is also used to show a uniformly bounded ground state energy gap 
for a large class of model Hamiltonians. For the case of closed chains, the 
insertion of a flux can close this gap and implement a non-trivial 
$\mathbb{Z}_2$-valued spectral flow.
The Kitaev chain with twisted boundary conditions is such an example.

\vspace{0.1cm}

Higher order interactions on  finite chains are studied in Section \ref{sec:higher_order}. 
A $\mathbb{Z}_2$-index for higher order interactions cannot be directly defined, but 
one can instead consider the ground state parity or Hamiltonians 
that can be connected to quadratic systems by a $C^1$-path with a uniformly bounded ground state gap.
We mostly focus on the solvable Kitaev Hamiltonian with a quartic interaction 
studied in~\cite{KST}.
We consider a closed chain, where a local $\pi$-flux will induce a $\mathbb{Z}_2$-phase 
change of ground states with a uniformly bounded ground state energy gap.

\vspace{0.2cm}

Section \ref{sec:quasifree_intro}  considers infinite systems 
and ground states of the CAR algebra that come from quasifree dynamics, where 
equivalence of quasifree states is determined by a Hilbert-Schmidt condition. 
This condition is used to derive a $\mathbb{Z}_2$-index map for Bogoliubov 
transformations between systems with different quasifree dynamics. This infinite $\mathbb{Z}_2$-index 
gives a natural generalisation of the $\mathbb{Z}_2$-index defined for finite quadratic chains. 
As in the finite-dimensional case, 
the $\mathbb{Z}_2$-valued spectral flow gives a topological obstruction for two ground states 
to have the same index. In particular, a non-trivial $\mathbb{Z}_2$-valued spectral flow between 
gapped quasifree ground states will cause the ground state gap of the infinite GNS Hamiltonian to close.

\vspace{0.1cm}

Finally, a $\mathbb{Z}_2$-index is defined in Section \ref{sec:general_Z2} for a class of pure and parity-invariant states of 
the CAR algebra of a one-dimensional lattice. 
We first review the Jordan--Wigner transform and show how for quasifree states the 
$\mathbb{Z}_2$-index is connected to the purity of the ground state of the Pauli algebra of spins. 
This example then motivates our more general definition of the $\mathbb{Z}_2$-phase label, 
which we show is well-defined for pure states satisfying the split property. 
The new $\mathbb{Z}_2$-index does not arise as a skew-adjoint Fredholm operator with a 
$\ZM_2$-index in general, but the two 
indices coincide when they are both defined. 
Elementary
properties of this new $\mathbb{Z}_2$-index are then shown, in particular that the 
ground state gap must close on paths 
connecting ground states of differing phase label. We conclude with some comments on 
future research directions.

\section{Preliminaries} \label{sec:prelim}

\subsection{Ground states of fermionic systems}

We will assume some familiarity with the $C^*$-algebraic approach to quantum statistical 
mechanics. A standard 
reference is~\cite{BR1, BR2}. An overview of modern techniques can be found in~\cite{NSY18}. 
We first recall the CAR algebra for general (potentially infinite) systems. 
Let $\mathcal{H}$ be a separable Hilbert space. The CAR algebra 
$A^\mathrm{car}(\calH)$ is the $C^*$-algebra generated by the 
identity and elements $\fraka(v)$, $v\in \calH$ such that 
$v\mapsto \fraka(v)$ is anti-linear and with anti-commutation relations 
\begin{align*}
 &\{ \fraka(v_1) , \fraka(v_2) \} \;=\; 0 \; , 
 &&\{ \fraka(v_1), \fraka(v_2)^* \} \;=\; \langle v_1, v_2 \rangle_{\calH} \; .
\end{align*}
If $\calH = \ell^2(\Lambda)$ for $\Lambda$ a countable set, then by taking the 
standard basis $\{\delta_j\}_{j \in \Lambda}$ of $\ell^2(\Lambda)$, we can 
simplify the  definition of $A^\mathrm{car}_\Lambda =A^\mathrm{car}(\ell^2(\Lambda))$ 
as the universal $C^*$-algebra  generated by the elements $\{ \fraka_{j}\}_{j\in \Lambda}$ with 
$\{\fraka_j , \fraka_k\} = 0$ and 
$\{\fraka_j , \fraka_k^*\} = \delta_{j,k} \,\one$, see~\cite[Section 5.2.2]{BR2} for example.

\vspace{0.2cm}

If $\Lambda' \subset \Lambda$ there is a natural embedding 
$A^\mathrm{car}_{\Lambda'} \subset A^\mathrm{car}_{\Lambda}$. 
In particular, if we let $\mathcal{P}_0(\Lambda)$ denote the set of 
finite subsets of $\Lambda$, there is the quasilocal structure 
$$
 A^\mathrm{car}_\Lambda \;\cong \; \overline{ (A^\mathrm{car}_\Lambda)_\mathrm{loc.} }^{\,C^*} \; , \qquad 
 (A^\mathrm{car}_\Lambda)_\mathrm{loc.} \;=\; \bigcup_{X \in \mathcal{P}_0(\Lambda)} A^\mathrm{car}_X \;.
$$ 
The CAR algebra $A^\mathrm{car}(\calH)$ comes equipped with the parity automorphism $\Theta$ defined by
$$
\Theta(\fraka(v))\; =\; -\fraka(v)\;,
\qquad
\Theta(\fraka(v)^*) \;=\; -\fraka(v)^*\;,
\qquad v\in \calH\;.
$$
One has $\Theta^2= \mathrm{Id}$.  If $\calH = \ell^2(\Lambda)$, then by the 
quasilocal structure $\Theta$ is the unique extension of the 
automorphism $\Theta_X$, $X \in \mathcal{P}_0(\Lambda)$, such that 
$$
  \Theta_X(a) \;=\; \Pp \, a\,  \Pp\; , \qquad \Pp = (-1)^{\sum_{j\in X} \fraka_j^* \fraka_j } \;
$$
for all $a \in A^\mathrm{car}_X$, see Section \ref{subsec:finite_parity}. The parity gives a 
decomposition $A^\mathrm{car}(\calH) \cong A^\mathrm{car}(\calH)^0 \oplus A^\mathrm{car}(\calH)^1$, 
where $\Theta(a) = (-1)^j a$ for $a \in A^\mathrm{car}(\calH)^j$. Elements in 
$A^\mathrm{car}(\calH)^0$ and $A^\mathrm{car}(\calH)^1$ are called even and odd respectively.

\vspace{0.2cm}

Let us now restrict our attention to $\calH = \ell^2(\Lambda)$ and $A^\mathrm{car}_\Lambda$. 
An \emph{interaction} $\Phi$ for a fermionic lattice is a map 
$\Phi : \mathcal{P}_0(\Lambda) \to A_\Lambda^\mathrm{car}$ such that 
$\Phi(X)^* = \Phi(X)$ for all $X \in \mathcal{P}_0(\Lambda)$. An interaction is called even
if its range is in $(A_\Lambda^\mathrm{car})^0$.
{Even} interactions are 
much better behaved with respect to Lieb--Robinson bounds, see~\cite{BSP,NSY17}.

\vspace{0.2cm}

Given an interaction $\Phi$ and a finite set $X$, one can define the 
local Hamiltonian 
$$
   \mathbf{H}_X^\Phi \;=\; \sum_{Y \subset X} \Phi(Y) \; .
$$
An even interaction $\Phi$ is called frustration-free if $\Phi$ 
has finite range and for all $X \in\mathcal{P}_0(\Lambda)$ 
$$
  \inf\,\sigma(\mathbf{H}_X^\Phi) \;=\; \sum_{Y \subset X} \inf\, \sigma(\Phi(Y)) \; .
$$
That is, the ground state of $\mathbf{H}_X^\Phi$ is simultaneously a ground state 
of all $\Phi(Y)$, $Y\subset X$.

\vspace{0.2cm}

While one can only define the Hamiltonian of an interaction on finite subsets, 
the infinite system can be studied by examining the dynamics generated by the Hamiltonian 
$$
  \beta_t^X( a) \;=\;    e^{\ii t \mathbf{H}_X} a e^{-\ii t \mathbf{H}_X} \; , \qquad t \in \mathbb{R} \,,\;\;
  a \in A^\mathrm{car}_X \; .
$$
As $X $ converges to $\Lambda$, one can guarantee that $\beta_t^X$ converges to a strongly continuous automorphism 
$\beta_t \in \mathrm{Aut}(A^\mathrm{car}_\Lambda)$ for all $t\in\mathbb{R}$ if the interaction $\Phi$ satisfies the 
(fermionic) Lieb--Robinson bound~\cite{NSY17,BSP}. To obtain such bounds, we require the set 
$\Lambda$ to have a metric and our interaction to have mild decay properties as the distance between 
points increases. If $\Lambda = \mathbb{Z}^\nu$ and 
the  interaction is finite range with a uniform bound on the coefficients, 
then the automorphism $\beta_t$ exists 
for all $t\in \mathbb{R}$.

\vspace{0.2cm}

Let us now fix an infinite dynamics, {\it i.e.} a strongly continuous map 
$\beta: \mathbb{R} \to \mathrm{Aut}(A^\mathrm{car}_\Lambda)$. 
A state  is a positive and continuous linear functional 
$\omega: A^\mathrm{car}_\Lambda\to \mathbb{C}$ such that 
$\omega( \one_{A^\mathrm{car}_\Lambda} ) = \one_\mathbb{C}$. 
Let $\delta$ be the generator of the dynamics $\beta$. Then 
$\omega$ is by definition a ground state on $A^\mathrm{car}_\Lambda$ with respect to $\beta$ if 
\begin{equation} \label{eq:GS_defn}
     -\ii \,  \omega\big( a^* \delta(a) \big) \;\geq\; 0\; ,   \quad 
    a \in \mathrm{Dom}(\delta) \; . 
\end{equation}
The set of ground states with respect to a fixed action $\beta$ forms a convex and 
compact set with respect to the weak $\ast$-topology.

\vspace{0.1cm}

One can also consider the GNS triple $( \pi_\omega, \mathfrak{h}_\omega, \Omega_\omega)$ 
associated to a ground state $\omega$. Equation \eqref{eq:GS_defn} implies that 
$\omega\circ \beta_t = \omega$ for all $t\in \mathbb{R}$. Therefore, there is a unitary 
operator $U_{\beta_t}$ on $\mathfrak{h}_\omega$ such that 
$\pi_\omega\circ \beta = \mathrm{Ad}_{U_{\beta_t}} \circ \pi_\omega$. Hence we obtain 
a $1$-parameter group of unitaries acting on $\mathfrak{h}_\omega$.  Thus, 
applying Stone's theorem, there is a self-adjoint operator $h_\omega$ such that 
$$
  e^{\ii t h_\omega} \pi_\omega(a) e^{-\ii th_\omega} \;=\; \pi_\omega( \beta_t(a) ) \; , \qquad 
  e^{\ii t h_\omega} \Omega_\omega \;=\; \Omega_\omega \; ,
$$
which implies that $\Omega_\omega$ is a $0$-energy eigenvector for $h_\omega$. 
Furthermore, Equation \eqref{eq:GS_defn} implies that $h_\omega \geq 0$ so $\Omega_\omega$ is 
a minimal eigenvector for $h_\omega$.

\begin{defini}
A ground state $\omega$ on $(A^\mathrm{car}_\Lambda, \beta)$ is called gapped if 
there is a constant $\gamma>0$ such that $\sigma(h_\omega)\cap (0,\gamma) = \emptyset$.
\end{defini}

For a unique ground state $\omega$, the property of being gapped  is equivalent (see {\it e.g.}~\cite{Matsui13}) to the condition that there is a $\gamma>0$ such that 
$$
  -\ii \,  \omega\big( a^* \delta(a) \big)  \;\geq\;  \gamma 
   \big( \omega(a^*a) - | \omega(a)|^2 \big) \; ,  \qquad a \in (A^\mathrm{car}_\Lambda)_\mathrm{loc.} \; . 
$$

\begin{proposi}[\cite{NSY17}] \label{prop:finite_uniform_gap_gives_GNS_gap}
Let $X\in \mathcal{P}_0(\Lambda)$ and $\mathbf{H}_X^\Phi$ be a finite-range Hamiltonian 
satisfying a Lieb--Robinson bound. If the spectral gap between lowest-energy eigenvalue of 
$\mathbf{H}_X^\Phi$ and the next-lowest eigenvalue is \emph{uniformly bounded} in $|X|$, 
then the weak $\ast$-limit of the finite-volume ground states is a 
gapped ground state on $A^\mathrm{car}_\Lambda$.
\end{proposi}

Suppose that $\omega$ is a $\Theta$-invariant state on $A^\mathrm{car}_\Lambda$, 
namely $\omega \circ \Theta = \omega$.
Then there exists a self-adjoint unitary $\Sigma$ on the GNS space $\mathfrak{h}_\omega$ with 
the properties 
\begin{align*}
  &\Sigma \pi_\omega(a)\Sigma \;=\; \pi_\omega(\Theta(a)) \; , 
  &&\Sigma \,\Omega_\omega \;=\; \Omega_\omega \; .
\end{align*}
Furthermore, we can decompose the  GNS space 
$$
  \mathfrak{h}_\omega \;=\; \mathfrak{h}_\omega^0 \oplus \mathfrak{h}_\omega^1 \; , \qquad 
  \mathfrak{h}_\omega^i \;=\; \frac{1}{2}(1+(-1)^i \Sigma)\mathfrak{h}_\omega \;=\; 
      \overline{\pi_\omega((A^\mathrm{car}_\Lambda)^i) \Omega_\omega } \; .
$$
If the  system is finite and $\omega_X$ on $A^\mathrm{car}_X$ is given by 
$\omega_X(a) = \langle \psi | a | \psi \rangle$, then  $\omega_X$ 
is parity invariant if $|\psi\rangle$ is even or odd under $\Pp$. 
In particular, a parity-invariant 
state on $A^\mathrm{car}_X$ need not come from only even lowest-energy eigenvectors.

\subsection{The \texorpdfstring{$\mathbb{Z}_2$}--valued spectral flow} \label{sec:Z_2_prelims}

We now review the $\mathbb{Z}_2$-valued spectral flow defined in~\cite{CPS} as a 
real analogue of the $\mathbb{Z}$-valued spectral flow defined  by Atiyah--Patodi--Singer~\cite{APS} 
and developed by Phillips~\cite{Phillips}. The $\mathbb{Z}$-valued spectral flow gives a concrete 
expression for the isomorphism $\pi_1( \mathrm{Fred}_\ast^\mathrm{sa}(\calH_\mathbb{C})) \cong \mathbb{Z}$ with 
$\mathrm{Fred}_\ast^\mathrm{sa}(\calH_\mathbb{C})$ the self-adjoint Fredholm operators on a 
complex Hilbert space and with essential spectrum above and below $0$. In contrast, 
the $\mathbb{Z}_2$-valued spectral flow measures the isomorphism 
$\pi_1(\mathrm{Fred}_{\ast}^\mathrm{sk}(\calH_\mathbb{R})) \cong \mathbb{Z}_2$ with 
$\mathrm{Fred}_{\ast}^\mathrm{sk}(\calH_\mathbb{R})$ the skew-adjoint Fredholm operators 
on a real Hilbert space with essential spectrum above and below the real axis.

\vspace{0.2cm}

Unfortunately, the term `spectral flow' already appears in the study of stability 
properties of gapped ground states~\cite{NSY18}. This spectral flow is distinct from 
the spectral flow considered by Atiyah--Patodi--Singer and Phillips. 
In this work, we will only focus on the $\mathbb{Z}_2$-valued spectral flow and 
to reduce ambiguity will always include the $\mathbb{Z}_2$ in the 
terminology.

\subsection*{Finite dimensions}

Let $\mathbb{R}^N$ be a real finite-dimensional Hilbert space with 
$T_0$ and $T_1$ invertible skew-adjoint matrices. 
By standard results in linear algebra, there exists an invertible matrix 
$A \in \mathrm{GL}(\mathbb{R}^N)$ such that $T_1 = A T_0 A^*$. The 
$\mathbb{Z}_2$-valued spectral flow detects if the orientation of the 
eigenvectors are inverted along the straight-line path connecting $T_0$ to $T_1$. 

\begin{defini}
Let $T_0$ and $T_1$ be invertible skew-adjoint operators on a finite-dimensional real 
Hilbert space and let $T_1 = A T_0 A^*$ with invertible $A$. The $\mathbb{Z}_2$-valued spectral 
flow of the straight-line path is given by 
$$
  \mathrm{Sf}_2( t\in [0,1]\mapsto (1-t) T_0 + t T_1) \;=\; \mathrm{sgn}\,\mathrm{det}(A)
  \;\in\;\ZM_2=\{-1,1\} \; .
$$
It is also simply denoted by $\mathrm{Sf}_2( T_0,T_1)$.
\end{defini}


While the $\mathbb{Z}_2$-valued spectral flow is defined on a real Hilbert space, 
we can also consider operators on complex Hilbert spaces that respect a fixed real structure.

\begin{remark} \label{rk:Z2_spec_flow_explanation}
{\rm
Let us give more justification for the name $\mathbb{Z}_2$-spectral flow. In the case of 
a complex Hilbert space, the
$\mathbb{Z}$-valued spectral flow counts the eigenvalue crossings though $0$ (with 
sign) of 
paths of self-adjoint matrices or Fredholm operators. 
In the case of skew-adjoint matrices and Fredholm operators on real Hilbert spaces, 
there is a symmetry of the spectrum about the real axis, $\overline{\sigma(T)} = \sigma(T)$. 
In particular, any eigenvalue 
crossings through $0$ will be double degenerate and the $\mathbb{Z}$-valued spectral 
flow will vanish. Instead the $\mathbb{Z}_2$-valued spectral flow measures if  
there is a parity change of the eigenvectors at the double degenerate 
crossing points. See~\cite{CPS} for more information.
}
\hfill $\diamond$
\end{remark}

\subsection*{Infinite dimensions}

We follow the approach of~\cite[Section 5-6]{CPS}. Fix a separable and 
real Hilbert space $\calH_\mathbb{R}$. A \emph{complex structure} on a real Hilbert 
space is a skew-adjoint unitary 
$$
  J \in \mathcal{B}(\calH_\mathbb{R}) \; , \qquad J^* \;=\; -J \; ,  \qquad  J^2 \;=\; -\one_\calH \; .
$$
We define the $\mathbb{Z}_2$-valued spectral flow via a $\mathbb{Z}_2$-index map on 
pairs of skew-adjoint unitaries. To set notation, given the real Hilbert space 
$\calH_\mathbb{R}$, we let $\mathcal{O}(\calH_\mathbb{R})$ be the orthogonal operators 
on $\calH_\mathbb{R}$, 
$\mathcal{K}(\calH_\mathbb{R})$ be the compact operators and 
$\mathcal{Q} = \mathcal{B}(\calH_\mathbb{R})/ \mathcal{K}(\calH_\mathbb{R})$ the Calkin algebra.

\begin{proposi}[\cite{CPS}, Proposition 5.2] \label{prop:Z_2_unitary_map}
Consider the space 
$$
  \mathcal{J}(\calH_\mathbb{R}) \;=\; 
  \big\{ (J_0,J_1) \in \mathcal{O}(\calH_\mathbb{R}) \,:\, J_i^* \;=\; -J_i \, , \; \|J_0 - J_1\|_{\mathcal{Q}} < 2 \big\} \; 
$$
with the norm topology. The map 
$$
\mathcal{J}(\calH_\mathbb{R}) \ni (J_0,J_1) \; \mapsto \; 
  \mathrm{Ind}_2(J_0, J_1) \;=\; (-1)^{\frac{1}{2}\, \mathrm{dim}\,\mathrm{Ker}(J_0 + J_1)} 
\; \in \mathbb{Z}_2 \; 
$$
is continuous.
\end{proposi}

The above proposition is stated in~\cite{CPS} with the bound $\|J_0 - J_1\|_{\mathcal{Q}} < \tfrac{1}{2}$, 
but we note that the result holds for $\|J_0 - J_1\|_{\mathcal{Q}} <2$, see~\cite[Proposition 4.3]{BCLR} 
or~\cite[Section 5]{DSBW} for a proof.

\vspace{0.2cm}

If $\calH_\mathbb{R}$ is finite-dimensional, then any pair of complex structures 
$(J_0,J_1)$ is an element of $\mathcal{J}(\calH_\mathbb{R})$ and 
$$
  (-1)^{\frac{1}{2} \mathrm{dim}\,\mathrm{Ker}(J_0 + J_1)} \;=\; \mathrm{sgn}\,\mathrm{det}(A) \; , \qquad 
  J_1 \;=\; A J_0 A^* \;.
$$
Therefore the  $\mathbb{Z}_2$-index map recovers the finite-dimensional 
$\mathbb{Z}_2$-valued spectral flow.

\vspace{0.2cm}

Now consider a norm-continuous path $[0,1]\ni t\mapsto T_t \in \mathrm{Fred}_{\ast}^\mathrm{sk}(\calH_\mathbb{R})$ 
with $T_0$ and $T_1$ invertible. One can consider the path $J_t = T_t|T_t|^{-1}$, where if $T_{t_0}$ has a 
non-trivial kernel, $J_{t_0}$ is completed by an arbitrary complex structure on its kernel to give a path of complex 
structures in $\mathcal{B}(\calH_\mathbb{R})$. The path $J_t$ is not continuous 
 in $\mathcal{B}(\calH_\mathbb{R})$ but \emph{is}  continuous in $\mathcal{Q}$. The $\mathbb{Z}_2$-index 
map from Proposition \ref{prop:Z_2_unitary_map} is now used to define the $\mathbb{Z}_2$-valued spectral flow.

\begin{defini}
Let $\{T_t\}_{t\in [0,1]}$ be a norm-continuous path in $\mathrm{Fred}_\ast^\mathrm{sk}(\calH_\mathbb{R})$ 
with $T_0$ and $T_1$ invertible. Let $J_t = T_t|T_t|^{-1}$ and partition the interval $0=t_0 < t_1 < \cdots < t_n = 1$ 
such that 
$\| J_{t_j} - J_{t_{j-1}} \|_\mathcal{Q} < 2$. The $\mathbb{Z}_2$-valued spectral flow 
is given by 
$$
   \mathrm{Sf}_2(t\in [0,1]\mapsto T_t) \;=\; 
  (-1)^{\sum_{j=1}^n \frac{1}{2}\, \mathrm{dim}\,\mathrm{Ker}( J_{t_{j-1}} + J_{t_j} )}\;\in\;\ZM_2=\{-1,1\} \;. 
$$
\end{defini}

\vspace{0.1cm}

Let us list the key properties of the $\mathbb{Z}_2$-valued spectral flow.

\begin{theo}[\cite{CPS}]
\begin{enumerate}
  \item[{\rm (i)}] The map $\mathrm{Sf}_2$ is independent of the choice of partition in the definition.
  \item[{\rm (ii)}] {\rm (Concatenation)} If $\{T_t\}_{t\in[0,1]}$ and $\{T_t\}_{t\in [1,2]}$ are continuous paths in 
  $\mathrm{Fred}_\ast^\mathrm{sk}(\calH_\mathbb{R})$ with invertible endpoints, then  
  $$
     \mathrm{Sf}_2(t\in [0,2]\mapsto T_t) \;=\;  \mathrm{Sf}_2( t\in [0,1]\mapsto T_t)\, \times \,  \mathrm{Sf}_2( t\in [1,2]\mapsto T_t ) \; .
  $$
  \item[{\rm (iii)}] {\rm (Homotopy invariance)} Let $\{T_t\}_{t\in[0,1]}$ and $\{\tilde{T}_t\}_{t\in[0,1]}$ be continuous paths in 
   $\mathrm{Fred}_{\ast}^\mathrm{sk}(\calH_\mathbb{R})$ with invertible endpoints such that $T_0 = \tilde{T}_0$ and $T_1 = \tilde{T}_1$. 
   If the two paths are connected by a continuous homotopy leaving endpoints fixed, then 
   $\mathrm{Sf}_2(t\in [0,1]\mapsto T_t) = \mathrm{Sf}_2(t\in [0,1]\mapsto \tilde{T}_t)$. 
   \item[{\rm (iv)}] The map $\mathrm{Sf}_2$ on loops in $\mathrm{Fred}_{\ast}^\mathrm{sk}(\calH_\mathbb{R})$ is a homotopy 
   invariant and induces an isomorphism $\pi_1(\mathrm{Fred}_{\ast}^\mathrm{sk}(\calH_\mathbb{R})) \cong \mathbb{Z}_2$.
\end{enumerate}
\end{theo}

\vspace{0.1cm}

Lastly, let us note that there is also an isomorphism 
$\pi_1(\mathrm{Fred}_{\ast}^\mathrm{sk}(\calH_\mathbb{R}))\cong KO^{-2}(\mathrm{pt})$~\cite{ASSkew}. 
Hence the $\mathbb{Z}_2$-valued spectral flow also has a $K$-theoretic interpretation. 


\section{Finite quadratic chains} \label{sec:finite_quadratic}

\subsection{Basic setup} \label{sec:basic_setup}

In this section, $\Lambda$ will denote a finite set with cardinality $|\Lambda|$. 
We consider the fermionic Fock space 
$\Ff_\Lambda=\Ff(\mathbb{C}^{|\Lambda|})$ of antisymmetric tensors in the full Fock space 
$\bigoplus_n ( \mathbb{C}^{|\Lambda|})^{\otimes n}$. 
For any $j\in\Lambda$, the creation and annihilation operators, $\fraka_j^*$ and $\fraka_j$, 
satisfy  the anticommutation relations
$$
\{\fraka_j^*,\fraka_i\}\;=\;\delta_{i,j}\,\one
\;,
\qquad
\{\fraka_j,\fraka_i\}\;=\;0
\;.
$$
A standard way to rewrite the Fock space  is
$$
\Ff(\mathbb{C}^{|\Lambda|})
\;\cong\;
\hat{\otimes}_{j\in\Lambda}\Ff(\ell^2(\{j\}))
\;\cong\;
\CM^2\hat{\otimes}\cdots\hat{\otimes}\CM^2
\; .
$$
Here $\hat\otimes$ is the $\mathbb{Z}_2$-graded tensor product of $\mathbb{Z}_2$-graded vector 
spaces, where for $V \cong V^0 \oplus V^1$ and $W \cong W^0 \oplus W^1$, 
$V\hat\otimes W$ is $\mathbb{Z}_2$-graded with 
$$
  (V\hat\otimes W)^0 \;\cong\; V^0 \otimes W^0 \oplus V^1 \otimes W^1 \; , \qquad 
  (V\hat\otimes W)^1 \;\cong \; V^0 \otimes W^1 \oplus V^1 \otimes W^0 \; .
$$
Returning to the fermionic Fock space, 
$\Ff(\ell^2(\{j\}))=\CM^2$ consists of two states, one is the empty and one 
the occupied state given by $|\Omega_j\rangle$ and $\fraka_j^*|\Omega_j\rangle$ 
respectively. The vacuum of the whole chain is then $|\Omega\rangle=\hat{\otimes}_{j\in\Lambda}|\Omega_j\rangle$.

\vspace{.2cm}

For the time being, we will restrict ourselves to Hamiltonians on $\Ff_\Lambda=\Ff(\mathbb{C}^{|\Lambda|})$ 
that are quadratic in the creation and annihilation operators, {\it i.e.}
$$
  \mathbf{H}_\Lambda \;=\; \sum_{j,k \in \Lambda}  h_{j,k}\fraka_j^* \fraka_k \;+\; \tilde{h}_{j,k} \fraka_{j}\fraka_{k} \;+\; \mathrm{Adjoint}\;.
$$
There there is a Bogoluibov--de Gennes (BdG) representation of this Hamiltonian. 
Introducing the column vectors $\fraka=(\fraka_j)_{j\in\Lambda}$ and $\fraka^*=(\fraka_j^*)_{j\in\Lambda}$ 
one then has the formal equation
\begin{equation}
\label{eq-BdGRep}
\HH_\Lambda
\;=\;
\frac{1}{2}
 \begin{pmatrix} \fraka^* & \fraka \end{pmatrix} 
H_\Lambda
\begin{pmatrix}
\fraka \\ \fraka^*
\end{pmatrix}
\; +\; \mathrm{Tr}(h)\, \one_{\Ff_\Lambda}
\;.
\end{equation}
We will neglect the constant $\mathrm{Tr}(h)\, \one_{\Ff_\Lambda}$ as it is, at most, 
a shift in energy.
The BdG Hamiltonian $H_\Lambda$ acts on the particle-hole space $\Hh_\ph=\ell^2(\Lambda)\otimes \CM^2$ 
and automatically has the (even) particle-hole symmetry (PHS)
\begin{equation}
\label{eq-BdGsymmetry}
K^*\,\overline{H_\Lambda}\,K
\;=\;
-H_\Lambda \;, \qquad K=\one \otimes \sigma_1 
\;,
\end{equation}
This means, in particular, that the off-diagonal entry of the BdG Hamiltonian 
is an anti-symmetric matrix. 

Suppose that $\phi\in\Hh_\ph$ is a non-vanishing zero-energy eigenvector of $H_\Lambda$. 
Such a vector $\phi$  necessarily satisfies 
$K\overline{\phi}=\phi$ (after a phase was absorbed). Associated to this vector is  an operator
$$
\frakb_\phi
\;=\;
\phi^t
\begin{pmatrix}
\fraka \\ \fraka^*
\end{pmatrix}
\;,
$$
where $\phi^t=(\overline{\phi}\,)^*$ is the transpose. The operator $\frakb_\phi$  
is self-adjoint and squares to $\one$ if $\|\phi\|=1$. Thus $\frakb_\phi$ is a so-called Majorana operator. 
By construction, it commutes with $\HH_\Lambda$. For kernels with degeneracy, Majorana 
operators can be constructed for each zero-energy state. 

\subsection{Bogoliubov transformation}
\label{sec-Bogoliobov}

We recall methods for diagonalising  quadratic Hamiltonians by canonical transformations 
following standard treatments, {\it e.g.} \cite{BR} or \cite{DS3}. The  PHS \eqref{eq-BdGsymmetry} 
of the Hamiltonian can be interpreted as follows: $\ii\,H$ is in the Lie algebra of the group
$$
\Gg\;=\;\left\{A\in \mbox{\rm GL}(\Hh_{\mbox{\tiny\rm ph}})\, : \, K^*\, \overline{A}\,K=A\right\}
\;.
$$
Let $\Uu_\ph =\Gg\cap \mathcal{U}(\Hh_{\mbox{\tiny\rm ph}})$ denote the unitaries in this group: 
$$
\Uu_\ph \;=\;\left\{W\in \mbox{\rm GL}(\Hh_{\mbox{\tiny\rm ph}})\, : \,  W^*=W^{-1}\;,\;\;
K^*\,\overline{W}\,K=W\right\}
\;.
$$
We remark that the group $\Uu_\ph$ is naturally isomorphic to the orthogonal matrices 
on the \emph{real} Hilbert space $\calH_\ph^\mathbb{R} = \{ \psi \in \calH_\ph\,:\, K \ol{\psi} = \psi \}$. 
Namely, for $\mathcal{O}_n$ the set of $n\times n$ real and orthogonal matrices
\begin{equation} \label{eq:CU_phC_is_On}
C^*\,\Uu_\ph \,C\;=\;
 \mathcal{O}_{2L}
\;,
\qquad
C\;=\;
2^{-\frac{1}{2}}\begin{pmatrix}
\one & \ii\,\one \\
\one & -\ii\,\one
\end{pmatrix}
\;
\end{equation}
by means of relations $C^*=C^{-1}$ and $C^TC=K$ with $K$ as in \eqref{eq-BdGsymmetry}. 
Now, given $W\in\Uu_\ph$, one can define
\begin{equation}
\label{eq-Bogtrafo}
\begin{pmatrix} 
\frakd \\
 \frakd^* 
\end{pmatrix}
 \;=\;
 W
\begin{pmatrix} 
\fraka \\
 \fraka^* 
\end{pmatrix}
\;.
\end{equation}
The particular form of $W$ assures that $\frakd$ and $\frakd^*$ are indeed mutually adjoint and that the 
CAR relations for $\frakd$ and $\frakd^*$ hold. 
A standard question is now whether \eqref{eq-Bogtrafo} can be implemented by a unitary opertor 
${\bf U}_W$ on Fock space in the sense that 
$$
\frakd\;=\;{\bf U}_W^*\, \fraka\, {\bf U}_W
\;.
$$ 
(Note that ${\bf U}_W$ is not quadratic in $\fraka$.) For a finite system, this is always possible, 
but in infinite dimension one has to impose a condition. It is sufficient to require off-diagonal 
entries of $W$ to be Hilbert-Schmidt~\cite{ShaleStinespring,NNS}. Then the unitary ${\bf U}_W$ is called a 
Bogoliubov transformation, while $W$ is usually called the associated canonical transformation. 
Hence $\Uu_\ph$ is also called the group of canonical transformations. 

\vspace{.2cm}

Now, suppose that $|\Lambda| =L<\infty$ and $\mathbf{H}_\Lambda$ has the eigenvalues 
$\{E_1,E_2,\ldots, E_L\}$ with $0\leq E_1 \leq \cdots \leq E_L$ (taking a shift if necessary to 
ensure that all eigenvalues are non-negative). 
Then the BdG Hamiltonian $H_\Lambda$ can be diagonalised by a 
canonical transformation $W \in \Uu_\ph$, see e.g.~\cite{BR, DS3}, 
\begin{equation}
\label{eq-Hdiag0}
W\, H_\Lambda\, W^*
\;=\;
\left(\begin{array}{cc}
E & 0\\
   0 & -E
 \end{array}\right)
\;, \qquad 
E \;=\;  \begin{pmatrix} E_{1} & & \\ & \ddots & \\ & & E_{L} \end{pmatrix}\;.
\end{equation}
Using this particular canonical transformation, one has
\begin{align}
{\bf H}_\Lambda
\;
&=\;
\frac{1}{2}\;
\begin{pmatrix}
\fraka^* &  \fraka
\end{pmatrix}
W^*WH_\Lambda
W^*W
\begin{pmatrix}
\fraka \\ \fraka^*
\end{pmatrix}
\nonumber
\\
\label{eq-Hdiag}
& =\;
\frac{1}{2}\;
\begin{pmatrix}
\frakd^* &  \frakd
\end{pmatrix}
\begin{pmatrix}
E & 0\\
   0 & -E 
\end{pmatrix}
\begin{pmatrix}
\frakd \\ \frakd^*
\end{pmatrix}
\\
& =\;
\frac{1}{2}\;
\UU_W^*
\begin{pmatrix}
\fraka^* &  \fraka
\end{pmatrix}
\begin{pmatrix}
E & 0\\
   0 & -E 
\end{pmatrix}
\begin{pmatrix}
\fraka \\ \fraka^*
\end{pmatrix}
\UU_W
\;.
\nonumber
\end{align}
Rewriting Equation \eqref{eq-Hdiag} using the CAR operations,
$$
\mathbf{H}_\Lambda \;=\; \sum_{j\in\Lambda} E_j\left( \frakd_j^* \frakd_j - \frakd_j\frakd_j^* \right) \;=\; 
\sum_{j\in\Lambda} E_j \left(2\frakd_j^*\frakd_j - 1\right) \;.
$$
Therefore, because $E_j\geq 0$, any vector that is eliminated by all the $\frakd_j$ 
with $E_j > 0$  
is a ground state of $\mathbf{H}_\Lambda$.  In particular, if 
$\frakd_1\frakd_2\cdots\frakd_L | \psi \rangle$ is non-zero, then it is 
a non-trivial ground state of $\mathbf{H}_\Lambda$. Using Lemma \ref{lem-vanishing} 
below, it can be shown that such non-zero vectors exist. 

\subsection{Majorana representation}
Recall Equation \eqref{eq:CU_phC_is_On}, where 
$C^* \Uu_\ph C\;=\; \mathcal{O}_{2L}$.
Let us extend this idea slightly and include a phase factor. 
Define 
$$
\frakb_{2j-1}
\;=\;
e^{\ii\frac{\theta}{2}}\fraka_j+
e^{-\ii\frac{\theta}{2}}\fraka_j^*
\;,
\qquad
\frakb_{2j}
\;=\;
-\ii\,e^{\ii\frac{\theta}{2}}\fraka_j+
\ii\,e^{-\ii\frac{\theta}{2}}\fraka_j^*
\;,
$$
for all $j\in\Lambda$. They satisfy the Clifford relations
$$
\frakb_j^*\;=\;\frakb_j
\;,
\qquad
\{\frakb_j,\frakb_i\}\;=\;2\,\delta_{i,j}\,\one
\;,
$$
and one readily checks
\begin{align}
\label{eq-Crel}
&
\frakb_{2j-1}\frakb_{2j}
\;=\;
2\,
\ii(-\fraka_{j}^*\fraka_j+\tfrac{1}{2}\one)
\;,
\nonumber
\\
& 
\frakb_{2j}\frakb_{2j+1}-\frakb_{2j-1}\frakb_{2j+2}
\;=\;
2\,
\ii(\fraka_{j+1}^*\fraka_j+\fraka_{j}^*\fraka_{j+1})
\;,
\\
&
\frakb_{2j}\frakb_{2j+1}+\frakb_{2j-1}\frakb_{2j+2}
\;=\;
2\,
\ii(e^{\ii\theta}\fraka_{j+1}\fraka_{j}+e^{-\ii\theta}\fraka_{j}^*\fraka_{j+1}^*)
\;.
\nonumber
\end{align}
This also implies
\begin{equation}
\label{eq-Crel2}
\ii\,\frakb_{2j}\frakb_{2j+1}
\;=\;
-\fraka_{j+1}^*\fraka_j-\fraka_{j}^*\fraka_{j+1} + e^{\ii\theta}\fraka_{j}\fraka_{j+1}+ e^{-\ii\theta}\fraka_{j+1}^*\fraka_{j}^*
\;.
\end{equation}

We can now write any quadratic Hamiltonian using the operators $\{\frakb_j\}$. 
Let $\frakb_\ev=(\frakb_{2j})_{j\geq 1}$ 
and $\frakb_\odd=(\frakb_{2j-1})_{j\geq 1}$ be the column vectors of Majorana's with even and 
odd index respectively with $\frakb=\binom{\frakb_\odd}{\frakb_\ev}$. Then
$$
\frakb
\;=\;
2^{\frac{1}{2}}\,C_\theta^*\,
\begin{pmatrix}
\fraka \\ \fraka^*
\end{pmatrix}
\;,
\qquad
C_\theta^*
\;=\;
2^{-\frac{1}{2}}
\begin{pmatrix}
e^{\ii\frac{\theta}{2}} & e^{-\ii\frac{\theta}{2}}
\\
-\ii \,e^{\ii\frac{\theta}{2}} & \ii\,e^{-\ii\frac{\theta}{2}}
\end{pmatrix}
\;=\;
C^*
\begin{pmatrix}
e^{\ii\frac{\theta}{2}} & 0
\\
0 & e^{-\ii\frac{\theta}{2}}
\end{pmatrix}
\;.
$$
One now obtains the Majorana representation of the Hamiltonian
\begin{equation}
\label{eq-HamMajoRep}
\HH_\Lambda
\;=\;  \sum_{j,k=1}^{2L} \alpha_{j,k} \frakb_j \frakb_k \;=\; 
\tfrac{\ii}{2}\,
\frakb^t\,A_\Lambda\,\frakb
\;,
\end{equation}
where the transpose $\frakb^t$ is a row vector and $A_\Lambda= -\frac{\ii}{2}\,C^*_\theta\,H_\Lambda\,C_\theta$ 
is real and skew-symmetric. 

\vspace{0.2cm}

Let us consider the diagonalisation of the operator $A_\Lambda= -\frac{\ii}{2}\,C^*_\theta\,H_\Lambda\,C_\theta$. 
Following standard treatments, \textit{e.g.}~\cite{BR}, 
there is an orthogonal 
matrix $V \in \calO_{2L}$, $V = C_\theta^* W C_\theta$ for $W \in \Uu_\ph$ such that 
\begin{equation}
\label{eq-AVdiag}
   V A_\Lambda V^* \;=\; \ \begin{pmatrix} 0 & E \\ -E & 0 \end{pmatrix} \; .
\end{equation}
Then 
\begin{align*}
  {\bf H}_\Lambda
\;=\; \frac{\ii}{2}\, \frakb^t V^*V A_\Lambda V^* V \frakb 
 \;=\; \frac{\ii}{2}\, \frakb^t V^*  \begin{pmatrix} 0 & E \\ -E & 0 \end{pmatrix}  V \frakb 
 \;=\; \frac{\ii}{2} \,\frakbtilde^t  \begin{pmatrix} 0 & E \\ -E & 0 \end{pmatrix}   \frakbtilde \;,
\end{align*}
where $\frakbtilde = V \frakb$ and $\{\frakbtilde_j\}_{j=1}^{2L}$ also satisfy the Clifford relations. Hence
$$
{\bf H}_\Lambda \;=\; \ii\, \sum_{j=1}^L E_j  \frakbtilde_{2j-1}\frakbtilde_{2j} \; 
$$
and the ground state space of ${\bf H}_\Lambda$ is  determined by the $-1$ eigenspaces 
of the commuting self-adjoint unitaries $\{\ii \frakbtilde_{2j-1} \frakbtilde_{2j} \}_{j=1}^L$. 
These eigenstates can be written out similar to the end of Section \ref{sec-Bogoliobov}.
Furthermore, we note that 
$$
   \mathrm{dim} \, \Ker(\mathbf{H}_\Lambda) \;=\; \tfrac{1}{2}\, \mathrm{dim}\, \Ker( A_\Lambda) \; .
$$

\subsection{Kitaev's \texorpdfstring{$\mathbb{Z}_2$}--index for finite quadratic Hamiltonians}
\label{sec:Z_2_index_def}

\begin{defini}[\cite{Kit}] \label{def:Z_2_index_def}
The Kitaev index of a strictly positive quadratic Hamiltonian $\mathbf{H}_\Lambda = \tfrac{\ii}{2} \frakb^t A_\Lambda \frakb$ is defined as the sign of the Pfaffian
$$
  j(\mathbf{H}_\Lambda)  \;=\;  \mathrm{sgn}\, \mathrm{Pf}(A_\Lambda) \; . 
$$
\end{defini}

Diagonalising the Hamiltonian as in \eqref{eq-AVdiag} and using properties of the Pfaffian,
$$
   \mathrm{Pf}(A_\Lambda)  \;=\;   \mathrm{det}(V) \,\, \mathrm{Pf} \begin{pmatrix} 0 & E \\ -E & 0 \end{pmatrix} 
   \;=\;   \mathrm{det}(V)  \, \prod_{j=1}^L E_j \; .
$$
If $\mathbf{H}_\Lambda$ is strictly positive (so $H_\Lambda$ has a 
spectral gap around $0$), then the Pfaffian is well-defined and its sign is determined by 
the sign of $\mathrm{det}(V)$. Furthermore, since $V = C_\theta^*WC_\theta$ for 
$W \in \Uu_\ph$ as in \eqref{eq-Hdiag0},
\begin{equation} \label{eq:Finite_Z2_index_formulae}
j(\mathbf{H}_\Lambda) \;=\;  \mathrm{sgn}\, \mathrm{Pf}(A_\Lambda)  \;=\;  \mathrm{sgn}\,\mathrm{det}(V)  
  \;=\; \mathrm{sgn}\,\mathrm{det}(W) \; ,
\end{equation}
which implies that $j(\mathbf{H}_\Lambda)$ is independent of the parameter $\theta$.

\begin{remark}
{\rm 
The Kitaev index is connected to the $\mathbb{Z}_2$-valued spectral flow in finite dimensions by 
\begin{equation}
\label{rk:finite_chain_index_as_sf}
j(\mathbf{H}_\Lambda) \;=\;  \mathrm{Sf}_2 \big(\ii H_\Lambda, W \ii H_\Lambda W^* \big) 
\;=\; \mathrm{Sf}_2 \big( A_\Lambda , V A_\Lambda V^* \big)\; ,
\end{equation}
as $\ii H_\Lambda$ is an invertible operator on the 
\emph{real} Hilbert space $\calH_\ph^\mathbb{R} = \{\psi \in \calH_\ph \,:\, K \ol{\psi} = \psi \}$.
\hfill $\diamond$
}
\end{remark}

\begin{proposi} \label{prop:sf_as_Z2_obstruction}
Let $\mathbf{H}_\Lambda(0)$ and $\mathbf{H}_\Lambda(1)$ be quadratic 
and strictly positive Hamiltonians on $\mathcal{F}(\mathbb{C}^L)$. Then 
$j(\mathbf{H}_\Lambda(0)) = j(\mathbf{H}_\Lambda(1))$ if and only if 
$\mathrm{Sf}_2( A_\Lambda(0) , A_\Lambda(1)) = 1$.
\end{proposi}
\noindent {\bf Proof.} 
Recall that $j(\mathbf{H}_\Lambda(k)) = \mathrm{sgn}\, \mathrm{det}(V_k) = 
\mathrm{sgn}\, \mathrm{Pf}(A_\Lambda(k))$ with $V_k$ the orthogonal 
matrix that diagonalises $A_\Lambda(k)$ for $k=0,\,1$. Because 
$\mathbf{H}_\Lambda(0)$ and $\mathbf{H}_\Lambda(1)$ are strictly positive, 
we can homotopy each $E_j$ to $1$ without changing the sign of the Pfaffian. 
Having flattened the spectrum of the Hamiltonians, both $A_\Lambda(0)$ and 
$A_\Lambda(1)$ will have the diagonal form 
$V_k A_\Lambda(k) V_k^* = J =  \one_L \otimes i\,\sigma_2$, $k=0,\,1$.
Thus the concatenation property of $\mathbb{Z}_2$-valued spectral flow implies that 
$$
  \mathrm{Sf}_2( A_\Lambda(0), A_\Lambda(1)) \;=\; \mathrm{Sf}_2( A_\Lambda(0), J) \,\mathrm{Sf}_2(J, A_\Lambda(1)) \; .
$$
Because 
$\mathrm{Sf}_2( A_\Lambda(k), J) = \mathrm{sgn}\,\mathrm{det}(V_k) =  j(\mathbf{H}_\Lambda(k))$ 
for $k=0,\,1$, {\it cf.} Equation~\eqref{rk:finite_chain_index_as_sf}, 
the $\mathbb{Z}_2$-valued spectral flow is non-trivial if and only if  $j(\mathbf{H}_\Lambda(0)) \neq j(\mathbf{H}_\Lambda(1))$.
\hfill $\Box$

\vspace{0.2cm}

We therefore see that the (finite-dimensional) $\mathbb{Z}_2$-valued spectral flow 
gives a topological obstruction for two Hamiltonians to have the same $\mathbb{Z}_2$-phase.

\begin{proposi} \label{prop:Z2_flow_and_degeneracy}
Let $\mathbf{H}_\Lambda(0)$ and $\mathbf{H}_\Lambda(1)$ be quadratic and strictly 
positive Hamiltonians and suppose $\mathrm{Sf}_2( A_\Lambda(0) , A_\Lambda(1)) = -1$. 
Then along the path $[0,1] \ni t \mapsto \mathbf{H}_\Lambda(t)$ connecting the Hamiltonians,  
there is some $t_0 \in (0,1)$ such that 
$\mathbf{H}_\Lambda(t_0)$ has a $0$-energy  state.
\end{proposi}
\noindent {\bf Proof.} 
To every $t \in [0,1]$ there is an $A_\Lambda(t)$ associated to $\mathbf{H}_\Lambda(t)$ 
by Equation \eqref{eq-HamMajoRep}
and the $\mathbb{Z}_2$-valued spectral flow is determined by the path $A_\Lambda(t)$. 
If the $\mathbb{Z}_2$-valued spectral flow is non-trivial, then   
there is some $t_0$ such that $A_\Lambda(t_0)$ has at least a  double degenerate 
$0$-eigenvalue (see Remark \ref{rk:Z2_spec_flow_explanation}). 
Because the eigenvalues of $A_\Lambda$ determine the spectrum of 
$\mathbf{H}_\Lambda$, in particular 
$\mathrm{dim} \, \Ker(\mathbf{H}_\Lambda) = \tfrac{1}{2}\, \mathrm{dim}\, \Ker( A_\Lambda)$, 
 it follows that $\mathbf{H}_\Lambda(t_0)$ has at least one $0$-energy state.
\hfill $\Box$

\vspace{0.2cm}

Combining the two previous propositions, if follows that if $j(\mathbf{H}_\Lambda(0)) \neq j(\mathbf{H}_\Lambda(1))$, 
then the two Hamiltonians cannot be continuously connected without the appearance of a Majorana 
operator from a zero-energy state. We will give an example of a non-trivial $\mathbb{Z}_2$-spectral flow 
via a flux insertion in Section \ref{sec:closed_chain_flux}.

\subsection{The parity operator} \label{subsec:finite_parity}
The (fermionic and not spatial) parity operator is defined by
$$
\Pp
\;=\;
(-1)^\NN
\;,
$$
where $\NN=\sum_{j=1}^L\fraka_j^*\fraka_j$ is the fermionic number operator on the chain 
$\Lambda=[1,L]$. It is a self-adjoint unitary:
$$
\Pp^2\;=\;\one\;,
\qquad \Pp^*\;=\;\Pp
\;,
$$
and hence introduces a grading on the Fock space. Any Hamiltonian that is 
an even polynomial in the in the creation and annihilation operators $\fraka_j^*$ and $\fraka_j$ 
will commute with the parity operator and be of even degree.  This includes 
higher-order interactions. Indeed, using 
$$
(-1)^{\fraka_k^*\fraka_k}
\;=\;
e^{\ii\pi \fraka_k^*\fraka_k}
\;=\;
e^{\ii\pi (1-\fraka_k\fraka_k^*)}
\;=\;
-e^{-\ii\pi \fraka_k\fraka_k^*}
\;=\;
-e^{\ii\pi \fraka_k\fraka_k^*}\;,
$$ 
one obtains
$$
\Pp\,\fraka_j\,\Pp\;=\;-\,\fraka_j
\;.
$$
In this form, the parity symmetry is a subgroup of the U$(1)$-charge conservation symmetry. 
As $\frakd_j$, 
$\frakb_j$ and $\frakbtilde_j$ are all linear combinations of $\fraka$ and $\fraka^*$'s, one also has
$$
\Pp\,\frakd_j\,\Pp\;=\;-\,\frakd_j\;,
\qquad
\Pp\,\frakb_j\,\Pp\;=\;-\,\frakb_j
\;,
\qquad
\Pp\,\frakbtilde_j\,\Pp\;=\;-\,\frakbtilde_j\;,
$$
Using \eqref{eq-Crel}, we can  express
\begin{equation}
\label{eq-ParRep}
\Pp
\;=\;
\prod_{j=1}^L\,(-1)^{\fraka_j^*\fraka_j}
\;=\;
\prod_{j=1}^L\,(-1)^{\frac{1}{2}(\one+\ii\,\frakb_{2j-1}\frakb_{2j})}
\;=\;
\prod_{j=1}^L\,(-\ii\,\frakb_{2j-1}\frakb_{2j})
\;,
\end{equation}
where in the last step it was used that the $\ii\,\frakb_{2j-1}\frakb_{2j}$ are commuting self-adjoint unitaries.

\subsection{The Kitaev model on an open chain}

Let us fix a finite chain $\Lambda = \{1,\ldots,L\}$ and consider the Hamiltonian on 
 $\Ff_\Lambda$ given by
\begin{equation} \label{eq:Kitaev_model}
\HH_\Lambda^\mathrm{Kit}
\;=\;
\sum_{j=1}^{L-1}\big(-w\,(\fraka_j^*\fraka_{j+1}+\fraka_{j+1}^*\fraka_{j})
+\Delta\,\fraka_j\fraka_{j+1}+\overline{\Delta}\,\fraka_{j+1}^*\fraka_{j}^*
\big) + \mu \sum_{j=1}^L (\fraka_j^*\fraka_{j}-\tfrac{1}{2})
\;.
\end{equation}
Here $w,\mu\in\RM$ and $\Delta=|\Delta|e^{\ii\theta}\in\CM$.
As the operator $\HH_\Lambda^\mathrm{Kit}$ is quadratic, we can write 
the associated BdG Hamiltonian 
$H_\Lambda$ on the particle-hole space $\Hh_\ph= \CM^L \otimes \CM^2$:
\begin{equation} \label{eq:BdGKitaev_model}
H_\Lambda^\mathrm{Kit}
\;=\;
\begin{pmatrix}
-w(S+S^*)-\mu & 
\overline{\Delta}(S^*-S) \\ \Delta(S-S^*) & w(S+S^*)+\mu
\end{pmatrix}
\;.
\end{equation}
Here $S$ is the right shift on $\CM^{L}$ with open boundary conditions: 
$$
S
\;=\;
\sum_{j=1,\ldots,L-1}
\,|j+1\rangle\langle j|
\;=\;
\begin{pmatrix}
0 & & & 
\\
1 & \ddots & & 
\\
& \ddots & \ddots &
\\
 & & 1 & 0
\end{pmatrix}
\;.
$$
The BdG Hamiltonian shows that $\mathbf{H}_\Lambda^\mathrm{Kit}$ models a $p$-wave 
interaction.

\subsection*{Case: $w=\Delta=0$ (trivial chain)}
Let us study the Kitaev chain in a few cases where the solutions are explicit.
First, we consider the case $w=\Delta=0$ and so 
$$
 \mathbf{H}_\Lambda^\mathrm{Kit} \;=\; \mu \sum_{j=1}^L (\fraka_j^*\fraka_{j}-\tfrac{1}{2})
\;=\;
\frac{\mu}{2}\;
\sum_{j=1}^{L} \frakb_{2j-1}\frakb_{2j}
\;.
$$
If $\mu\geq 0$, then the energy of $\mathbf{H}_\Lambda^\mathrm{Kit}$ is minimized by any 
state $| \psi\rangle$ such that $\fraka_j |\psi\rangle = 0$. Therefore, if $\mu >0$, 
fermionic vacuum $|\Omega\rangle$ gives the unique  ground state.

\subsection*{Case: $\mu=0$, $w=|\Delta|$ (non-trivial chain and quantum Ising model)}
In the case $\mu=0$ and $\Delta=e^{\ii \theta}w$,  the Hamiltonian takes the particularly simple form 
in the Majorana representation, namely with \eqref{eq-Crel2}
\begin{equation}
\label{eq-HamMajoChain}
\HH_\Lambda^\mathrm{Kit}
\;=\;  w \sum_{j=1}^{L-1} \big( - \fraka_j^*\fraka_{j+1}-\fraka_{j+1}^*\fraka_{j} 
 + e^{\ii \theta}\fraka_j\fraka_{j+1}+ e^{-\ii \theta}\fraka_{j+1}^*\fraka_{j}^* \big)  \;=\;
\ii \,w
\sum_{j=1}^{L-1} \frakb_{2j}\frakb_{2j+1}
\;.
\end{equation}

The Kitaev Hamiltonian with $w=|\Delta|$ can be directly mapped to the quantum Ising chain via 
the Jordan--Wigner transform. Namely, using the notation $\sigma_k^{x/y/z}$ to 
denote operators 
analogous to the Pauli matrices at site $k \in \{1,\ldots,L\}$, we define
$$
   \sigma_j^x \;=\; \Big( e^{-i\pi \sum_{k=1}^{j-1} \fraka_k^* \fraka_k} \Big) \fraka_j^* \; , \qquad 
   \sigma_j^y \;=\; \Big( e^{ i\pi \sum_{k=1}^{j-1} \fraka_k^* \fraka_k} \Big) \fraka_j \; , \qquad 
   \sigma_j^z \;=\; 2 \fraka_j^* \fraka_j - \one \; .
$$
Then for $J_x =w$ and $h= \tfrac{\mu}{2}$, the  Hamiltonian becomes
$$
    \mathbf{H}_\Lambda^\text{spin}  \;=\; -J_x \sum_{j=1}^{L-1} \sigma_j^x \sigma_{j+1}^x - h \sum_{j-1}^L \sigma_j^z\; .
$$
The Hamiltonian $\mathbf{H}_\Lambda^\text{spin}$ describes a quantum Ising chain.
For completeness, we also recall the inverse Jordan--Wigner transform, 
$$
   \frakb_{2j-1} \;=\; \Big( \prod_{k=1}^{j-1} \sigma_k^z \Big) \sigma_j^x \; , \qquad 
   \frakb_{2j} \;=\; \Big( \prod_{k=1}^{j-1} \sigma_k^z \Big) \sigma_j^y\; 
$$
which gives the fermionic (Majorana) representation.

\vspace{0.2cm}

Expressing $\HH_\Lambda^\mathrm{Kit}$ in the Majorana representation, we see that 
 only Majorana operators on different sites are coupled. 
Moreover, each of the summands $\ii \,\frakb_{2j}\frakb_{2j+1}$ in \eqref{eq-HamMajoChain} 
is a self-adjoint unitary and thus allows to introduce a self-adjoint projection on Fock space
\begin{equation}
\label{eq-PjDef}
\PP_j
\;=\;
\tfrac{1}{2}(\one+\ii \,\frakb_{2j}\frakb_{2j+1})
\;.
\end{equation}
These projections commute $[\PP_j,\PP_i]=0$ and the Hamiltonian can be written as
\begin{equation}
\label{eq-ProjRepMajo}
\HH_\Lambda^\mathrm{Kit}
\;=\;
w
\sum_{j=1}^{L-1}
(2\,\PP_j-\one)
\;.
\end{equation}

Another way to write the Hamiltonian is to build a new pair of creation and annihilation 
operators $\{\frakd_j\}_{j=1}^{L-1}$ from the pair $\frakb_{2j}$ and $\frakb_{2j+1}$:
\begin{equation}
\label{eq-frakaDef}
\frakd_j\;=\;
\tfrac{1}{2}
(\frakb_{2j}+\ii\,\frakb_{2j+1})\;,
\qquad
\frakd_j^*\;=\;
\tfrac{1}{2}(\frakb_{2j}-\ii\,\frakb_{2j+1})\;,
\end{equation}
or more explicitly
\begin{align}
&
\frakd_j\;=\;
\tfrac{\ii}{2}\big(-e^{\ii\frac{\theta}{2}}\,\fraka_{j}+e^{-\ii\frac{\theta}{2}}\fraka_{j}^* +e^{\ii\frac{\theta}{2}}\,\fraka_{j+1}+e^{-\ii\frac{\theta}{2}}\fraka_{j+1}^*  \big)
\;,
\label{eq-frakaDef1}
\\
&
\frakd_j^*\;=\;
\tfrac{\ii}{2}\big(-e^{\ii\frac{\theta}{2}}\,\fraka_{j}+e^{-\ii\frac{\theta}{2}}\fraka_{j}^* -e^{\ii\frac{\theta}{2}}\,\fraka_{j+1}-e^{-\ii\frac{\theta}{2}}\fraka_{j+1}^*  \big)
\;.
\label{eq-frakaDef2}
\end{align}
These operators satisfy again the CAR's:
$$
\{\frakd_j^*,\frakd_i\}\;=\;\delta_{i,j}\,\one
\;,
\qquad
\{\frakd_j,\frakd_i\}\;=\;0
\;,
$$
and using 
\begin{equation}
\label{eq-CTildeAtilde}
\ii\,\frakb_{2j}\frakb_{2j+1}
\;=\;
2\,\frakd^*_j\frakd_j-
\one 
\end{equation}
allow to write the Hamiltonian as
\begin{equation}
\label{eq-HamOsci}
\HH_\Lambda^\mathrm{Kit}
\;=\;
w
\sum_{j=1}^{L-1}
(2\,\frakd_j^*\frakd_j-\one)
\;, \qquad 
\PP_j
\;=\;
\frakd_j^*\frakd_j
\;.
\end{equation}
Let us refer to this as the quantum Ising Kitaev Hamiltonian.
Another key property of $\HH_\Lambda^\mathrm{Kit}$ in the non-trivial region 
are the two ``dangling'' Majorana operators $\frakb_1$ and $\frakb_{2L}$ on the finite chain $\Lambda=[1,L]$, 
which influence the degeneracy of the spectrum. 
We set 
$$
\frakd_\bound\;=\;
\tfrac{1}{2}(\frakb_{2L}+\ii\,\frakb_{1})\;,
\qquad
\frakd_\bound^*\;=\;
\tfrac{1}{2}(\frakb_{2L}-\ii\,\frakb_{1})\;.
$$
which also satisfy the CAR's (together with the other $\frakd_j$). In terms of the initial creation and annihilation operators,
\begin{align*}
&
\frakd_\bound\;=\;
\tfrac{\ii}{2}\big(-e^{\ii\frac{\theta}{2}}\,\fraka_{L}+e^{-\ii\frac{\theta}{2}}\fraka_{L}^* +e^{\ii\frac{\theta}{2}}\,\fraka_{1}+e^{-\ii\frac{\theta}{2}}\fraka_{1}^*  \big)
\;,
\\
&
\frakd_\bound^*\;=\;
\tfrac{\ii}{2}\big(-e^{\ii\frac{\theta}{2}}\,\fraka_{L}+e^{-\ii\frac{\theta}{2}}\fraka_{L}^* -e^{\ii\frac{\theta}{2}}\,\fraka_{1}-e^{-\ii\frac{\theta}{2}}\fraka_{1}^*  \big)
\;.
\end{align*}
Again one can define $\PP_\bound=\frakd_\bound^*\frakd_\bound$ and, as in \eqref{eq-CTildeAtilde},
\begin{equation}
\label{eq-CTildeAtildeBound}
\ii\,\frakb_{2L}\frakb_{1}
\;=\;
2\,\frakd^*_\bound\frakd_\bound-\one 
\;.
\end{equation}

Turning our attention to the ground state space, we see that for $w\geq 0$, 
$\frakd_1\cdots \frakd_{L-1}| \Omega\rangle$ will minimize the energy. 
However, if 
$\frakd_\bound \frakd_1\cdots \frakd_{L-1}| \Omega\rangle$ is non-zero, then 
it is also a ground state. Furthermore, as these states have different parity 
(as $\frakd_\bound$ is odd), then this shows the ground state space will have a double 
degeneracy.
We will show  that for every $L$, either 
$\frakd_\bound^* \frakd_1\cdots \frakd_{L-1}| \Omega\rangle$ 
or $\frakd_\bound \frakd_1\cdots \frakd_{L-1}| \Omega\rangle$ is non-zero 
and, along with $\frakd_1\cdots \frakd_{L-1}| \Omega\rangle$, completely 
characterise the ground state space.

\subsection*{An orthonormal basis in Fock space}
\label{sec-ONB}
Let us now use the the new CAR operators $\{ \frakd_j\}_{j\in \Lambda}$ to 
characterise a basis for the fermionic Fock space $\mathcal{F}_\Lambda$ that solves 
the quantum Ising/Kitaev Hamiltonian \eqref{eq-HamOsci}.

\vspace{0.2cm}

First let us rewrite the parity operator using 
$\{\frakd_j\}_{j\in\Lambda}$. Starting from 
Equation \eqref{eq-ParRep}, 
$$
\Pp
\;=\;
(\ii\,\frakb_{2L}\frakb_{1})\prod_{j=1}^{L-1}\,(-\ii\,\frakb_{2j}\frakb_{2j+1})
\;=\;
(\ii\,\frakb_{2L}\frakb_{1})\prod_{j=1}^{L-1}\,(-1)^{\frakd_j^*\frakd_j}
\;=\;
(\ii\,\frakb_{2L}\frakb_{1})\prod_{j=1}^{L-1}\,(\one-2\,\frakd_j^*\frakd_j)
\;,
$$
and finally using \eqref{eq-CTildeAtildeBound}
\begin{equation}
\label{eq-ParityRewrite}
\Pp
\;=\;
-(\one-2\,\frakd_\bound^*\frakd_\bound)\prod_{j=1}^{L-1}\,(\one-2\,\frakd_j^*\frakd_j)
\;.
\end{equation}
It ought to be stressed that for this to hold one has to use $\frakd_\bound=\tfrac{1}{2}(\frakb_{2L}+\ii\,\frakb_{1})$ and is {\it not} allowed to exchange $\frakb_{2L}$ and $\frakb_{1}$, which is equivalent to exchanging $\frakd_\bound$ with $\frakd_\bound^*$. This would produce a sign change.
For occupation numbers $i_\bound,i_1,\ldots,i_{L-1}\in\{0,1\}$, let us introduce the states 
\begin{equation}
\label{eq-OccStates0}
| 0;i_1,\ldots,i_{L-1}\rangle
\;=\;
2^{\frac{L-1}{2}}\,
\frakd_1^{(i_{1})}\cdots\frakd_{L-1}^{(i_{L-1})} | \Omega\rangle
\;,
\end{equation}
where
$$
\frakd_j^{(0)}\;=\;\frakd_j\;,
\qquad
\frakd_j^{(1)}\;=\;\frakd_j^*\;,
$$
for $j=1,\ldots,L-1$. The $0$ in the first entry indicates that neither $\frakd_\bound$ 
nor $\frakd_\bound^*$ is involved. This will be modified later on. The parity of these 
states is easily read off of $\Pp\,\frakd_j\,\Pp=-\,\frakd_j$ and $\Pp|\Omega\rangle= | \Omega\rangle$
\begin{equation}
\label{eq-OccStatesPar}
\Pp\,| 0;i_1,\ldots,i_{L-1}\rangle
\;=\;
(-1)^{L-1}| 0;i_1,\ldots,i_{L-1}\rangle
\;.
\end{equation}
Now one can obtain states of parity $(-1)^L$ by either applying $\frakd_\bound$ or $\frakd_\bound^*$ to these states. However, the following result shows that one of the outcomes vanishes. 

\begin{lemma}
\label{lem-vanishing}
\begin{itemize}
\item[{\rm (i)}] $\langle 0;i_1,\ldots,i_{L-1}| 0;i'_1,\ldots,i'_{L-1}\rangle=\delta_{i_1,i'_1}\cdots \delta_{i_{L-1},i'_{L-1}}$

\item[{\rm (ii)}]
If $L+\sum_{j=1}^{L-1}i_j=0\;\mbox{\rm mod}\,2$, then 
$$
\frakd_\bound | 0;i_1,\ldots,i_{L-1}\rangle\,=\,0\;,
\qquad
\|\frakd^*_\bound | 0;i_1,\ldots,i_{L-1}\rangle\|\,=\,1
\;.
$$

\item[{\rm (iii)}]
If $L+\sum_{j=1}^{L-1}i_j=1\;\mbox{\rm mod}\,2$, then 
$$
\frakd_\bound^* | 0;i_1,\ldots,i_{L-1}\rangle\,=\,0
\;,
\qquad
\|\frakd_\bound | 0;i_1,\ldots,i_{L-1}\rangle\|\,=\,1
\;.
$$

\end{itemize}
\end{lemma}

\noindent {\bf Proof.} (i) We focus on the diagonal case $i_j=i'_j$. Then let us start with the following algebraic manipulation:
\begin{align*}
\|\,| 0;i_1,\ldots,i_{L-1}\rangle\|^2
&
\;=\;
2^{L-1}\,
\langle
\frakd_1^{(i_{1})}\cdots\frakd_{L-1}^{(i_{L-1})} \Omega| 
\frakd_1^{(i_{1})}\cdots\frakd_{L-1}^{(i_{L-1})}\Omega\rangle
\\
&
\;=\;
2^{L-1}\,
\langle
\Omega| 
(\frakd_1^{(i_{1})})^*\frakd_1^{(i_{1})}\cdots(\frakd_{L-1}^{(i_{L-1})})^*\frakd_{L-1}^{(i_{L-1})}\Omega\rangle
\;,
\end{align*}
because each $(\frakd_j^{(i_{j})})^*\frakd_j^{(i_{j})}$ commutes with $\frakd_k^{(i_{k})}$. 
Now due to \eqref{eq-HamOsci}, each factor $(\frakd_j^{(i_{j})})^*\frakd_j^{(i_{j})}$ is 
either $\PP_j$ or $\one-\PP_j$, pending on whether $i_j=0$ or $i_j=1$. Hence let us set 
$\PP_j^{(0)}=\PP_j$ and $\PP_j^{(1)}=\one-\PP_j$. Then
$$
\|\,| 0;i_1,\ldots,i_{L-1}\rangle\|^2
\;=\;
2^{L-1}\,
\langle
\Omega| 
\PP_1^{(i_1)}\cdots \PP_{L-1}^{(i_{L-1})}
| \Omega\rangle
\;.
$$
Now these projections commute and one can check using \eqref{eq-frakaDef1} and \eqref{eq-frakaDef2}
\begin{equation}
\label{eq-PjDef1}
\PP_j^{(i_j)}| \Omega\rangle
\;=\;
\tfrac{1}{2}(\one+(1-2i_j)\,e^{-\ii\theta} \,\fraka_{j}^*\fraka_{j+1}^*)| \Omega\rangle
\;
\end{equation}
and so $\langle \Omega | \PP_j^{(i_j)}| \Omega\rangle = \tfrac{1}{2}$ independently of 
the value of $i_j$. Iterating on this idea, $\|\,| 0;i_1,\ldots,i_{L-1}\rangle\|^2 = 1$, 
which shows the claim.

\vspace{.1cm}

(ii) On the one hand, one has \eqref{eq-OccStatesPar} so that
$$
\Pp\,\frakd_\bound | 0;i_1,\ldots,i_{L-1}\rangle
\;=\;
(-1)^L\,\frakd_\bound | 0;i_1,\ldots,i_{L-1}\rangle
\;.
$$
On the other hand, due to the CAR's, 
$\frakd_j^*\frakd_j\frakd_j^{(i_j)} = i_j\,\frakd_j^{(i_j)}$
and using Equation \eqref{eq-ParityRewrite}
\begin{align*}
\Pp\,
\frakd_\bound | 0;i_1,\ldots,i_{L-1}\rangle
&
\;=\;
-\,(\one-2\,\frakd_\bound^*\frakd_\bound)\prod_{j=1}^{L-1}\,(\one-2\,\frakd_j^*\frakd_j)
\frakd_\bound | 0;i_1,\ldots,i_{L-1}\rangle
\\
&
\;=\;
-\,(\one-2\,\frakd_\bound^*\frakd_\bound)\frakd_\bound
\prod_{j=1}^{L-1}\,(\one-2\,\frakd_j^*\frakd_j)
 | 0;i_1,\ldots,i_{L-1}\rangle
\\
&
\;=\;
-\,(\one-2\,\frakd_\bound^*\frakd_\bound)\frakd_\bound \prod_{j=1}^{L-1}\,(-1)^{i_j}
 | 0;i_1,\ldots,i_{L-1}\rangle
\\
&
\;=\;
-\,\frakd_\bound
(-1)^{\sum_{j=1}^{L-1}i_j}
 | 0;i_1,\ldots,i_{L-1}\rangle
\;.
\end{align*}
Hence if $L+\sum_{j=1}^{L-1}i_j$ is even, $\frakd_\bound | 0;i_1,\ldots,i_{L-1}\rangle=0$. Now
\begin{align*}
\|\,\frakd_\bound^* | 0;i_1,\ldots,i_{L-1}\rangle\|^2
& 
\;=\;
\langle 0;i_1,\ldots,i_{L-1} |  \frakd_\bound \frakd_\bound^*  | 0;i_1,\ldots,i_{L-1}\rangle
\\
&
\;=\;
\langle 0;i_1,\ldots,i_{L-1} |  (\one-\frakd_\bound^*\frakd_\bound)  | 0;i_1,\ldots,i_{L-1}\rangle
\\
&
\;=\;
\|\, | 0;i_1,\ldots,i_{L-1}\rangle\|^2
\;.
\end{align*}
The claim (iii) follows in the same manner. 
\hfill $\Box$

\vspace{.2cm}

Given the above lemma, let us now define the states
\begin{equation}
\label{eq-OccStates1}
|1;i_1,\ldots,i_{L-1}\rangle
\;=\;
\left\{
\begin{array}{cc}
\frakd^*_\bound\,|0;i_1,\ldots,i_{L-1}\rangle\;\;
&
\mbox{\rm if }L+\sum_{j=1}^{L-1}i_j\;\mbox{\rm even}\;,
\\
\frakd_\bound\,|0;i_1,\ldots,i_{L-1}\rangle\;\;
&
\mbox{\rm if } L+\sum_{j=1}^{L-1}i_j\;\mbox{\rm odd}\;.
\end{array}
\right.
\end{equation}
The parity of these states is given by
\begin{equation}
\label{eq-OccStatesPar1}
\Pp\,|1;i_1,\ldots,i_{L-1}\rangle
\;=\;
(-1)^L
\,|1;i_1,\ldots,i_{L-1}\rangle
\;.
\end{equation}
Comparing with \eqref{eq-OccStatesPar}, one sees that the first entry $i_\bound$ in 
$|i_\bound;i_1,\ldots,i_{L-1}\rangle$ indicates a parity change.

\begin{proposi}  \label{prop:Kit_open_eigenbasis}
The set $\big\{|i_\bound;i_1,\ldots,i_{L-1}\rangle\,:\,i_\bound,i_1,\ldots,i_{L-1}\in\{0,1\}\big\}$ is an orthogonal basis of $\Ff_\Lambda$.
\end{proposi}

\noindent {\bf Proof.} Due to Lemma~\ref{lem-vanishing}, it only remains to prove the following orthogonality relations:
$$
\langle 1;i'_1,\ldots,i'_{L-1} | 0;i_1,\ldots,i_{L-1}\rangle
\;=\;0
\;,
\qquad
\langle 1;i'_1,\ldots,i'_{L-1} | 1;i_1,\ldots,i_{L-1}\rangle
\;=\;\delta_{i_1,i'_1}\cdots \delta_{i_{L-1},i'_{L-1}}
\;.
$$
The first claim follows because the two states have different parity. The second one is based on Lemma~\ref{lem-vanishing}(i) and an argument as in the proof of Lemma~\ref{lem-vanishing}(ii).
\hfill $\Box$

\vspace{0.2cm}

Let us also note that by the relation $(2\frakd_j^* \frakd_j - \one) \frakd_j^{(i_j)} = (-1)^{ i_j + 1}\frakd_j^{(i_j)}$ with 
$i_j \in \{0,1\}$ the occupation number, we deduce from Equation \eqref{eq-HamOsci} that 
\begin{equation*} 
   \mathbf{H}_\Lambda^\mathrm{Kit} | i_\bound;i_1,\ldots,i_{L-1}\rangle \;=\;  
   w \Big( \sum_{j=1}^{L-1} (-1)^{i_j + 1} \Big) | i_\bound;i_1,\ldots,i_{L-1}\rangle \;. 
\end{equation*}
Therefore, the orthonormal basis 
$\big\{|i_\bound;i_1,\ldots,i_{L-1}\rangle\,:\,i_\bound,i_1,\ldots,i_{L-1}\in\{0,1\}\big\}$ 
diagonalises the quantum Ising/Kitaev Hamiltonian \eqref{eq-HamOsci}. In particular, the 
ground state space of $\mathbf{H}_\Lambda^\mathrm{Kit}$ is spanned by 
$| 0; 0,\ldots,0 \rangle$ and $| 1; 0,\ldots,0 \rangle$.

\subsection{The Kitaev model on a closed chain}
\label{sec-KitaevClosedChain}

The  previous analysis on the Kitaev Hamiltonian was for systems with open boundary 
conditions. We can close up the chain with periodic or anti-periodic boundary conditions by heuristically 
choosing $\fraka_{L+1} = \pm \fraka_1$. Let us now consider the case of periodic and anti-periodic boundary 
conditions. This leads to the Hamiltonian 
\begin{align*}
  \mathbf{H}^\mathrm{Kit}_\Lambda(\pm) 
  \;&=\;
\sum_{j=1}^{L-1}\big(-w\,(\fraka_j^*\fraka_{j+1}+\fraka_{j+1}^*\fraka_{j})
+\Delta\,\fraka_j\fraka_{j+1}+\overline{\Delta}\,\fraka_{j+1}^*\fraka_{j}^*
\big) + \mu \sum_{j=1}^L (\fraka_j^*\fraka_{j}-\tfrac{1}{2})  \\
  &\qquad  \pm \big( -w(\fraka_L^*\fraka_1 + \fraka_1^*\fraka_L) + \Delta\fraka_L\fraka_1 +  \ol{\Delta}\fraka_1^*\fraka_L^* \big)
\;.
\end{align*}
Clearly in the `trivial phase' $w=\Delta =0$, then the  Hamiltonian is the same as the 
trivial Hamiltonian with open boundary conditions and, hence, has the ground state 
$| \Omega\rangle$ for $\mu >0$.

\vspace{0.2cm}

In the non-trivial regime $\mu=0$ and $\Delta=e^{\ii \theta}w$, the 
Majorana representation of $\mathbf{H}_\Lambda^\mathrm{Kit}(\pm)$ is as 
in \eqref{eq-HamMajoChain} with the supplementary summand $\ii w\frakb_{2L}\frakb_{1}$ 
which has to be evaluated as in \eqref{eq-Crel2}: 
\begin{equation} \label{eq:Per_aPer_Kitaev}
 \mathbf{H}_\Lambda^\mathrm{Kit}(\pm) \;=\; 
 \ii w\sum_{j=1}^{L-1} \frakb_{2j}\frakb_{2j+1} \;\pm \;\ii w \, \frakb_{2L}\frakb_{1} \; .
\end{equation}
Assuming non-negative $w$, 
the ground state space of $\mathbf{H}_\Lambda^\mathrm{Kit}(\pm)$ is 
built from the $-1$ eigenstates of the commuting 
even self-adjoint unitaries $\{\ii \,\frakb_{2j}\frakb_{2j+1}\}_{j=1}^{L-1}$ and the 
$\mp 1$ eigenstate of $\ii\, \frakb_{2L}\frakb_1$, 
$$
   \calH_\mathrm{GS}^\pm \;\cong\; 
   \frac{1}{2}(1\mp \ii \frakb_{2L}\frakb_1) \prod_{j=1}^{L-1} \frac{1}{2}(1- \ii \frakb_{2j}\frakb_{2j+1}) \cdot \mathcal{F}(\mathbb{C}^L) \; .
$$
Like the open chain, we can characterise the ground state space by the new CAR operators 
\begin{align*}
  &\frakd_j \;=\;  \frac{1}{2}( \frakb_{2j} +\ii \frakb_{2j+1}) \; , 
  &&\frakd_{\bound}^\pm \;=\;  \frac{1}{2}( \frakb_{2L} \pm \ii \frakb_{1}) \; , \\
  &\ii \,\frakb_{2j}\frakb_{2j+1} \;=\;  2\frakd_j^*\frakd_j - 1 \; ,  
  &&\pm \ii \, \frakb_{2L}\frakb_1 \;=\;  2(\frakd_{\bound}^\pm)^* \frakd_{\bound}^\pm - 1 \; .
\end{align*}
In particular $\Ran(\frakd_j)$ is a subspace of the $-1$ eigenspace of 
$\ii \,\frakb_{2j}\frakb_{2j+1}$ and $\Ran(\frakd_{\bound}^\pm)$ is a subspace 
of the $\mp 1$ eigenspace of $\ii \,\frakb_{2L}\frakb_1$. To ensure that  
the ground state space is characterised, we just need to make sure these spaces are 
non-trivial. But indeed
\begin{align*}
  &\frakd_j  \; = \;  \frac{\ii }{2}( -e^{\ii \frac{\theta}{2}}\fraka_j + e^{-\ii \frac{\theta}{2}}\fraka_j^* + 
         e^{\ii \frac{\theta}{2}}\fraka_{j+1} + e^{-\ii \frac{\theta}{2}}\fraka_{j+1}^*) \; ,  
 &&\frakd_{\bound}^\pm  \; = \; \frac{\ii }{2}( -e^{\ii \frac{\theta}{2}}\fraka_L + e^{-\ii \frac{\theta}{2}}\fraka_L^* \pm 
         e^{\ii \frac{\theta}{2}}\fraka_{1} \pm e^{-\ii \frac{\theta}{2}}\fraka_{1}^*) \; ,
\end{align*}
and so $\frakd_j | \Omega\rangle$ and $\frakd_{\bound}^\pm | \Omega\rangle$ are non-zero. 
Like the open chain, we again need to account for the parity operator, where the following 
lemma plays an analogous role to Lemma \ref{lem-vanishing}.

\begin{lemma} \label{lem:closed_chain_basis_cases}
\begin{enumerate}
 \item[{\rm (i)}] If $L$ is even, then $\frakd_\bound^+ \frakd_1\cdots\frakd_{L-1}|\Omega\rangle = 0$ and 
 $\frakd_\bound^- \frakd_1\cdots\frakd_{L-1}|\Omega\rangle \neq 0$.
 \item[{\rm (ii)}] If $L$ is odd, then $\frakd_\bound^- \frakd_1\cdots\frakd_{L-1}|\Omega\rangle = 0$ and 
 $\frakd_\bound^+ \frakd_1\cdots\frakd_{L-1}|\Omega\rangle \neq 0$.
\end{enumerate}
\end{lemma}
\noindent {\bf Proof.} 
Let us consider the vectors $\frakd_\bound^\pm \frakd_1\cdots\frakd_{L-1}|\Omega\rangle$. 
Because $\frakd_j$ and $\frakd_\bound^\pm$ are odd operators, it follows that 
$$
  \Pp\,  \frakd_\bound^\pm \frakd_1\cdots \frakd_{L-1} | \Omega\rangle \;=\; 
  (-1)^{L} \,\frakd_\bound^\pm \frakd_1\cdots \frakd_{L-1} \Pp | \Omega\rangle \;=\; 
  (-1)^{L} \,  \frakd_\bound^\pm \frakd_1\cdots \frakd_{L-1}  | \Omega\rangle \; .
$$
On the other hand, let us recall 
$$
 \Pp \;=\; \prod_{j=1}^{L}(-\ii  \frakb_{2j-1}\frakb_{2j}) \;=\; (\ii  \frakb_{2L}\frakb_1) \prod_{j=1}^{L-1} ( -\ii  \frakb_{2j}\frakb_{2j+1}) 
 \;=\; \pm \big( 2(\frakd_\bound^\pm)^* \frakd_\bound - 1\big) \prod_{j=1}^{L-1}\big( 1- 2\frakd_j^* \frakd_j\big) \; . 
$$
Computing the parity, 
\begin{align*}
 \Pp \, \frakd_\bound^\pm \frakd_1\cdots \frakd_{L-1} | \Omega\rangle \;&=\; 
 \pm \big( 2(\frakd_\bound^\pm)^* \frakd_\bound - 1\big) \prod_{j=1}^{L-1}\big( 1- 2\frakd_j^* \frakd_j\big) \,
 \frakd_\bound^\pm \frakd_1\cdots \frakd_{L-1} | \Omega\rangle \\
 &=\; \pm \big( 2(\frakd_\bound^\pm)^* \frakd_\bound - 1\big) \frakd_\bound^\pm \frakd_1\cdots \frakd_{L-1} | \Omega\rangle \\
 &=\; \mp\, \frakd_\bound^\pm \frakd_1\cdots \frakd_{L-1} | \Omega\rangle \; .
\end{align*}
Therefore if $L$ is even, then we have that 
$\frakd_\bound^+ \frakd_1\cdots\frakd_{L-1}|\Omega\rangle$ is both even and 
odd. Thus it must be $0$. Similarly, if $L$ is odd, 
$\frakd_\bound^- \frakd_1\cdots\frakd_{L-1}|\Omega\rangle$ is even and odd 
and so must vanish.
\hfill $\Box$ 

\vspace{0.2cm}

Lemma \ref{lem:closed_chain_basis_cases} can  be used to prove the following special case of Proposition \ref{prop:closed_chain_GS_generic} below.

\begin{proposi} \label{prop:per_aper_Kit_GS}
If $L$ is even, a ground state of $\mathbf{H}_\Lambda^\mathrm{Kit}(\pm)$ is given by 
$$
  | \psi_\pm \rangle \;=\; \begin{cases} \frakd_1\cdots \frakd_{L-1} |\Omega\rangle\, , & \fraka_{L+1} \,=\, \fraka_1\, , \\
     \frakd_\bound^- \frakd_1\cdots \frakd_{L-1}| \Omega\rangle  \, , & \fraka_{L+1} \,=\, -\fraka_1 \, \end{cases}.
$$
If $L$ is odd, a ground state of $\mathbf{H}_\Lambda^\pm$ is given by 
$$
  | \psi_\pm \rangle \;=\; \begin{cases} \frakd_\bound^+ \frakd_1\cdots \frakd_{L-1} |\Omega\rangle\, , & \fraka_{L+1} \,=\, \fraka_1\, , \\
   \frakd_1\cdots \frakd_{L-1}|\Omega\rangle \, , & \fraka_{L+1}  \,=\, -\fraka_1 \, \end{cases}.
$$
In particular, $\Pp |\psi_\pm \rangle = \mp | \psi_\pm \rangle$.
\end{proposi}

It is true that for $w>0$ the ground states specified in Proposition \ref{prop:per_aper_Kit_GS} are 
unique, see~\cite{TwistedKitaev} for example. To prove such a statement 
requires constructing an eigenbasis as in Proposition \ref{prop:Kit_open_eigenbasis}.

\subsection*{Connection to index on canonical transformations}

Unlike the case of open boundary conditions, the Kitaev model on the closed chain 
does not have a double degenerate ground state. However, one can still differentiate 
between different `phases' using the $\mathbb{Z}_2$-index from Definition \ref{def:Z_2_index_def}.

\hspace{0.2cm}

First consider the trivial Hamiltonian, namely $w=0$:
$$
  \HH_\Lambda^\mathrm{Kit}(\pm) \;=\; \mu \sum_{j=1}^L  (\fraka_j^*\fraka_j  -\tfrac{1}{2}) 
  \;=\;  \frac{1}{2} \begin{pmatrix} \fraka^* & \fraka \end{pmatrix} \begin{pmatrix} \mu & 0 \\ 0 & -\mu \end{pmatrix} 
  \begin{pmatrix} \fraka \\ \fraka^* \end{pmatrix} 
 \; . 
$$
Hence the BdG Hamiltonian $H_\Lambda^\mathrm{Kit}(\pm)$ is already in diagonal form and it does not depend 
on the sign, so the canonical 
transformation is $W = \one_{2L}$ and
$$
   j( \HH_\Lambda^\mathrm{Kit}(\pm) ) \;=\; \mathrm{sgn}\,\mathrm{det}(\one) \;=\; 1 \; ,
\qquad
\mbox{for }w=0\;.
$$

\vspace{0.1cm}

Consider now the (orthogonal) shift operator 
\begin{equation}  \label{eq:per_antiper_shift}
  (V_\pm \frakb)_j = \begin{cases} \frakb_{j+1}, & 1\leq j\leq 2L-1\,, \\ \pm \frakb_1, & j=2L\,, \end{cases} \qquad 
  \mathrm{det}(V_\pm) = \mp 1 \; .
\end{equation}
Recall the Kitaev Hamiltonian with periodic or anti-periodic boundary conditions 
from Equation \eqref{eq:Per_aPer_Kitaev}. We compute that 
$$
 {\bf H}_\Lambda^\mathrm{Kit}(\pm) \;=\; 
 \ii w \sum_{j=1}^{L-1}   \frakb_{2j}\frakb_{2j+1}  \pm \ii w \, \frakb_{2L}\frakb_1 \\
 \;=\; \frac{\ii w}{2} \, \frakb^t V_\pm^*  \begin{pmatrix} 0 & \one \\ -\one & 0 \end{pmatrix}  V_\pm \frakb
$$
Therefore, we see that $V_\pm$ diagonalises the skew-symmetric matrix $A_\Lambda(\pm)$ 
in the Kitaev chain with periodic or anti-periodic boundary conditions.
%
Because $\mathrm{det}(V_\pm) = \mp 1$, we see that the periodic and anti-periodic 
chains have different phase labels. 
\begin{equation}
\label{eq-KitaevIndNonTrivial}
 j( \HH_\Lambda^\mathrm{Kit}(\pm) ) \;=\; \mathrm{det} (V_\pm) \;=\; \mp 1 \; ,
\qquad
\mbox{for }\mu=0\;.
\end{equation}
Furthermore, this
$\mathbb{Z}_2$-index can be detected by the 
parity of the ground state $|\psi_\pm \rangle$ from Proposition \ref{prop:per_aper_Kit_GS}.
The matrix $V_-$ can be connected 
to the identity via a continuous path. This path can then be used to connect the 
anti-periodic Kitaev chain to the trivial chain.

\subsection{Other examples} \label{subsec:Other_examples}

Here we study some non-translation invariant interactions and ground states. 
This also prepares the ground for the study of a flux insertion through a chain, which 
merely consists of a modification of a few matrix elements.

\subsection*{Double-sided chain}

The basic Hamiltonian is the following
\begin{align*}
  \mathbf{H}_{[-L,L]} \; &= \;  \sum_{j=-L}^{L-1} w_j \big[ -(\fraka_j^* \fraka_{j+1} + \fraka_{j+1}^*\fraka_j ) + 
     (e^{i\theta}\fraka_j \fraka_{j+1} + e^{-i\theta}\fraka_{j+1}^* \fraka_j^*) \big] 
 \; + \; \sum_{j=-L}^{L}  \mu_j ( \fraka_j^*\fraka_j - \tfrac{1}{2})  \\
   &=\;  \sum_{j=-L}^{L-1}   w_j \,\ii\, \frakb_{2j} \frakb_{2j+1} + \sum_{j=-L}^{L}  \tfrac{\mu_j}{2}  \,\ii\, \frakb_{2j-1}\frakb_{2j} \; ,
    \qquad w_j \, , \, \mu_j \in \mathbb{R} \, \text{ for all } \, j \; .
\end{align*}

One can roughly think of $\{\ii \, \frakb_{2j} \frakb_{2j+1}\}_{j=-L}^{L-1}$  as 
playing the role of a spin site and $\{\ii \, \frakb_{2j-1}\frakb_{2j}\}_{j=-L}^L$ specifying 
an external field. In particular, for $|\mu_j|$ small, the sign of $w_j$  
determines the `spin-orientation' of the ground state space at site $j$.

 \vspace{0.2cm} 
  
\noindent  {\bf Case: $w_j = 0$ for all $j$} \\
If there are only the diagonal terms $\mu_j ( \fraka_j^*\fraka_j - \tfrac{1}{2})$, the ground state 
space is determined 
by the sign of $\mu_j$ at each site. If $\mu_j >0$, then the vacuum $| \Omega_j\rangle$ at site $j$ will be the ground state 
of $\mu_j ( \fraka_j^*\fraka_j - \tfrac{1}{2})$. If $\mu_j < 0$, then $\fraka_j^*| \Omega_j\rangle$ is the 
ground state with energy $\frac{\mu_j}{2}$. One can describe the total ground state as a product of the ground state at each site. 
To write this down, we assume $\mu_j \neq 0$ and 
introduce $s_{\mu_j} = 0$ if $\mu_j >0$ and $s_{\mu_j} = 1$ if $\mu_j <0$. 
Then the ground state is 
$$
  | \psi \rangle \; = \; \prod_{j=-L}^L  (\fraka_j^*)^{s_{\mu_j}} | \Omega \rangle \; .
$$
If $\mu_{k_1}= \cdots = \mu_{k_m} =0$ for some $m\geq 1$, then 
$$
\big\{ \fraka_{k_j}^* |\psi\rangle\big\}_{j=1}^m
\;,
\qquad
\mbox{with }\;  
| \psi \rangle \; = \; \prod_{\substack{ j=-L, \\ j \neq k_l}}^L  (\fraka_j^*)^{s_{\mu_j}} | \Omega \rangle \;, 
$$
are all ground states and so there is an extra degeneracy.

\vspace{0.2cm}

\noindent  {\bf Case: $\mu_j = 0$ and $w_j\neq 0$ for all $j$} \\
This corresponds to the non-periodic Kitaev (quantum Ising) chain 
$$
  \mathbf{H}_{[-L,L]} \; = \;  \sum_{j=-L}^{L-1}   w_j \,\ii\, \frakb_{2j} \frakb_{2j+1} \; , \qquad 
  [ \mathbf{H}_{[-L,L]},  \frakb_{2L} ] \; =\;  [ \mathbf{H}_{[-L,L]},  \frakb_{-2L-1} ] \;=\; 0 \; .
$$
Let us assume for the time being that $w_j \neq 0$ for all $j$. Then the ground state space at site $j$ is spanned 
by the $\pm 1$ eigenspace of the self-adjoint unitary $i \frakb_{2j} \frakb_{2j+1}$ depending on the 
sign of $w_j$. Using again $s_{w_j}$ to be $0$ or $1$ if $w_j$ is positive or negative, 
one can write down ground states explicitly via the operators $\{\frakd_j\}_{j=-L}^{L-1}$, 
\begin{align*}
  &\frakd_j \;=\; \tfrac{1}{2}( \frakb_{2j} + (-1)^{s_{w_j}} \ii  \frakb_{2j+1} )\; ,   
  &&(2 \frakd_j^* \frakd_j - 1) \; =\; (-1)^{s_{w_j}} \ii  \frakb_{2j}\frakb_{2j+1} \; , \\
  &\{ \frakd_i^*, \frakd_j\} \;=\; \delta_{i,j}\, \one\; ,  &&\{\frakd_i, \frakd_j\} \;=\; 0\; .
\end{align*}
Indeed, one has 
\begin{equation}  \label{eq:dbl_sided_spin_simple}
  \mathbf{H}_{[-L,L]} \; = \;  \sum_{j=-L}^{L-1}  (-1)^{s_{w_j}} w_j \, (2\frakd_j^*\frakd_j -1) \; , 
\end{equation}
where all coefficients in the sum are now positive. Analogous to the case of the 
Kitaev chain on the one-sided chain with open boundary conditoins, the vector 
$$
   | \psi \rangle = \prod_{j=-L}^{L-1} \frakd_j | \Omega \rangle 
$$
is a non-zero ground state with energy $\sum_{j=1}^{L-1} (-1)^{s_{w_j}+1} w_j$. 
Because $\frakd_j$ is odd for all $j$, 
we have that $\mathcal{P}| \psi \rangle = | \psi \rangle$.
Now $\mathbf{H}_{[-L,L]}$ commutes with $\frakb_{-2L-1}$ and $\frakb_{2L}$
and this leads to a degeneracy of the ground state space that will be investigated next.
Let us consider the boundary operator 
$\frakd_\bound = \frac{1}{2}( \frakb_{2L} + \ii  \frakb_{-2L-1})$ which satisfies the CAR relations 
with the other $\frakd_j$ operators. Either $\frakd_\bound | \psi \rangle$ or 
$\frakd_\bound^* | \psi \rangle$ is also a ground state of the Hamiltonian 
({\it cf.} Lemma \ref{lem-vanishing}) that is, moreover, odd. 
To determine which one should be used, let us first note that 
$$
  \mathcal{P} \;=\; \prod_{j=-L}^L (-\ii  \frakb_{2j-1} \frakb_{2j} ) 
   \;=\; \ii  \frakb_{2L} \frakb_{-2L-1} \prod_{j=-L}^{L-1} (-\ii  \frakb_{2j} \frakb_{2j+1}) 
   \;=\; (2 \frakd_\bound^* \frakd_\bound -1) \prod_{j=-L}^{L-1} (-1)^{s_{w_j}} (1 - 2\frakd_j^* \frakd_j) \; .
$$
Let $i_\bound \in\{0,1\}$ be the occupancy number $\frakd_\bound$, {\it i.e.} 
$\frakd_\bound^{(0)} = \frakd_\bound$, $\frakd_\bound^{(1)} = \frakd_\bound^\ast$.
Then $\mathcal{P} \frakd_\bound^{(i_\bound)} | \psi \rangle = - \frakd_\bound^{(i_\bound)} |\psi \rangle$. 
This will be compared with 
\begin{align*}
  \mathcal{P}\, \frakd_\bound^{(i_\bound)} | \psi \rangle \;&=\;  
    (2 \frakd_\bound^* \frakd_\bound -1) \prod_{j=-L}^{L-1} (-1)^{s_{w_j}} (1 - 2\frakd_j^* \frakd_j) \,
       \frakd_\bound^{(i_\bound)} \frakd_{-L}\cdots \frakd_{L-1} | \Omega \rangle \\
  &=\;  (2 \frakd_\bound^* \frakd_\bound -1) \frakd_\bound^{(i_\bound)}  \prod_{j=-L}^{L-1} (-1)^{s_{w_j}} (1 - 2\frakd_j^* \frakd_j) \,
       \frakd_{-L}\cdots \frakd_{L-1} | \Omega \rangle \\
  &=\; (-1)^{1+ i_\bound} \frakd_\bound^{(i_\bound)} \Big( \prod_{j=1}^{L-1} (-1)^{s_{w_j}} \Big) \frakd_{-L}\cdots \frakd_{L-1} | \Omega \rangle \\
  &=\; (-1)^{1+ i_\bound} (-1)^{\sum_{j=1}^{L-1} s_{w_j} } \, \frakd_\bound^{(i_\bound)} | \psi \rangle \; .
\end{align*}
Suppose that there are $M$ sites with $w_j < 0$.
If $M$ is odd, then $\frakd_\bound^* |\psi \rangle$ is a ground state and 
$\frakd_\bound |\psi \rangle = 0$. If $M$ is even, then 
$\frakd_\bound |\psi \rangle$ is a ground state and $\frakd_\bound^* |\psi \rangle =0$. 
We then see that if we change the orientation of a \emph{single} spin site, $w_{j_0} \mapsto - w_{j_0}$, 
then the ground state space changes.

\vspace{0.3cm}

\noindent  {\bf Case: $\mu_j = 0$, $w_{j_1}=\cdots = w_{j_k}=0$ for $k< 2L$} \\
We now consider the more degenerate case, where some of the spin coefficients 
$\{w_{j_i}\}_{i=1}^k$ are zero with $k < 2L$. 
Let $Z = \{j_1, \ldots, j_k\}\subset [-L,L]\cap \mathbb{Z}$ be the set of labels for the $0$-coefficient 
spin-sites. Then the  Hamiltonian can be written 
$$
\mathbf{H}_{[-L,L]} \;=\;  \sum_{\substack{ j \in [-L,L]\cap \mathbb{Z}, \\ j \notin Z}}  w_j \,\ii\, \frakb_{2j}\frakb_{2j+1} \, .
$$
The techniques of the previous section still apply. 
In particular, we still have that 
\begin{align*}
  &\mathbf{H}_{[-L,L]} \; = \;  \sum_{\substack{ j \in [-L,L]\cap \mathbb{Z}, \\ j \notin Z}} (-1)^{s_{w_j}} w_j \, (2\frakd_j^*\frakd_j -1) \; , 
  &&\frakd_j \;=\; \frac{1}{2}( \frakb_{2j} + (-1)^{s_{w_j}} \ii  \frakb_{2j+1} )\; ,
\end{align*}
and the vector 
$$
   | \psi \rangle = \prod_{\substack{ j \in [-L,L]\cap \mathbb{Z}, \\ j \notin Z}} \frakd_j | \Omega \rangle 
$$
is a ground state. We now consider the extra degeneracy, where  
the commuting family of self-adjoint unitaries 
$\{\ii \, \frakb_{2j} \frakb_{2j+1}\}_{j\in Z}$ commute with the Hamiltonian and also the 
ground state projection. Therefore, the vectors 
$\big\{ \frac{1}{2} (\frakb_{2j} + \ii  \frakb_{2j+1} )| \psi \rangle \big\}_{j \in Z}$ are also  
a family of linearly independent ground states. 
As previously, either $\frakd_\bound | \psi \rangle$ or 
$\frakd_\bound^* | \psi \rangle$ is another ground state. Therefore in total we have 
a $(k+2)$-fold degeneracy with  $k = |Z|$.

\subsection*{Closed chain}

The Hamiltonian of study will again be the (non-trivial) Kitaev chain but without 
translation invariance of interactions,
\begin{align}  \label{eq:non-periodic-closed-Kitaev}
  \mathbf{H}_L \;&=\; \sum_{j=1}^{L-1} w_j \big[ -(\fraka_j^* \fraka_{j+1} + \fraka_{j+1}^*\fraka_j ) + 
     (e^{i\theta}\fraka_j \fraka_{j+1} + e^{-i\theta}\fraka_{j+1}^* \fraka_j^*) \big]   \nonumber \\
     &\qquad \quad  +
     w_L\big[ -(\fraka_L^* \fraka_{1} + \fraka_1^*\fraka_L)  + (e^{i\theta}\fraka_L\fraka_1 + e^{-i\theta}\fraka_1^*\fraka_L^*) \big]  \nonumber \\
   &=\; \sum_{j=1}^{L-1}  w_j \, \ii  \frakb_{2j}\frakb_{2j+1}  + 
     w_L\,  \ii \frakb_{2L} \frakb_1  \; .
\end{align}
We again let $s_{w_j}$ be such that $(-1)^{s_{w_j}} w_j$ is non-negative.
As previously, the 
ground state is given by the $(-1)^{s_{w_j}+1}$ eigenspaces of the commuting self-adjoint 
unitaries $\{\ii \, \frakb_{2j} \frakb_{2j+1}\}_{j=1}^{L-1}$ and $\ii \, \frakb_{2L} \frakb_1$.
We again characterise the ground state space by the operators 
$\{\frakd_j\}_{j=1}^{L-1}$ and $\frakd_\bound$, where 
\begin{align*}
  &\frakd_j \;=\; \frac{1}{2}( \frakb_{2j} + (-1)^{s_{w_j}} \ii  \frakb_{2j+1} )\; ,   
  &&\frakd_\bound \;=\; \frac{1}{2}( \frakb_{2L} + (-1)^{s_{w_L}} \ii  \frakb_{1} )\; , \\
  &(2 \frakd_j^* \frakd_j - 1) \; =\; (-1)^{s_{w_j}} \ii  \frakb_{2j}\frakb_{2j+1} \; , 
  &&(2 \frakd_\bound^* \frakd_\bound - 1) \; =\; (-1)^{s_{w_L}} \ii  \frakb_{2L}\frakb_{1} \; , 
\end{align*}
and
\begin{equation*} 
  \mathbf{H}_{L} \; = \;  \sum_{j=-L}^{L-1}  (-1)^{s_{w_j}} w_j (2 \frakd_j^*\frakd_j  - \one)
  + (-1)^{s_{w_L}}  w_L (2 \frakd_\bound^* \frakd_\bound - \one)  \; 
\end{equation*}
with each coefficient $\{(-1)^{s_{w_j}} w_j\}_{j=1}^L$ strictly positive.

\begin{proposi} \label{prop:closed_chain_GS_generic}
Let $s_P = \sum_{j=1}^L s_{w_j}$ be the number of spin sites with negative orientation. 
\begin{enumerate}
  \item[{\rm (i)}] If $L$ and $s_P$ have the same parity, then  
  $\frakd_0\cdots \frakd_{L-1}| \Omega\rangle$ is a ground state of $\mathbf{H}_L$.
  \item[{\rm (ii)}] If $L$ and $s_P$ have different parity, then 
  $\frakd_\bound \frakd_1 \cdots \frakd_{L-1}| \Omega\rangle$ is a ground 
  state of $\mathbf{H}_L$. 
\end{enumerate}
\end{proposi}
\noindent {\bf Proof.} 
Again let  $i_\bound\in \{0,1\}$ 
be the occupancy number, that is, $\frakd_\bound^{(0)} = \frakd_\bound$ and
$\frakd_\bound^{(1)} = \frakd_\bound^*$.
We note that $\frakd_\bound^{(i_\bound)} \frakd_1 \cdots \frakd_{L-1}| \Omega\rangle$ has 
parity $(-1)^{L}$. We also use that 
$$
 \Pp \;=\; \ii \frakb_{2L}\frakb_1 \prod_{j=1}^{L-1} (-\ii \,\frakb_{2j}\frakb_{2j+1}) \;=\; 
 (-1)^{s_{w_L}} (2\frakd_\bound^*\frakd_\bound - 1) \prod_{j=1}^{L-1} (-1)^{s_{w_j}} (1-2\frakd_j^*\frakd_j) \; ,
$$
so 
\begin{align*}
\Pp \, \frakd_\bound^{(i_\bound)} \frakd_1 \cdots \frakd_{L-1}| \Omega\rangle  \;&=\; 
 (-1)^{s_{w_L}} (2\frakd_\bound^*\frakd_\bound - 1) \prod_{j=1}^{L-1} (-1)^{s_{w_j}} (1-2\frakd_j^*\frakd_j) \, 
 \frakd_\bound^{(i_\bound)} \frakd_1 \cdots \frakd_{L-1}| \Omega\rangle \\
 &=\; \Big( (-1)^{s_{w_L}+i_\bound+1} \prod_{j=1}^{L-1} (-1)^{s_{w_j}} \Big) \, 
  \frakd_\bound^{(i_\bound)} \frakd_1 \cdots \frakd_{L-1}| \Omega\rangle  \\
 &=\; (-1)^{i_\bound+1+s_P} \, \frakd_\bound^{(i_\bound)} \frakd_1 \cdots \frakd_{L-1}| \Omega\rangle \; .
\end{align*}
Now, if $L$ and $s_P$ are even, then 
$\frakd_\bound \frakd_1\cdots \frakd_{L-1}|\Omega\rangle$ will have even and odd parity 
and so will vanish. Hence $\frakd_1\cdots \frakd_{L-1}|\Omega\rangle$ minimises 
the term $(-1)^{s_{w_L}}2w_L \,\frakd_\bound^*\frakd_\bound $ and gives a ground state.
If $L$ is even and $s_P$ odd, then 
$\frakd_\bound \frakd_1\cdots \frakd_{L-1}| \Omega\rangle$ has consistent 
parity (the term with $\frakd_\bound^*$ does not) and so will minimise $\mathbf{H}_L$.
If $L$ is odd and $s_P$ even, then 
$\frakd_\bound \frakd_1\cdots \frakd_{L-1}| \Omega\rangle$ is again non-zero and 
hence is a ground state.
If $L$ and $s_P$ are odd, then $\frakd_\bound \frakd_1\cdots \frakd_{L-1}|\Omega\rangle$ 
will have odd and even parity and so must be zero. Hence 
$\frakd_1\cdots \frakd_{L-1}|\Omega\rangle$ is a ground state.
\hfill $\Box$

\subsection{Ground state gap}

The Hamiltonians that we have considered so far are given by sums of commuting 
projections. For such models, it is relatively straight-forward to show that the 
Hamiltonian has  a uniformly 
bounded ground state energy gap. For more general situations, a common technique 
to show a uniformly bounded ground state energy gap is to employ the Martingale 
method~\cite[Section 5]{NSY17}.
In order to introduce the method, in this section we will show how it can be applied to the simple models 
we have considered thus far.

\subsection*{Double-sided chain}

Let us consider the case of the spin chain with nearest-neighbour interactions. For convenience, we would 
like the ground state energy to be $0$, so take the Hamiltonian
\begin{equation} \label{eq:spinH_normalised}
   \mathbf{H}_{[-L,L]}  \;=\;  \sum_{j=-L}^{L-1} \ii  w_j\, \frakb_{2j}\frakb_{2j+1}  -  E_G \,\one\; , \qquad 
   E_G \;=\; \sum_{j=-L}^{L-1} (-1)^{s_{w_j}+1} w_j \; , \qquad w_j \;\neq\; 0 \; .
\end{equation}
Let us first define a sequence of Hamiltonians $\{\mathbf{H}_n\}_{n=0}^L \subset (A_{[-L,L]\cap\mathbb{Z}}^\mathrm{car})^0$ 
where $\mathbf{H}_0 = 0$ and 
$$
   \mathbf{H}_n \;=\; \sum_{j=-n}^{n-1} w_j(   \ii  \, \frakb_{2j}\frakb_{2j+1}  + (-1)^{s_{w_j}} \one) \; .
$$
Thus we have a non-decreasing sequence of non-negative Hamiltonians such that the 
kernels $\mathcal{G}_n = \Ker( \mathbf{H}_n )$ form a non-increasing sequence of subspaces 
$$
  \mathcal{F}(\mathbb{C}^{2L+1}) \;=\; \mathcal{G}_0 \;\supset \;\mathcal{G}_1 \;\supset \cdots 
    \supset\; \mathcal{G}_L \;=\;  \mathcal{H}_\text{GS}  \; .
$$
Now let $h_n = \mathbf{H}_n - \mathbf{H}_{n-1}$ and let $g_n$ be the kernel projection 
of $h_n$. In this case, using Equation \eqref{eq:dbl_sided_spin_simple}, 
$$  
h_n \;=\;   2(-1)^{s_{w_{-n}}} w_{-n} \, \frakd_{-n}^* \frakd_{-n} + 2(-1)^{s_{w_{n-1}}} w_{n-1} \frakd_{n-1}^* \frakd_{n-1} \; .
$$
Hence $\frakd_{-n}\frakd_{n-1} \cdot \mathcal{F}(\mathbb{C}^{2L+1})$ is the ground state space of $h_n$. 
Alternatively, the kernel is determined by the $(-1)^{s_{w_{-n}}+1}$ and $(-1)^{s_{w_{n-1}}+1}$-eigenspaces 
of $i\frakb_{-2n}\frakb_{-2n+1}$ and $i\frakb_{2n-2}\frakb_{2n-1}$. Hence
\begin{align*}
  h_n \;&=\;  (-1)^{s_{w_{-n}}} w_{-n} \big(1 + (-1)^{s_{w_{-n}}} \ii  \frakb_{-2n}\frakb_{-2n+1} \big) 
          + (-1)^{s_{w_{n-1}}} w_{n-1} \big( 1+ (-1)^{s_{w_{n-1}}} \ii  \frakb_{2n-2}\frakb_{2n-1} \big)  \\
    &=\; (-1)^{s_{w_{-n}}} \frac{ w_{-n}}{2} P_{(-1)^{s_{w_{-n}}}} + (-1)^{s_{w_{n-1}}} \frac{  w_{n-1}}{2} P_{(-1)^{s_{w_{n-1}}}} \\
    &\geq \; \gamma_n ( \one - g_n )\; , \qquad\qquad\qquad \gamma_n = \mathrm{min}\big\{ \tfrac{|w_{-n}|}{2}, \tfrac{|w_{n-1}|}{2} \big\} \; ,
\end{align*}
where $P_{\pm 1}$ is the projection onto the $\pm 1$ eigenspace. 
If we take $\gamma = \mathrm{min}_j \big\{ \tfrac{|w_{-j}|}{2}\} >0$, then for any 
$0 \leq n \leq L$, $h_n \geq \gamma (\one - g_n )$.
Next let us introduce the projections 
\begin{align*}
  &E_n \;=\; \begin{cases} \one - P_{\Ker(\mathbf{H}_1)} \, , & n = 0 \, , \\ P_{\Ker(\mathbf{H}_n)} - P_{\Ker(\mathbf{H}_{n+1})} \, , & 1 \leq n \leq L-1 \, \\ 
        P_{\Ker(\mathbf{H}_L)} \, , & n = L \end{cases} \; ,   &&E_n E_m \;=\; \delta_{n,m} E_n \; , 
      &&\sum_{n=1}^L E_n \;=\; \one \; .
\end{align*}
In this case, one has  explicitly 
\begin{align*}
   E_n \; = \; \begin{cases} \one -   \frac{1}{2} \big( 1 - (-1)^{s_{w_{-1}}} \ii  \frakb_{-2}\frakb_{-1} \big) \, , & n = 0 \, , \\
      \one - \frac{1}{2}\big( 1- (-1)^{s_{w_{-n-1}}} \ii  \frakb_{-2n-2} \frakb_{-2n-1} \big) 
        \frac{1}{2} \big( 1- (-1)^{s_{w_n}} \ii \frakb_{2n}\frakb_{2n+1} \big) \, ,   &  1 \leq n \leq L-1 \, , \\
        \prod\limits_{j=-L}^{L-1} \frac{1}{2} \big( 1 - (-1)^{s_{w_j}} \ii \frakb_{2j}\frakb_{2j+1} \big) \, , &  n=L \end{cases} \; .
\end{align*}
Similarly, we have that $g_{n+1} = P_{\Ker(h_{n+1})}$ can be written as
$$
  g_{n+1} \;=\; \tfrac{1}{2} \big( 1 - (-1)^{s_{w_{-n-1}}} \ii  \frakb_{-2n-2}\frakb_{-2n-1} \big) \, 
    \tfrac{1}{2} \big( 1 - (-1)^{s_{w_{n}}} \ii  \frakb_{2n}\frakb_{2n+1} \big) \; . 
$$
One can then check that $[E_n, g_{n+1}]=0$ and $E_n g_{n+1} E_n = 0$ for 
$0\leq n \leq L-1$.
We therefore satisfy the hypothesis of~\cite[Theorem 5.1]{NSY17}, which 
implies the following result.

\begin{proposi} \label{prop:uniform_gap_quadratic}
The Hamiltonian from Equation \eqref{eq:spinH_normalised} 
{with $\mathrm{min}_{-L\leq j\leq L} \big\{ \tfrac{|w_{-j}|}{2}\} >0$} 
uniformly in $L$ has a spectral gap above 
the ground state energy that is uniform in the size of the chain $[-L,L]\cap \mathbb{Z}$.
\end{proposi}

Recalling Proposition \ref{prop:finite_uniform_gap_gives_GNS_gap}, 
Proposition \ref{prop:uniform_gap_quadratic} guarantees that the 
infinite volume GNS Hamiltonian $h_\omega$ coming from the weak-$\ast$ limit 
of the finite-volume ground states will have a spectral gap above $0$.

\vspace{0.3cm}

\noindent  {\bf Case: $\mu_j = 0$ and $w_j =0$ for $j\in Z$, a fixed finite set}  \\
Next we consider the case of extra degeneracy in the finite chains. To this end we fix a 
set of sites with $w_j=0$ \emph{that will not change as $L$ increases}. That is, we 
start with a sufficiently large $L$.
Given such a set $Z$, we enumerate the set $[-L,L]\cap \mathbb{Z} \setminus Z$ 
by $\{j_1, \ldots, j_{N}\}$ with $j_{i} < j_{i+1}$. This allows to define the sequence 
$$
  0\;=\; \mathbf{H}_0 \;\leq\; \mathbf{H}_1 \;\leq\;  \cdots \;\leq\; \mathbf{H}_N \;=\; \mathbf{H}_{[-L,L]} \; ,
$$
where
$$
   \mathbf{H}_n \;=\; \sum_{j = j_1}^{j_n} w_j \big( \ii \, \frakb_{2j}\frakb_{2j+1} + (-1)^{s_{w_j}} \one \big) \; .
$$
Again suppose that there is a strictly positive $0 < \gamma$ with $\gamma < \mathrm{min}\{ \frac{|w_j|}{2} \,:\, w_j \neq 0 \}$. 
As in the non-degenerate case,  
we define  $h_n  = \mathbf{H}_n - \mathbf{H}_{n-1}$, $g_n = P_{\Ker(h_n)}$ and 
\begin{align*}
  &E_n \;=\; \begin{cases} \one - P_{\Ker(\mathbf{H}_1)} \, , & n = 0 \, , \\ P_{\Ker(\mathbf{H}_n)} - P_{\Ker(\mathbf{H}_{n+1})} \, , & 1 \leq n \leq N-1 \, , \\ 
        P_{\Ker(\mathbf{H}_L)} \, , & n = N \, , \end{cases} \;    &&E_n E_m \;=\; \delta_{n,m} E_n \; , 
      &&\sum_{n=1}^N E_n \;=\; \one \; .
\end{align*} 
Note that in the degenerate picture, $P_{\Ker(\mathbf{H}_1)}$ is a larger projection than in the case 
$w_j \neq 0$ for all $j$. However, one can still follow the previous method of 
argument without issue, where
we have that $h_n  \geq \gamma (\one - P_{\Ker(h_n)})$, 
$[E_n, g_{n+1}]=0$ and $E_n g_{n+1} E_n = 0$ for 
$0\leq n \leq N-1$.
Therefore the Martingale method applies again, which will ensure that in the thermodynamic 
limit $L \to \infty$ (which implies $N\to \infty$), the infinite volume ground state is gapped. 

\vspace{0.1cm}

The system with $w_j = 0$ for a fixed finite set is the same as the  system with $w_j\neq 0$ 
up to a finite-rank operator. Hence the  GNS representations of the infinite volume 
ground states  will 
be unitarily equivalent ({\it cf.}~\cite[Example 6.2.56]{BR2}).

\subsection*{Closed chain}
Finally we study the ground state gap of the Hamiltonian 
\begin{align*}  
  \mathbf{H}_L \;&=\; 
    \sum_{j=1}^{L-1}  w_j \big( \ii  \frakb_{2j}\frakb_{2j+1} + (-1)^{s_{w_j}} \one \big) + 
     w_L \big(\ii  \frakb_{2L} \frakb_1 + (-1)^{s_{w_L}}\one \big) \; , 
  \quad s_{w_j} = \begin{cases} 0 \, ,  & w_j \,\geq\, 0 \, , \\ 1 \, ,  & w_j \,< \, 0\; , \end{cases} 
\end{align*}
where again $0 < \gamma \leq \frac{1}{2} |w_j|$ for all $j$.
Because the details of the proof are 
very similar to the case of the open chain, some details will be skipped.

\vspace{0.2cm}

We define the sequence of non-negative Hamiltonians $\{\mathbf{H}_n\}_{n=0}^L$ 
with $\mathbf{H}_0 = 0$, 
 $\mathbf{H}_L$, as before and 
$$
  \mathbf{H}_n \;=\; \sum_{j=1}^{n} w_j \big( \ii  \frakb_{2j}\frakb_{2j+1} + (-1)^{s_{w_j}} \one \big) \; , 
    \qquad 1 \; \leq n \; \leq \; L-1 \; .
$$
The operators of interest for the Martingale method are 
$h_n = \mathbf{H}_n - \mathbf{H}_{n-1}$, $g_n = P_{\Ker(h_n)}$, where in this case
$$
  h_n \;=\; \begin{cases} w_1\big( \ii \frakb_2\frakb_3 + (-1)^{s_{w_1}} \one \big) \, , & n\,=\, 1\, , \\ 
    w_1\big( \ii \frakb_{2n}\frakb_{2n+1} + (-1)^{s_{w_n}} \one \big) \, ,  &  2\,\leq\,n\,\leq\, L-1\, , \\
    w_L\big( \ii \frakb_{2L}\frakb_{1} + (-1)^{s_{w_L}} \one \big)\, , & n\,=\, L\, ,  \end{cases} \;  \qquad 
 g_n \;=\; \frac{1}{2} \big( \one - (-1)^{s_{w_n}} \ii \frakb_{2n} \frakb_{2n+1} \big) \; .
$$
By the Spectral Theorem, 
$$
  h_n \;=\;  \frac{w_n}{2} \big(1+ (-1)^{s_{w_n}} \ii  \frakb_{2n}\frakb_{2n+1} \big) 
  \;=\; \frac{w_n}{2} \big(1- P_{\Ker(h_n)} \big)  \;\geq \; \gamma \big( 1- g_n \big) \; 
$$
for $0<\gamma \leq \mathrm{min}_j \frac{|w_j|}{2}$.
We also have the family of projections 
\begin{align*}
  &E_n \;=\; \begin{cases} \one - P_{\Ker(\mathbf{H}_1)} \, , & n = 0 \, , \\ P_{\Ker(\mathbf{H}_n)} - P_{\Ker(\mathbf{H}_{n+1})} \, , & 1 \leq n \leq L-1 \, \\ 
        P_{\Ker(\mathbf{H}_L)} \, , & n = L \end{cases} \; ,   &&E_n E_m \;=\; \delta_{n,m} E_n \; , 
      &&\sum_{n=1}^L E_n \;=\; \one \; .
\end{align*}
Again 
$$
  E_n \;=\; \begin{cases}  \one - \frac{1}{2}\big( \one - (-1)^{s_{w_1}} \ii  \frakb_2 \frakb_3 \big) \,,\, & n\,=\, 0\, , \\
     P_{\Ker(\mathbf{H}_n)} \big( \one - g_{n+1} \big) \, ,   &  1\, \leq \, n \,\leq\, L-1 \, , \\
     \Big( \prod_{j=1}^{L-1} \big( \one - (-1)^{s_{w_j}} \ii \frakb_{2j}\frakb_{2j+1} \big) \Big) 
     \big( \one - (-1)^{s_{w_j}} \ii \frakb_{2L}\frakb_{1}\big) \, ,  &  n\,=\, L \,  \end{cases}\; 
$$
and it is straight-forward to check that $[E_n, g_{n+1}]=0$ and $E_n g_{n+1} E_n = 0$ for 
$0\leq n \leq L-1$. Thus the hypotheses of the Martingale method are satisfied and 
one has the following.

\begin{proposi} \label{prop:closed_chain_uniform_gap}
The Hamiltonian in Equation \eqref{eq:non-periodic-closed-Kitaev} has a spectral gap 
above the ground state energy that is uniform in the length $L$ of the chain.
\end{proposi}

\subsection{Flux insertion and \texorpdfstring{$\mathbb{Z}_2$}--valued spectral flow} \label{sec:closed_chain_flux}
Recall from \eqref{rk:finite_chain_index_as_sf} on page \pageref{rk:finite_chain_index_as_sf} 
that the $\mathbb{Z}_2$-index for 
quadratic chains can be interpreted as a (finite-dimensional) $\mathbb{Z}_2$-valued spectral  
flow between skew-symmetric matrix $A_\Lambda$ (or equivalently $\ii H_\Lambda$) and 
its diagonalisation. Here we further investigate such applications 
of the $\mathbb{Z}_2$-valued spectral flow by considering a flux insertion in 
closed fermionic chains.

\vspace{0.2cm}

Let us first note that we can immediately use the concatenation properties of the 
$\mathbb{Z}_2$-valued spectral flow to establish a path between the Kitaev (or quantum Ising) model with periodic and anti-periodic 
chains. Namely, for $V_\pm$ as in Equation \eqref{eq:per_antiper_shift},
$$
{\mathrm{Sf_2}(V_+ A_\Lambda V_+^*, \, V_- A_\Lambda V_-^*)} \;=\; 
{\mathrm{Sf_2}(V_+ A_\Lambda V_+^*, \, A_\Lambda)} \,
{\mathrm{Sf}_2( A_\Lambda, \, V_- A_\Lambda V_-) }
   \;=\; \mathrm{det}(V_+) \, \mathrm{det}(V_-) \;=\; -1\;,
$$
and so the $\mathbb{Z}_2$-valued spectral flow is non-trivial. This result is also immediate from 
Proposition \ref{prop:sf_as_Z2_obstruction},  though we would like 
to show this in a more physically meaningful way.

\vspace{0.2cm}

We insert a flux term into the  closed chain that plays the role of switching the  
boundary conditions from periodic to anti-periodic. Such a system was previously 
studied in~\cite{TwistedKitaev}.
The Hamiltonian is
\begin{align*}
  \mathbf{H}_\Lambda^\mathrm{Kit}(\alpha)
  \;&=\;
\sum_{j=1}^{L-1}\big(-w\,(\fraka_j^*\fraka_{j+1}+\fraka_{j+1}^*\fraka_{j})
+\Delta\,\fraka_j\fraka_{j+1}+\overline{\Delta}\,\fraka_{j+1}^*\fraka_{j}^*
\big) + \mu \sum_{j=1}^L (\fraka_j^*\fraka_{j}-\tfrac{1}{2})  \\
  &\qquad + \big( -w(e^{-\ii \alpha}\fraka_L^*\fraka_1 + e^{\ii \alpha}\fraka_1^*\fraka_L) + 
      \Delta e^{{\ii \alpha}}\fraka_L\fraka_1 + \ol{\Delta} e^{{-\ii \alpha}}\fraka_1^*\fraka_L^* \big)
\;.
\end{align*}
One clearly has that $\mathbf{H}_\Lambda^\mathrm{Kit}(0) = \mathbf{H}_\Lambda^\mathrm{Kit}(+)$ and 
$\mathbf{H}_\Lambda^\mathrm{Kit}(\pi) = \mathbf{H}_\Lambda^\mathrm{Kit}(-)$. 
In the case $w=\Delta=0$, the Hamiltonian is constant throughout the 
deformation of $\alpha$ and, hence, will have no $\mathbb{Z}_2$-valued spectral flow. 
In the case $\Delta=e^{\ii \theta}w$ and $\mu=0$, however, one can again 
re-write the Hamiltonian in the Majorana representation as 
$$
  \mathbf{H}_\Lambda^\mathrm{Kit}(\alpha) = \ii w \sum_{j=1}^{L-1} \frakb_{2j}\frakb_{2j+1} 
    + \ii w \cos(\alpha)\frakb_{2L}\frakb_1 - \ii w \sin(\alpha) \frakb_{2L}\frakb_2 \; ,
$$
where the following identity was used:
$$
\frakb_{2L}\frakb_2
\;=\;
\fraka_L^*\fraka_1 - \fraka_1^*\fraka_L - e^{\ii \theta}\fraka_L\fraka_1 
+e^{-\ii \theta} \fraka_1^*\fraka_L^*
\;.
$$
The following result also follows from \eqref{eq-KitaevIndNonTrivial} combined with Proposition \ref{prop:sf_as_Z2_obstruction}, but 
we provide a separate proof.

\begin{proposi} \label{prop:Maj_flux_flow}
The $\mathbb{Z}_2$-valued spectral flow defined by the path $\alpha\in[0,\pi]\mapsto\mathbf{H}_\Lambda^\mathrm{Kit}(\alpha)$ is 
non-trivial { in the case $\Delta=e^{\ii \theta}w$ and $\mu=0$.}
\end{proposi}
\noindent {\bf Proof.} 
Recalling that the Majorana operators are ordered in column vector 
$\frakb=\binom{\frakb_\odd}{\frakb_\ev}$, 
the skew-adjoint matrix from $ \mathbf{H}_\Lambda^\mathrm{Kit}(\alpha)$ is given by 
$$
  A_\Lambda(\alpha) \;=\; 
  \frac{w}{2} \begin{pmatrix}  & & & & & -\cos(\alpha) \\  & & &  -1 & &  \\ & & & & -\one_{L-2} & \\ 
& 1 & & & & \sin(\alpha) \\ & & \one_{L-2} & & &  \\ 
 \cos(\alpha) & & &  -\sin(\alpha)  & & \end{pmatrix}\; .
$$
In particular, one can connect $A_\Lambda(\pi) = V A_\Lambda(0) V^*$, where 
$$
V 
\;=\; 
\begin{pmatrix}  & & & 1 \\ & & U & \\ & -U & & \\ 1 & & & \end{pmatrix} \;, 
\qquad 
U 
\;=\; 
\begin{pmatrix}  & & 1 \\ &  \iddots & \\  1 & & \end{pmatrix} \in \calO_{L-1}\; .
$$
Then
$$
  \mathrm{Sf}_2(\alpha\in[0,\pi]\mapsto A_\Lambda(\alpha)) 
  \;=\; 
  \mathrm{sgn}\,\mathrm{det}(V) \;=\; -1 \;,
$$
as required.
\hfill $\Box$

\vspace{0.2cm}

As the $\mathbb{Z}_2$-valued spectral flow is non-trivial, one expects a double degenerate level crossing at the 
midpoint of the path. Indeed, the Hamiltonian is 
$$
   \mathbf{H}_\Lambda^\mathrm{Kit}(\tfrac{\pi}{2}) \;=\; \ii w\sum_{j=1}^{L-1}\frakb_{2j}\frakb_{2j+1}  -\ii w \frakb_{2L} \frakb_2 
  \;=\; \ii w\sum_{j=2}^{L-1} \frakb_{2j}\frakb_{2j+1} + \ii w\frakb_2(\frakb_3 + \frakb_{2L})\; .
$$
One can then check that $\mathbf{H}_\Lambda^\mathrm{Kit}(\tfrac{\pi}{2})$ commutes with the anti-commuting 
self-adjoint unitaries $\frakb_1$ and $\tfrac{1}{\sqrt{2}}(\frakb_3 - \frakb_{2L})$. Hence 
if $|\psi\rangle$ is a ground state of $\mathbf{H}_\Lambda^\mathrm{Kit}(\tfrac{\pi}{2})$, then so is 
$\frakb_1| \psi\rangle$ and $\tfrac{1}{\sqrt{2}}(\frakb_3 - \frakb_{2L})|\psi \rangle$.

\vspace{0.2cm}

By Proposition \ref{prop:closed_chain_uniform_gap},  
$\mathbf{H}_L(0)$ and $\mathbf{H}_L(\pi)$ are known to have a uniformly bounded ground state gap. 
Therefore, the ground state energy gap of $\mathbf{H}_L(\alpha)$ goes to $0$ as $\alpha \to \frac{\pi}{2}$.

\begin{remark} \label{rk:Z2_flow_in_any_flux}
{\rm 
We can readily extend Proposition \ref{prop:Maj_flux_flow} to say that a flux insertion that changes 
the orientation of any single spin site $i\frakb_{2j}\frakb_{2j+1}$ will give a non-trivial $\mathbb{Z}_2$-spectral flow. 
Thus, while there are many examples of Hamiltonians on the closed chain with a uniformly bounded 
ground state gap (Proposition \ref{prop:closed_chain_uniform_gap}), this gap can be closed by a 
local perturbation.  One reason for this behaviour is that  
a fermionic Hamiltonian on a closed chain becomes 
highly non-local under the Jordan--Wigner transformation, which is often utilized in proofs of 
the stability of the ground state energy gap.
}
\hfill $\diamond$
\end{remark}

\noindent  {\bf Flux insertion in two cells}   \\
Here we briefly show that adding a magnetic flux through two unit cells does 
not  substantially change the  system. 
The Hamiltonian is
\begin{align*}
  \tilde{\mathbf{H}}_{L}(\alpha) \;&=\; w\sum_{j = 2}^{L-1} \big[ -( \fraka_j^* \fraka_{j+1} + 
  \fraka_{j+1}^* \fraka_j) +    \fraka_j \fraka_{j+1} + \fraka_{j+1}^* \fraka_{j+1}  \big]  \\
    &\quad + w\big[ -(e^{\ii \alpha} \fraka_L^* \fraka_1 + e^{-\ii \alpha}\fraka_1^* \fraka_L) +
        e^{\ii \alpha}\fraka_L \fraka_1 + e^{-\ii \alpha} \fraka_1^* \fraka_L^* \big] \\
    &\quad + w\big[ -(e^{-\ii \alpha} \fraka_1^* \fraka_2 + e^{\ii \alpha}\fraka_2^* \fraka_1) +
        e^{\ii \alpha}\fraka_1 \fraka_2 + e^{-\ii \alpha} \fraka_2^* \fraka_1^* \big]  \; ,
 \end{align*}
 where for simplicity we have set the phase factor $\theta=0$.
In the Majorana representation
 \begin{align*}
 \tilde{\mathbf{H}}_{L}(\alpha) \;&=\;  w\sum_{j=2}^{L-1} \ii \frakb_{2j}\frakb_{2j+1} 
       + w\big(\cos(\alpha) \ii \frakb_L\frakb_1 - \sin(\alpha)\ii \frakb_L \frakb_2\big) 
       + w \big( \cos(\alpha) \ii \frakb_2\frakb_3 + \sin(\alpha) \ii \frakb_1 \frakb_3 \big) \\
     &=\; w\sum_{j=2 }^{L-1} \ii \frakb_{2j}\frakb_{2j+1}  
        + \ii w\frakb_1\big( \sin(\alpha)\frakb_3 - \cos(\alpha)\frakb_L\big) 
        + \ii w \frakb_2 \big( \cos(\alpha) \frakb_3 + \sin(\alpha) \frakb_L \big) \; .
\end{align*}
A careful check shows that for any $\alpha$ the operators $\ii \frakb_1 ( \sin(\alpha)\frakb_3 - \cos(\alpha)\frakb_L)$ 
and $\ii  \frakb_2 ( \cos(\alpha) \frakb_3 + \sin(\alpha) \frakb_L )$ are commuting 
self-adjoint unitaries that also commute with the other terms $\ii \frakb_{2j}\frakb_{2j+1}$ in 
the Hamiltonian. Hence the ground state space can be explicitly characterised by the 
$-1$ eigenstate of each self-adjoint unitary in the sum.

\vspace{0.2cm}

We can again define the CAR operators 
$$
  \frakd_j \;=\; \begin{cases} 
       \frac{1}{2} \big( \frakb_1 + \ii (\sin(\alpha)\frakb_3 - \cos(\alpha) \frakb_L) \big) \, , &  j \,=\, 1 \, ,  \\  
    \frac{1}{2} \big( \frakb_{2j} + \ii  \frakb_{2j+1} \big) \, ,  & 2 \,\leq\, j \,\leq\, L-1\, ,\,    \\
   \frac{1}{2} \big( \frakb_2 + \ii (\cos(\alpha)\frakb_3 + \sin(\alpha)\frakb_L) \big) \, , &  j \,=\, L \, 
    \end{cases} \; .
$$
Then the  Hamiltonian can once again be written as 
$$
   \tilde{\mathbf{H}}_{L}(\alpha) \;=\; w\sum_{j=1}^{L}  (2 \frakd_j^* \frakd_j - 1)
$$
and so any ground state must look like $\prod_j  \frakd_j | \psi \rangle$.

\vspace{0.2cm}

While the specific characterisation of the ground state space depends $\alpha$, the key spectral 
properties of $\tilde{\mathbf{H}}_{L}(\alpha)$ do not. In particular, 
the Martingale method used to show the ground state gap of 
$\tilde{\mathbf{H}}_{L}(0)$ and $\tilde{\mathbf{H}}_{L}(\pi)$ in 
Proposition \ref{prop:closed_chain_uniform_gap} also remains valid 
along the deformation. 

\vspace{0.2cm}

In this case, we have that $j(\tilde{\mathbf{H}}_{L}(0)) = j(\tilde{\mathbf{H}}_{L}(\pi)) = -1$ 
and the ground state gap remains uniformly bounded along 
the path $\tilde{\mathbf{H}}_L(\alpha)$ connecting the two Hamiltonians. 
As is perhaps to be expected of a $\mathbb{Z}_2$-invariant, changing 
the orientation of a single spin site will cause a $\mathbb{Z}_2$-phase change. But simultaneously changing 
the orientation of two spin sites can be done without closing the ground state gap.
{This also follows by inserting the two fluxes consecutively and applying 
the concatenation property of the $\ZM_2$-valued spectral flow.}


\section{Higher order interactions on finite chains} \label{sec:higher_order}

\subsection{Gapped ground states in finite volume systems}

Let us now turn or attention to even interactions on finite chains  that need not be quadratic. 
We say that two finite-volume Hamiltonians 
$\mathbf{H}_\Lambda(0)$ and $\mathbf{H}_\Lambda(1)$ are in the same gapped phase if 
there is a $C^1$-path of finite volume Hamiltonians 
$s\in[0,1]\mapsto \{\mathbf{H}_\Lambda(s)\}_\Lambda$ connecting 
$\mathbf{H}_\Lambda(0)$ and $\mathbf{H}_\Lambda(1)$ and with the property that there is a 
spectral gap above the ground state energy of $\mathbf{H}_\Lambda(s)$ for all $s$ that is
\emph{uniform} in $|\Lambda|$.

\vspace{0.2cm}

In this section, we consider Hamiltonians with higher-order interactions and paths 
between gapped Hamiltonians where the ground state gap may close, indicating 
that such Hamiltonians have distinct topological phase labels. As in the case of 
quadratic interactions, one way we will induce such gap closings is via a local flux insertion.

\subsection*{Parity and gap closing}

We note a result that is mathematically simple but 
has important physical consequences.

\begin{lemma} \label{lem:opp_parity_gives_gap_closing}
Let $\mathbf{H}_\Lambda(0)$ and $\mathbf{H}_\Lambda(1)$ be 
parity-symmetric Hamiltonians on the fermionic Fock space $\mathcal{F}_\Lambda$ 
with $\Lambda$ finite. Suppose $\mathbf{H}_\Lambda(0)$ and $\mathbf{H}_\Lambda(1)$ 
have unique ground states with opposite parity. Then the ground state gap 
will close along any continuous 
path $\mathbf{H}_\Lambda(s)$  connecting  $\mathbf{H}_\Lambda(0)$ and $\mathbf{H}_\Lambda(1)$ 
with the property that $\Pp \mathbf{H}_\Lambda(s) \Pp = \mathbf{H}_\Lambda(s)$ for all $s \in[0,1]$.
\end{lemma}
\noindent {\bf Proof.} 
Provided that we include multiplicity, we can take a continuous enumeration of 
the eigenvalues $\{\lambda_j(s)\}$  of $\mathbf{H}_\Lambda(s)$, where each 
$\lambda_j:[0,1]\to \mathbb{R}$ is continuous~\cite[Chapter 2, {\S}5]{Kato}. 
Because the ground state eigenvalues of the Hamiltonians at the end points of 
the path have opposite parity and we restrict to parity-symmetric paths, 
$\Pp \mathbf{H}_\Lambda(s) \Pp = \mathbf{H}_\Lambda(s)$,  there must 
be at least one $s_0 \in (0,1)$ such that the lowest energy eigenvalue projection at $s_0$
is discontinuous. Such a discontinuity must come from a double degeneracy or 
crossing of eigenvalues.
\hfill $\Box$

\vspace{0.2cm}

We remark that while the previous statement is mathematically trivial, it can be 
applied to finite volume Hamiltonians with arbitrarily large interaction terms. 
The much more non-trivial question for finite volume systems is to find 
a physically interesting pair of Hamiltonians with unique  ground states 
 with opposite parity. 
A large and important class of such Hamiltonians can be constructed using 
fermionic matrix product states of even and odd parity~\cite{BWH}.
Another more involved question is to what extent 
an index derived from the parity of ground state eigenvectors still makes sense in the infinite volume limit.

\subsection{The interacting Kitaev chain}
Here we summarise the key results of~\cite{KST}. Starting from the 
Kitaev Hamiltonian $\mathbf{H}_\Lambda^\mathrm{Kit}$ from Equation \eqref{eq:Kitaev_model}, 
one adds a quartic interaction term with damping parameter $K\geq 0$,
\begin{align*}
  \mathbf{H}_\Lambda^\mathrm{int} \;&=\; \sum_{j=1}^{L-1}\big[-w(\fraka_j^* \fraka_{j+1} + \fraka_{j+1}^*\fraka_j) + 
  \Delta\,\fraka_j\fraka_{j+1}+\overline{\Delta}\,\fraka_{j+1}^*\fraka_{j}^* \big] \\
   &\quad -  \frac{1}{2}\sum_{j=1}^L \mu_j (\fraka_j^* \fraka_j - 1) + K \sum_{j=1}^{L-1}(2\fraka_j^*\fraka_j -1)(2\fraka_{j+1}^*\fraka_{j+1}-1)\; .
\end{align*}
We note that the term $\frac{1}{2}\sum_{j} \mu_j (\fraka_j^* \fraka_j - 1)$ is now negative. We do 
this to better align our results with~\cite{KST} as the map $\mu_j \to -\mu_j$ does change the 
spectrum  of the Hamiltonian (though the ground state in the trivial case 
$K=w = \Delta=0$ is now spanned by the occupied state rather than the vacuum).

\vspace{0.2cm}

We can again consider the spin-chain analogue of the  interacting chain. 
Recalling the Jordan--Wigner transformation,
\begin{align*}
   &\sigma_j^x \;=\; \Big( e^{-i\pi \sum_{k=1}^{j-1} \fraka_k^* \fraka_k} \Big) \fraka_j^* \; , \qquad 
   \sigma_j^y \;=\; \Big( e^{ i\pi \sum_{k=1}^{j-1} \fraka_k^* \fraka_k} \Big) \fraka_j \; , \qquad 
   \sigma_j^z \;=\; 2 \fraka_j^* \fraka_j - \one \; , \\
 &\qquad \qquad \qquad 
   \frakb_{2j-1} \;=\; \Big( \prod_{k=1}^{j-1} \sigma_k^z \Big) \sigma_j^x \; , \qquad 
   \frakb_{2j} \;=\; \Big( \prod_{k=1}^{j-1} \sigma_k^z \Big) \sigma_j^y\; ,
\end{align*}
the interacting chain maps to the XYZ chain in a magnetic field
$$
  \mathbf{H}_\Lambda^\text{spin} \;=\; \sum_{j=1}^{L-1}\big( -J_x \sigma_j^x \sigma_{j+1}^x - J_y \sigma_j^y\sigma_{j+1}^y 
    + J_z \sigma_j^z \sigma_{j+1}^z \big) - \frac{1}{2} \sum_{j=1}^L \mu_j \sigma_j^z \; ,
$$
with $J_x = (w+\Delta)/2$, $J_y = (w-\Delta)/2$, $J_z = K$. 
See~\cite{Baxter} for properties and analysis on the XYZ chain and related models.

\vspace{0.2cm}

One of the key achievements of~\cite{KST} is that on a certain line in the parameter space, 
the Hamiltonian $\mathbf{H}_\Lambda^\mathrm{int}$ becomes frustration-free (that is, the ground state 
simultaneously minimises each interaction term).

\begin{theo}[\cite{KST}] \label{thm:Interacting_GS}
Let $\Delta \in \mathbb{R}$, $\mu_2=\mu_3=\cdots =\mu_{L-1} = \mu_e$ and $\mu_1=\mu_L = \tfrac{\mu_e}{2}$ with 
$\mu_e = 4\sqrt{K^2+wK + \frac{w^2-\Delta^2}{4}}$. Then
\begin{enumerate}
    \item[{\rm (i)}] $\mathbf{H}_\Lambda^\mathrm{int}$ has an explicit frustration-free and double degenerate ground state.
  \item[{\rm (ii)}] There is a $C^1$-path $\mathbf{H}_\Lambda(t)$ such that 
  $\mathbf{H}_\Lambda(0) = \mathbf{H}_\Lambda^\mathrm{Kit}$, the quadratic Hamiltonian 
  from Equation \eqref{eq:Kitaev_model}, 
  and $\mathbf{H}_\Lambda(2K) = \mathbf{H}_\Lambda^\mathrm{int}$, the quartic Hamiltonian. 
  \item[{\rm (iii)}] For all $t\geq 0$, $\mathbf{H}_\Lambda(t)$ has a double degenerate  ground state.
  \item[{\rm (iv)}] For all $t\geq 0$, $\mathbf{H}_\Lambda(t)$ has a spectral gap above the ground state 
  energy that is uniform in $|\Lambda|$.
\end{enumerate}
\end{theo}

 As noted in~\cite{KST}, the equation for $\mu_e$ from Theorem \ref{thm:Interacting_GS} 
 that ensures the interacting Kitaev chain 
has a frustration-free ground state has a direct analogue for the XYZ chain in a magnetic 
field, {\it cf.}~\cite{KTM82}.

\subsection{Flux insertion and gap closing in the closed chain} \label{sec:closed_interacting_chain}

Let us now insert a local flux into the interacting Kitaev chain.
Our analysis closely follows~\cite[Appendix D]{TwistedKitaev}, who considered the 
interacting Kitaev chain with 
twisted boundary conditions. 
We add periodic boundary conditions to the  Hamiltonian with a local flux,
\begin{align*}
  \mathbf{H}_\Lambda^\mathrm{int}(\alpha) \; &=\; -w(e^{-\ii \alpha}\fraka_1^* \fraka_2 + e^{\ii \alpha} \fraka_2^* \fraka_1) 
    + w( e^{\ii  \alpha} \fraka_1 \fraka_2 + e^{-\ii \alpha} \fraka_2^* \fraka_1^*) \\
    &\quad + \sum_{j=2}^{L-1} \big( -w( \fraka_j^* \fraka_{j+1} + \fraka_{j+1}^* \fraka_j) + w( \fraka_j \fraka_{j+1} + \fraka_{j+1}^* \fraka_{j+1} ) \big)  \\
    &\quad -w(\fraka_L^*\fraka_1 + \fraka_{1}^*\fraka_L) + w(\fraka_L\fraka_1 + \fraka_1^*\fraka_L^*)
      - \frac{1}{2} \sum_{j=1}^L \mu_j (\fraka_j^* \fraka_j -1)  \\
    &\quad + K \sum_{j=1}^{L-1}(2\fraka_j^*\fraka_j -1)(2\fraka_{j+1}^*\fraka_{j+1}-1) 
      + K(2\fraka_L^* \fraka_L - 1)(2\fraka_1^*\fraka_1 - 1) \; .
\end{align*}
We choose a local flux to emphasise 
that highly local perturbations in closed chains are capable of closing a 
uniformly bounded ground state gap in the closed chain. This is in direct contrast to typical properties 
of ground states with Local Topological Quantum Order, where small perturbations  
will not close the ground state gap~\cite{MZ,MN18}.

\vspace{0.2cm}

We write the Hamiltonian as a sum $\mathbf{H}_\Lambda^\mathrm{int}(\alpha) = \sum_{j=1}^L h_j(\alpha)$,  
where 
\begin{align*}
 h_1(\alpha) \;&=\;   w(-e^{-\ii \alpha}\fraka_1^* \fraka_2 - e^{\ii \alpha} \fraka_2^* \fraka_1 
     + e^{\ii  \alpha} \fraka_1 \fraka_2 + e^{-\ii \alpha} \fraka_2^* \fraka_1^*) \\
   &\quad - \frac{\mu_e}{2}( \fraka_1^* \fraka_1 + \fraka_{2}^*\fraka_{2}-1) +
     K ( 2\fraka_1^*\fraka_1 -1)(2\fraka_{2}^* \fraka_{2} -1 ) \; , \\
  h_j \;&=\;   w( -\fraka_j^* \fraka_{j+1} - \fraka_{j+1}^* \fraka_j +  \fraka_j \fraka_{j+1} + \fraka_{j+1}^* \fraka_{j}^* ) \\
   &\quad - \frac{\mu_e}{2}( \fraka_j^* \fraka_j + \fraka_{j+1}^*\fraka_{j+1}-1) +
     K ( 2\fraka_j^*\fraka_j -1)(2\fraka_{j+1}^* \fraka_{j+1} -1 ) \; , \quad  2 \leq j \leq l-1 \; 
\end{align*}
and lastly  
\begin{align*}
 h_L \;&=\;   w( -\fraka_L^* \fraka_{1} - \fraka_{1}^* \fraka_L +  \fraka_L \fraka_{1} + \fraka_{1}^* \fraka_{L}^* ) \\
   &\quad - \frac{\mu_e}{2}( \fraka_L^* \fraka_L + \fraka_{1}^*\fraka_{1}-1) +
     K ( 2\fraka_L^*\fraka_L -1)(2\fraka_{1}^* \fraka_{1} -1 ) \; .
\end{align*}

To study the flux insertion, we first explicitly solve the ground state space of 
$\mathbf{H}_\Lambda^\mathrm{int}(\alpha)$ at the end points $\alpha = 0$ and $\alpha=\pi$.
To assist our computations on the closed chain, we first determine the ground state 
space of the Hamiltonian with open boundary conditions 
$\sum_{j=1}^{L-1} h_j(\alpha)$.

\vspace{0.2cm}

When $\alpha = 0$ and $\mu_j$ are as in Theorem \ref{thm:Interacting_GS}, the ground states of 
the open chain are 
computed in~\cite{KST} as the pair 
$$
 A_{L,\alpha=0}^{\pm} | \Omega \rangle \;  :=\; (1 \pm \beta\fraka_1^*)(1 \pm \beta\fraka_2^*) \cdots (1 \pm \beta\fraka_L^*) | \Omega\rangle  \; , 
  \quad \beta^2 \;=\; \cot(\tfrac{\theta}{2}) \; , \; \; \theta \;=\; \arctan(\tfrac{2w}{\mu_e}) \in [0,\pi] \; .
$$
We note that 
$\mathcal{P} A_{L,\alpha=0}^{\pm} | \Omega \rangle =A_{L,\alpha=0}^{\mp} | \Omega \rangle$ for 
$\mathcal{P}$ the parity operator.

\vspace{0.2cm}

For the case of $\alpha = \pi$.
We can follow the same basic analysis as in~\cite{KST}, where we find that any state of the form 
$$
(1\pm \beta \fraka_1^*)( 1 \mp \beta\fraka_2^*)\, p(\fraka_3^*, \ldots, \fraka_{L}^*) | \Omega \rangle 
$$
is a ground state of $h_1(\pi)$, where $p(\fraka_3^*, \ldots, \fraka^{*}_{L})$ is a (non-zero) polynomial. 
Because the states $(1 \pm \beta \fraka_2^*)\cdots(1\pm \beta \fraka_L^*)| \Omega\rangle$ 
minimise $\{h_j\}_{j=2}^{L-1}$, we have that 
$$
 A_{L,\alpha=\pi}^{\pm} | \Omega \rangle \; :=\; (1 \mp \beta \fraka_1^*)(1 \pm \beta \fraka_2^*)( 1 \pm \beta \fraka_3^*) \cdots 
     (1 \pm \beta \fraka_L^*) | \Omega \rangle
$$
gives a double degenerate and frustration-free ground state for 
$\sum_{j=1}^{L-1} h_j(\pi)$ 
with $\mathcal{P} A_{L,\alpha=\pi}^{\pm} | \Omega \rangle = A_{L,\alpha=\pi}^{\mp} | \Omega \rangle$.

\vspace{0.2cm}

Let us now consider the closed chain $\sum_{j=1}^L h_j(\alpha)$. 
We first note that any state of the form 
$(1\pm \beta \fraka_L^*)(1\pm \fraka_1^*) p(\fraka_2^*,\ldots,\fraka_{L-1}^*)| \Omega\rangle$ will 
minimise $h_L$.
When $\alpha =0,\pi$, the ground state can be solved provided that we take 
the coefficients $\mu_1=\mu_2=\ldots=\mu_L = \mu_e$ from Theorem \ref{thm:Interacting_GS}. 

\vspace{0.2cm}

\noindent  {\bf Ground state at $\alpha=0$} \\
We first note that any (normalised) linear combination of 
$A_{L,\alpha=0}^{\pm} | \Omega \rangle$ will also give a ground state of $\sum_{j=1}^{L-1}h_j(0)$.
Therefore, we compute
\begin{align*}
  A_{L,\alpha=0}^+ - A_{L,\alpha=0}^- \;&=\; \big(A_{L-1,\alpha=0}^{+} + A_{L-1,\alpha=0}^{-}\big) (1+\beta \fraka_L^*) 
    - 2A_{L-1,\alpha=0}^-  \\
    &=\; (1+\beta \fraka_L^*) \big(A_{L-1,\alpha=0}^{+} + A_{L-1,\alpha=0}^{-}\big) - 2A_{L-1,\alpha=0}^- \\
    &=\; (1+\beta \fraka_L^*)(1+\beta \fraka_1^*)\cdots (1+\beta\fraka_{L-1}^*) - 
      (1-\beta \fraka_L^*) (1-\beta \fraka_1^*) \cdots (1-\beta\fraka_{L-1}^*) \; ,
\end{align*}
which shows that (the normalisation of) 
$A_{L,\alpha=0}^{+} | \Omega \rangle - A_{L,\alpha=0}^{-} | \Omega \rangle$ is a 
frustration-free ground state of 
$\mathbf{H}_\Lambda^\mathrm{int}(0)$ on the closed chain. The linearly independent vector 
$A_{L,\alpha=0}^{+} | \Omega \rangle + A_{L,\alpha=0}^{-} | \Omega \rangle$ is not a ground state as it 
does not minimise $h_L$, something that is verified by direct computation. 
In particular, 
$\Pp (A_{L,\alpha=0}^{+} | \Omega \rangle - A_{L,\alpha=0}^{-} | \Omega \rangle) 
= -(A_{L,\alpha=0}^{+} | \Omega \rangle - A_{L,\alpha=0}^{-} | \Omega \rangle)$ and the ground state is odd.

\vspace{0.2cm}

\noindent  {\bf Ground state at $\alpha=\pi$} \\
Again we consider normalised linear combinations 
of $A_{L,\alpha=\pi}^{\pm} | \Omega \rangle$, where we have that
\begin{align*}
  A_{L,\alpha=\pi}^+ + A_{L,\alpha=\pi}^{-}  \;&=\; \big( A_{L-1,\alpha=\pi}^+ - A_{L-1,\alpha=\pi}^-\big)(1+\beta \fraka_L^*) 
      + 2 A_{L-1,\alpha=\pi}^{-} \\
   &=\; (1- \beta \fraka_L^*)\big( A_{L-1,\alpha=\pi}^+ - A_{L-1,\alpha=\pi}^-\big) + 2 A_{L-1,\alpha=\pi}^{-} \\
   &=\; (1-\beta\fraka_L^*)(1-\beta\fraka_1^*)(1+\beta\fraka_2^*)\cdots(1+\beta\fraka_{L-1}^*) \\
     &\quad + (1+\beta\fraka_L^*)(1+\beta\fraka_1^*)(1-\beta\fraka_2^*)\cdots (1-\beta \fraka_{L-1}^*) \; .
\end{align*}
Hence, the normalisation of 
$A_{L,\alpha=\pi}^{+} | \Omega \rangle + A_{L,\alpha=\pi}^{-} | \Omega \rangle$ is a 
frustration-free ground 
state of $\mathbf{H}_\Lambda^\mathrm{int}(\pi)$ on the closed chain with even parity. 
In contrast, the 
vector $A_{L,\alpha=\pi}^{+} | \Omega \rangle - A_{L,\alpha=\pi}^{-} | \Omega \rangle$ does not 
minimise $h_L(\pi)$ and so is not a ground state.

\vspace{0.2cm}

Because the interacting Kitaev chain with flux has a unique   
ground state at the endpoints $\alpha=0,\pi$ but with 
opposite parity, we can apply Lemma \ref{lem:opp_parity_gives_gap_closing} and 
obtain that the ground state gap closes along the path 
$\mathbf{H}_\Lambda(\alpha)$. As we will show in Proposition \ref{prop:closed_chain_GS_path}, 
this is despite the fact that the endpoints have a uniformly bounded ground state gap and 
we take a local flux only. 

\subsection*{Connection to Kitaev's $\mathbb{Z}_2$-index}


\begin{proposi} \label{prop:closed_chain_GS_path}
The Hamiltonians $\mathbf{H}_\Lambda^\mathrm{int}(0)$ and $\mathbf{H}_\Lambda^\mathrm{int}(\pi)$ can be connected 
to quadratic Hamiltonians by a $C^1$-path along which the ground state gap is uniformly bounded.
\end{proposi}
\noindent {\bf Proof.} 
Similar to the case of Theorem \ref{thm:Interacting_GS}, we write an explicit path 
connecting the interacting and 
non-interacting Hamiltonians.

For $\alpha=0$, recalling the constant 
$\theta = \arctan(\tfrac{2w}{\mu_e})$ and using the notation 
$\frakn_j = \fraka_j^* \fraka_j$, we take the following path 
$\mathbf{H}_\Lambda(0,t)=\sum_{j=1}^L h_j(0,t)$, 
where for $1\leq j \leq L-1$, 
\begin{align*}
   h_j(0,t) \;&=\; -\fraka_j^*\fraka_{j+1} - \fraka_{j+1}^*\fraka_j + (1+t)\sin(\theta)(\fraka_j \fraka_{j+1} + \fraka_{j+1}^* \fraka_j^*) 
     - (1+t) \cos(\theta)(1- \frakn_j - \frakn_{j+1})  \\
       &\quad + \frac{t}{2} (2 \frakn_j - 1)(2 \frakn_{j+1} - 1) + 1+ \frac{t}{2} \; 
\end{align*}
and 
\begin{align*}
   h_L(0,t) \;&=\; -\fraka_L^*\fraka_{1} - \fraka_{1}^*\fraka_L + (1+t)\sin(\theta)(\fraka_L \fraka_{1} + \fraka_{1}^* \fraka_L^*) 
     - (1+t) \cos(\theta)(1- \frakn_L - \frakn_{1})  \\
       &\quad + \frac{t}{2} (2 \frakn_L - 1)(2 \frakn_{1} - 1) + 1+ \frac{t}{2} \; .
\end{align*}
We see that $\mathbf{H}_\Lambda(0,0)$ is the quadratic Hamiltonian on a closed chain 
studied in Section \ref{subsec:Other_examples} (up to a scaling of the 
constants) and $\mathbf{H}_\Lambda(0,2K)$ is the 
interacting chain with ground state energy shifted to $0$.
A direct computation gives that 
\begin{align*}
   h_j( 0, t) \;&=\; Q_jQ_j^* + (1+t) Q_j^* Q_j  \\
     Q_j \;&=\; \begin{cases} 
       \cos(\tfrac{\theta}{2})\big( -\fraka_j^*(1-\frakn_{j+1}) + \fraka_{j+1}^*(1-\frakn_j) \big)  
        - \sin(\tfrac{\theta}{2}) \big( \fraka_j\frakn_{j+1} + \fraka_{j+1} \frakn_j \big)\, ,  &  j \,\leq \, L-1\, \\
         \cos(\tfrac{\theta}{2})\big( -\fraka_L^*(1-\frakn_{1}) + \fraka_{1}^*(1-\frakn_L) \big)  
        - \sin(\tfrac{\theta}{2}) \big( +\fraka_L \frakn_{1} + \fraka_{1} \frakn_L \big)\,      &  j\,=\,L  
         \end{cases} \; , 
\end{align*}
which implies that $\mathbf{H}_\Lambda(0,t) \geq \mathbf{H}_\Lambda(0,0)$ for all $t\geq 0$. 
Furthermore, 
$Q_j A_{L,\alpha=0}^{\pm} | \Omega \rangle = Q_j^* A_{L,\alpha=0}^{\pm} | \Omega \rangle =0$, so 
$A_{L,\alpha=0}^{+} | \Omega \rangle - A_{L,\alpha=0}^{-} | \Omega \rangle$ is a 
$0$-energy ground state 
throughout the path. Because the ground state energy gap is uniformly bounded at $t=0$ 
by Proposition \ref{prop:closed_chain_uniform_gap}, the inequality 
$\mathbf{H}_\Lambda(0,t) \geq \mathbf{H}_\Lambda(0,0)$ then ensures that the ground 
state gap is uniformly bounded for all $t\geq 0$.

\vspace{0.2cm}

The case of $\alpha=\pi$ follows the same argument. In particular, we take 
$h_j(\pi,t) = h_j(0,t)$ for $j\geq 2$ and 
\begin{align*}
   h_1(\pi,t) \;&=\; \fraka_1^*\fraka_2 + \fraka_2^*\fraka_1 + (1+t)\sin(\theta)(-\fraka_1\fraka_2 - \fraka_2^* \fraka_1^*) 
     - (1+t) \cos(\theta)(1- \frakn_1 - \frakn_2)  \\
       &\quad + \frac{t}{2} (2 \frakn_1 - 1)(2 \frakn_2 - 1) + 1+ \frac{t}{2} \; .
\end{align*}
Similarly, we take 
$$
  Q_1 \;=\; \cos(\tfrac{\theta}{2})\big( \fraka_1^*(1-\frakn_2) + \fraka_2^*(1-\frakn_1) \big) 
        + \sin(\tfrac{\theta}{2}) \big( \fraka_1\frakn_2 - \fraka_2 \frakn_1 \big) \; .
$$
and $Q_j$ the same as $\alpha=0$ for $j\geq 2$.
\hfill $\Box$

\vspace{0.2cm}

Using the homotopy from Proposition \ref{prop:closed_chain_GS_path}, we can consider the path 
\begin{equation}  \label{eq:easy_sf_path}
  \mathbf{H}_\Lambda(0,2K) \; \xrightarrow{t} \; \mathbf{H}_\Lambda(0,0) \; \xrightarrow{\alpha} \; 
  \mathbf{H}_\Lambda(\pi, 0) \;\xrightarrow{t} \; \mathbf{H}_\Lambda(\pi,2K) \; 
\end{equation}
which connects $\mathbf{H}_\Lambda^\mathrm{int}(0)$ and $\mathbf{H}_\Lambda^\mathrm{int}(\pi)$
on the closed chain. Because there is no changes in the ground state space along 
the paths indexed by $t$, there can not be any $\mathbb{Z}_2$-valued spectral 
flow along these paths. By contrast, the path indexed by $\alpha$ will have a non-trivial 
$\mathbb{Z}_2$-valued spectral flow as discussed in Remark \ref{rk:Z2_flow_in_any_flux}. 
Therefore by the  
concatenation properties of the $\mathbb{Z}_2$-valued spectral flow, we 
obtain the following.  

\begin{proposi} \label{prop:closed_chain_taut_spec_flow}
The path of Hamiltonians given in Equation \eqref{eq:easy_sf_path} gives rise to 
a non-trivial 
$\mathbb{Z}_2$-valued spectral flow. In particular, the ground state 
gap closes at a point along the path and the 
ground state becomes doubly degenerate.
\end{proposi}


\section{Quasifree ground states of the infinite CAR algebra} \label{sec:quasifree_intro}

The remainder of the paper considers infinite systems and ground states of the CAR algebra 
$A^\mathrm{car}(\calH)$ over an infinite dimensional separable Hilbert space $\calH$. 
We are particularly interested 
in \emph{pure} ground states, which cannot be written as a convex combination of 
other states. A state $\omega$ is pure if and only if its 
GNS representation $\pi_\omega$ is irreducible \cite[Theorem 2.3.19]{BR1}. A key difference to CAR algebras 
over finite dimensional $\calH$ is that different pure states $\omega_0$ and $\omega_1$ of 
$A^\mathrm{car}(\calH)$ can give \emph{inequivalent} GNS representations, so there is 
no unitary $U:\mathfrak{h}_{\omega_0} \to \mathfrak{h}_{\omega_1}$ intertwining the 
representations. This can be used to distinguish pure ground states.

\vspace{0.2cm}

For this section, we will restrict to quasifree states on $A^\mathrm{car}(\calH)$ as they are 
more simple to work with. To determine criteria for pure quasifree states to be equivalent, 
it is useful to work with the \emph{self-dual} CAR algebra introduced by Araki~\cite{Araki}, 
where equivalence of representations of quasifree states can be reduced to a Hilbert-Schmidt condition 
({\it cf.} Theorem \ref{thm:Araki_quasifree} below). A more detailed introduction to quasifree states 
of the CAR algebra and their basic properties can be found in~\cite[Chapter 6]{EvansKawahigashi}.

\subsection{Quasifree states of the self-dual CAR algebra}
\label{sec-QFonSFCAR}

Let us fix a separable complex Hilbert space 
$\mathcal{H}$ and a real structure $\Gamma$, namely an anti-unitary involution. Typically we will be interested in the 
case that $\mathcal{H} = \calH_{\ph} = \ell^2(\Lambda) \otimes \mathbb{C}^2$ is a Nambu space with 
$\Lambda$ countable and particle-hole involution $\Gamma = \mathcal{C}(\one \otimes \sigma_1)$ with  
$\mathcal{C}$ complex conjugation. 
The self-dual CAR algebra $A^\mathrm{car}_\mathrm{sd}(\calH,\Gamma)$ 
is the $C^*$-algebra generated by $\one$ and $\frakc(v)$ for $v\in \calH$ such that 
$v\mapsto \frakc(v)$ is linear and with relations 
$$
  \frakc(v)^* \;=\; \frakc(\Gamma v) \; ,  \qquad \{\frakc(v)^*,\frakc(w)\} \;=\; \langle v,w \rangle_{\calH}\; .
$$
The self-dual CAR algebra is also graded with parity automorphism $\Theta$ such 
that $\frakc(v)$ is odd for all $v\in \calH$.
One recovers the more familiar CAR algebra by means of a \emph{basis projection}, 
which is a projection $E$ on $\calH$ such that $E + \Gamma E \Gamma = \one_\calH$. 
Given a basis projection, there is a graded isomorphism 
$\phi: A^\mathrm{car}(E\calH) \rightarrow A^\mathrm{car}_\mathrm{sd}(\calH,\Gamma)$ which on 
generators is given by 
\begin{equation}
\label{eq-fraks}
  \fraka^*(Ev)   \; \mapsto \frakc(Ev) \; \; ,  \qquad  \fraka(Ev)  \;\mapsto\; \frakc(\Gamma Ev) \; .
\end{equation}
In the case $\calH_{\ph} \cong \ell^2(\Lambda)\otimes \mathbb{C}^2\cong \ell^2(\Lambda)\oplus \ell^2(\Lambda)$ with $\Gamma = \mathcal{C}(\one \otimes \sigma_1)$, then 
analogous to the case of the usual CAR algebra, we can choose the canonical basis 
of $\calH_\ph$ and so $A^\mathrm{car}_\mathrm{sd}(\calH_{\ph},\Gamma )$ is 
the universal $C^*$-algebra generated by the 
elements $\{\frakc(j,k)\}_{(j,k)\in \Lambda\times\Lambda}$ satisfying the relations 
$$
    \frakc(j,k)^* \;=\; \frakc(k,j) \; ,   \qquad  \{\frakc(j_1,k_1)^*,\frakc(j_2,k_2)\} \;=\; 2\,\delta_{j_1,j_2} \, \delta_{k_1,k_2} \; .
$$
In the case of $\calH_\ph$, the basis projection $\tilde E(u_1, u_2) = u_1$ is of particular interest 
as  $\tilde{E}\,\calH_{\ph}=\ell^2(\Lambda)$ 
and \eqref{eq-fraks} leads to a concrete form of the isomorphism 
$\phi: A^\mathrm{car}_\Lambda\to  A^\mathrm{car}_\mathrm{sd}(\calH_{\ph},\Gamma )$ given by
\begin{equation}
\label{eq-fraks2}
  \phi(\fraka_j)  \;=\; \frakc(j,0) \;, \qquad 
  \phi^{-1}(\frakc(j,k)) \;=\; \fraka_j + \fraka_k^* \;.
\end{equation}

\begin{theo}[\cite{Araki}] \label{thm:Araki_quasifree}
Let $E$ be a basis projection on $\calH$. 
\begin{enumerate}
  \item[{\rm (i)}] There is a quasifree state $\omega_E$ on $A^\mathrm{car}_\mathrm{sd}(\calH,\Gamma)$ 
with  
$$
  \omega_E \big(\frakc(u)^* \frakc(v)\big) \;=\;  \langle u, Ev \rangle_{\calH} \; 
$$
which is extended to $A^\mathrm{car}_\mathrm{sd}(\calH,\Gamma)$ by the formulas 
\begin{align*}
  &\omega_E (\frakc(v_1)\cdots \frakc(v_{2n+1})) \;=\; 0 \; ,   \\
  &\omega_E (\frakc(v_1)\cdots \frakc(v_{2n})) \;=\; (-1)^{n(n-1)/2} \sum_{\sigma} (-1)^{\sigma} 
      \prod_{j=1}^n \omega_E \big( \frakc(v_{\sigma(j)}) \frakc(v_{\sigma(j+n)}) \big) \; 
\end{align*}
where the sum is over permutations $\sigma$ such that 
$$
  \sigma(1) < \sigma(2) < \ldots < \sigma(n) \; , \qquad \sigma(j) < \sigma(j+n) \; , \qquad  j\;=\; 1,\ldots, n \; .
$$
  \item[{\rm (ii)}] The state $\omega_E$ is pure and $\Theta$-invariant. In particular, 
  the GNS representation $(\mathfrak{h}_E, \pi_E, \Omega_E)$ associated to $\omega_E$ is irreducible.
\item[{\rm (iii)}] 
Let $E_0$ and $E_1$ be basis projections on $\calH$. The following statements 
  are equivalent:
  \begin{enumerate}
    \item[{\rm (1)}] The states $\omega_{E_0}$ and $\omega_{E_1}$ are unitarily equivalent. 
    \item[{\rm (2)}] The operator $E_0 - {E}_1$ is in the ideal of Hilbert-Schmidt operators.
  \end{enumerate}
\end{enumerate}
\end{theo}

The state $\omega_E$  is called the \emph{Fock state} associated to a basis projection $E$.
 From the state $\omega_{E}$ on $A^\mathrm{car}_\mathrm{sd}(\calH,\Gamma )$, 
we can use the isomorphism  
$\phi: A^\mathrm{car}(E\calH) \rightarrow A^\mathrm{car}_\mathrm{sd}(\calH,\Gamma )$  
from  Equation \eqref{eq-fraks} to 
get a state $\omega_E \circ \phi$ on $A^\mathrm{car}(E\calH)$.
In a slight abuse of notation, we will also denote this state by $\omega_{E}$ and call it a quasifree state on 
$A^\mathrm{car}(E\calH)$.
Given two basis projections $E_0$ and $E_1$ on a separable and infinite dimensional $\mathcal{H}$,  
the corresponding CAR algebras 
$A^\mathrm{car}(E_0\calH)$ and $A^\mathrm{car}(E_1\calH)$ are abstractly 
isomorphic by the universal property of the infinite CAR algebra~\cite[Theorem 5.2.5]{BR2}. 
Theorem \ref{thm:Araki_quasifree} then gives a sufficient and necessary condition 
for the irreducible GNS representations $\pi_{E_0}$ and $\pi_{E_1}$ to be unitarily equivalent.

\vspace{0.2cm}

Following~\cite[Section 6.6]{EvansKawahigashi}, let us us give some some further 
justification as to why $\omega_E$ is called a Fock state. 
Given a basis projection $E$ on $\calH$, 
let $(\mathfrak{h}_E, \pi_E, \Omega_E)$ be the GNS triple of $A^\mathrm{car}(E\calH)$.
Setting $\bigwedge^0 E\calH =\mathbb{C}\Omega_E$, the one-dimensional space spanned by the 
cyclic vector $\Omega_E$, one can identify 
\begin{equation} \label{eq:GNS_Fock}
    \mathfrak{h}_E \;\cong \;  \bigoplus_{n=0}^\infty \bigwedge\nolimits^{\! n} E\calH \; .
\end{equation}
Under this equivalence, the GNS representation of $A^\mathrm{car}(E\calH)$ can be written as 
$$
  \pi_E( \fraka^*(v) )\, u_1 \wedge \cdots \wedge u_n \;=\;  v \wedge u_1 \wedge \cdots \wedge u_n \; , \quad 
    \qquad v \, , \; u_j \in E\calH \; .
$$
That is, the cyclic vector $\Omega_E$ acts as the fermionic vacuum in the GNS space.

\subsection{Quasifree dynamics and BdG Hamiltonians}
\label{sec-QuasifreeDynBdG}

Let us now consider ground states $\omega$ on $A^\mathrm{car}_\mathrm{sd}(\calH,\Gamma)$ 
with respect to a strongly continuous $\mathbb{R}$-action
$\beta:\mathbb{R} \to \mathrm{Aut}(A^\mathrm{car}_\mathrm{sd}(\calH,\Gamma))$ with 
generator $\delta$, namely states satisfying  $-\ii\, \omega( a^* \delta(a)) \geq 0$ for all $a\in \mathrm{Dom}(\delta)$. 

\begin{defini}
The dynamics $\beta:\mathbb{R} \to \mathrm{Aut}(A^\mathrm{car}_\mathrm{sd}(\calH,\Gamma))$ 
is called quasifree  if $\beta_t(\frakc(v)) = \frakc( e^{\ii H t}v)$ for 
any $\frakc(v) \in A^\mathrm{car}_\mathrm{sd}(\calH,\Gamma)$ and where $H=H^*$ is 
an operator on $\calH$ such that $\Gamma H \Gamma = -H$. 
\end{defini}

The self-adjoint operator $H$ on $\calH$ that generates the quasifree dynamics 
$\beta$ plays the role of the Bogoliubov--de Gennes Hamiltonian in infinite 
systems, and will be referred to as the BdG Hamiltonian. 
Again, this operator comes with a natural particle-hole symmetry. Thus quasifree 
dynamics play an analogous role to quadratic interactions.

\begin{proposi}[\cite{EvansKawahigashi}, Proposition 6.37] \label{prop:gappedBdG_gappedGS}
Let $\beta:\mathbb{R} \to \mathrm{Aut}(A^\mathrm{car}_\mathrm{sd}(\calH,\Gamma))$ be a 
quasifree dynamics with BdG Hamiltonian $H$. If $0 \notin \sigma(H)$, then the 
Fock state $\omega_{E}$ associated to the spectral projection $E=\chi_{(0,\infty)}(H)$ is 
the unique ground state for the dynamics $\beta$. Furthermore, the 
GNS Hamiltonian $h_\omega$ on $\mathfrak{h}_{E}$ 
has a spectral gap above $0$.
\end{proposi}
\noindent {\bf Proof.} 
The particle-hole symmetry of $H$ implies that $\Gamma E \Gamma = \one_\calH - E$, 
so $E$ is a basis projection. 
The proof that $\omega_{E}$ is a ground state comes from the cited proposition. To show 
the spectral gap, we use the presentation of $\mathfrak{h}_{E}$ as 
a Fock space from Equation \eqref{eq:GNS_Fock}. In particular, the 
GNS Hamiltonian $h_\omega$ can be written as the the second quantisation 
of the BdG Hamiltonian $H$ restricted to antisymmetric tensors on $E \calH$. 
Because there is a strictly positive spectral gap around $0$ of  $\sigma(H)$ and  
$h_\omega$ comes from the restriction of $H$ to the positive spectral projection $E$, 
its second quantisation is  strictly positive. Hence there is 
some $\gamma >0$ such that $\sigma(h_\omega)\cap (0,\gamma) = \emptyset$.
\hfill $\Box$

\vspace{0.2cm}

Proposition \ref{prop:gappedBdG_gappedGS} shows that any BdG Hamiltonian 
$H$ on the Nambu space $(\calH,\Gamma)$ with a spectral gap at $0$ gives rise to a 
 basis projection $E = \chi_{(0,\infty)}(H)$ and a gapped pure ground state 
$\omega_{E}$ on $A^\mathrm{car}_\mathrm{sd}(\calH,\Gamma)\cong A^\mathrm{car}(E \calH)$. 
This process is reversible: given a basis projection  $E$ on $\calH$, one can define a
gapped BdG Hamiltonian $H = 2E - \one$. The quasifree state $\omega_E$ will then 
be the unique ground state for the quasifree dynamics generated by $H$.

\vspace{0.2cm}

Given two quasifree actions $\beta^{(0)}$ and $\beta^{(1)}$ on 
$A^\mathrm{car}_\mathrm{sd}(\calH,\Gamma)$ arising from gapped BdG Hamiltonians 
$H_0$ and $H_1$ on $\calH$, one has two basis projections $E_0$ and 
$E_1$. It is known that the two ground state representations $\pi_{E_0}$ 
and $\pi_{E_1}$ are equivalent if and only if $E_0 - E_1$ is 
Hilbert-Schmidt. Let us further investigate this issue by 
defining the skew-adjoint real unitary operators
$$
  J_k \;=\; \ii  H_k |H_k|^{-1} \;=\; \ii (2 E_k - \one ) \; , \qquad 
  J_k^* \;=\; -J_k  \; , \qquad J_k^2 \;=\; -\one \; , \qquad \Gamma J_k \Gamma \;=\; J_k \; . 
$$
Hence, the operators $J_k$ define a complex structure on the real Hilbert 
space $\calH^\Gamma_\mathbb{R} = \{ v \in \calH \, :\, \Gamma v = v \}$. 

\vspace{0.2cm}

We now make use of the following elementary fact.
\begin{lemma} \label{lem:On_transitive_on_J}
The orthogonal group acts transitively on the complex structures on a real Hilbert space.
\end{lemma}

\noindent {\bf Proof.} If $J$ is a complex structure on a real Hilbert space 
$\Hh_\RM$, it extends by linearity to a complex linear operator on the 
complexification $\Hh_\CM=\Hh_\RM\otimes\CM$ which is denoted 
by the same letter $J$. It is real in the sense that $J$ is equal to $\overline{J}=\Cc J\Cc$. 
This complexifation is skew-adjoint and unitary, 
so that $i J$ is a selfadjoint unitary on $\Hh_\CM$. By the spectral theorem 
and the reality of $J$, there is thus a projection $P$ on $\Hh_\CM$ such 
that $i J=2P-\one$ and $\overline{P}=\one-P$. 
Let $\Phi:\ell^2(\NM)\to\Hh_\CM$ be a frame for $P$, namely 
$\Phi\Phi^*=P$ and $\Phi^*\Phi=\one$. Here $\ell^2(\NM)$ is a complex 
Hilbert space equipped with complex conjugation $\Cc$ (entry-wise). 
Then $J=i(\Phi\Phi^*-\overline{\Phi}\,\overline{\Phi}^*)$ and 
$\Phi^*\overline{\Phi}=0$. Now set 
$V=2^{-\frac{1}{2}}\big(\Phi+\overline{\Phi},i\overline{\Phi}-i\Phi\big)$ 
which is real and unitary from $\ell^2(\NM)\oplus\ell^2(\NM)$ to $\Hh_\CM$. 
Moreover, one checks
$$
J
\;=\;
V\begin{pmatrix}
  0 & \one \\ -\one & 0
 \end{pmatrix}
V^*
\;.
$$
This is a normal form for $J$. Now given two complex structures 
$J_0,J_1$, there are two associated orthogonals 
$V_0,V_1:\ell^2(\NM)\oplus\ell^2(\NM)\to\Hh_\CM$. Set $W=V_1V_0^*$. 
This is an orthogonal on $\Hh_\CM$ which hence restricts to $\Hh_\RM$ as a 
linear opertor. One then has $J_1=WJ_0W^*$ which implies the claim.
\hfill $\Box$

\vspace{0.2cm}

Applying Lemma \ref{lem:On_transitive_on_J} to the complex structures 
$J_k = i(2E_k - \one)$ with $k=0,\,1$,
there exists a unitary $W \in \mathcal{U}(\calH)$ with properties 
$$
  J_1 \;=\; W J_0 W^* \; , \qquad W^*W \;=\; WW^* \;=\; \one \; , \qquad \Gamma W \Gamma = W \; .
$$
Hence $W$ is the infinite dimensional analogue of  the canonical transformations 
in Section \ref{sec-Bogoliobov} and so we continue to call such unitaries canonical 
transformations. 
In order for $W$ to give a Bogoluibov transformation on the second quantised Fock space 
$\mathcal{F}(E_0\calH) \to \mathcal{F}(E_1 \calH)$, the representations $\pi_{E_0}$ 
and $\pi_{E_1}$ must be equivalent,  
which occurs if and only if 
$[J_0, W]$ is Hilbert-Schmidt.


\begin{example}   \label{ex:inf_Kitaev_chain}
{\rm
(Kitaev chain): 
Let us briefly show how Theorem \ref{thm:Araki_quasifree} applies to the Kitaev chain on 
the infinite lattice $\Lambda = \mathbb{Z}$. To make the formulas a little simpler and as 
a preparation for another example in Section~\ref{sec:Z_2_index_as_sf},  let us choose 
the parameter $\Delta=-\ii w$. Then  the Kitaev Hamiltonian on a 
finite region $[a,b]\cap \mathbb{Z}$ becomes
\begin{equation}
\label{eq-KitaevMuw1}
\mathbf{H}_{[a,b]}^\mathrm{Kit}(\mu,w) 
\;=\; -\,w\sum_{j=a}^{b-1} \big[ \fraka_j^* \fraka_{j+1} + \fraka_{j+1}^*\fraka_j 
    \;+\;\ii \fraka_{j}\fraka_{j+1} -\ii \fraka_{j+1}^* \fraka_j^* \big] + \mu \sum_{j=a}^b \big( \fraka_j^* \fraka_j - \tfrac{1}{2}\big) \; . 
\end{equation}
The local Hamiltonians $\mathbf{H}_{[a,b]}^\mathrm{Kit}(\mu,w)$ give the infinite Kitaev chain  
which will be studied via 
the quasifree dynamics generated by BdG Hamiltonian $H_\mathbb{Z}$ defined on 
$\calH_\ph = \ell^2( \mathbb{Z}) \otimes \mathbb{C}^2$. As in \eqref{eq:BdGKitaev_model},  
\begin{equation}
\label{eq-KitaevMuw2}
    H^\mathrm{Kit}_\mathbb{Z}(\mu,w) \;=\; \begin{pmatrix} -w(S+S^*) - \mu & -\ii w(S^*-S)  \\ -\ii w(S^*-S) & w(S+S^*) + \mu \end{pmatrix} \; , 
\end{equation}
with $S$ the unilateral shift operator on $\ell^2(\mathbb{Z})$.

\vspace{0.1cm}

As in the case of finite chains, one expects a difference between the trivial region $w=0$ and 
the non-trivial region $\mu=0$. To compare these systems let us consider the unitary 
$$
  W \;=\;  \frac{\ii }{2} \begin{pmatrix} (\one +S) & \ii(\one -S) \\ \ii (\one -S) & -(\one +S) \end{pmatrix}\; , \qquad W^*W \;=\; WW^* \;=\; \one  \; , 
  \qquad \Gamma W \Gamma \;=\; W \; ,
$$
which has the property
$$
   W \begin{pmatrix} -\mu & 0 \\ 0 & \mu \end{pmatrix} W^* \;=\; 
     -\frac{\mu}{2}  \begin{pmatrix} (S+S^*) & \ii(S^* -S) \\ \ii(S^*- S) & -(S+S^*) \end{pmatrix} \; . 
$$
Hence $W$ maps the trivial system  
$H^\mathrm{Kit}_\mathbb{Z}(\mu, 0)$ to the non-trivial Hamiltonian $H^\mathrm{Kit}_\mathbb{Z}(0,\tfrac{\mu}{2})$. 
Passing to the spectrally flattened complex structures, $W (-\ii \sigma_z) W^* = J$ with 
$J$ the complex structure associated to the Kitaev chain 
$H^\mathrm{Kit}_\mathbb{Z}(0,\tfrac{1}{2})$. We note that $\mathrm{Ad}_W$ plays the role of the Kramers--Wannier 
automorphism in the quantum Ising chain.

\vspace{0.1cm}

By Theorem \ref{thm:Araki_quasifree}, the ground states for parameters $(\mu,0)$ and $(0,\frac{\mu}{2})$
are equivalent if and only if 
$[-\ii \sigma_z,W]$ is Hilbert-Schmidt. But 
this is clearly not the case as $S-\one $ is not Hilbert-Schmidt. Hence, by studying the 
GNS representations of the quasifree ground states, one can distinguish 
between the trivial and non-trivial region of the infinite Kitaev chain.
\hfill $\diamond$
}
\end{example}

\subsection{Quasifree ground states on the even subalgebra}

The algebras $A^\mathrm{car}_\mathrm{sd}(\calH,\Gamma)$ and $A^\mathrm{car}(E\calH)$ are 
naturally graded by the parity automorphism $\Theta$. We are most interested in 
ground states arising from $\Theta$-invariant interactions, so it is 
also natural to consider of representations of Fock 
states restricted to the even subalgebra of the CAR algebra.

\vspace{0.2cm}

To align our approach with standard texts, {\it e.g.}~\cite{EvansKawahigashi,ASS}, we set some notation. 
If $E_0$, $E_1$ are basis projections with 
$E_0-E_1$ Hilbert-Schmidt, let $E_0\wedge (1-E_1)$ be
the spectral projection $\chi_{\{1\}}(E_0-E_1)$, which is finite-rank by the Hilbert-Schmidt 
hypothesis~\cite{ArakiEvans}.

\begin{theo}[\cite{ArakiEvans}, Theorem 4] \label{theo:even_GS_condition}
Let $E_0,\,E_1 \in \calB(\calH)$ be basis projections with corresponding Fock states 
 $\omega_{E_0}$ and $\omega_{E_1}$. 
The restrictions of $\omega_{E_0}$ and $\omega_{E_1}$ to the even 
subalgebra $A^\mathrm{car}(E_i \calH)^0$ give 
rise to equivalent representations if and only if $E_0 - E_1$ is Hilbert-Schmidt and 
$\mathrm{dim}\big( E_0 \wedge (\one-E_1) \big)$ is even. 
\end{theo}

Let $\beta$ be a quasifree dynamics with BdG Hamiltonian $H=-\Gamma H \Gamma $ on $\calH$ with $0\notin\sigma(H)$. 
By Proposition \ref{prop:gappedBdG_gappedGS},  
the Fock state $\omega_{E}$ for $E = \chi_{(0,\infty)}(H)$ 
is the unique ground state on $A^\mathrm{car}_\mathrm{sd}(\calH,\Gamma)$ 
relative to $\beta$. We now consider the restriction of $\omega_{E}$ to 
$A^\mathrm{car}_\mathrm{sd}(\calH,\Gamma)^0 \cong A^\mathrm{car}(E\calH)^0$.

\begin{theo}[\cite{EvansKawahigashi}, Theorem 6.38] \label{thm:even_CAR_GS}
Let $\beta$ be a quasifree dynamics with BdG Hamiltonian $H$ such that $0 \notin \sigma(H)$. 
There exists a unique ground state for 
$(A^\mathrm{car}(E\calH)^0,\beta)$ if and only if the infimum of the positive part of the spectrum 
of $H$ is not an eigenvalue of $H$. If this is the case, the 
restriction of $\omega_{E}$ to $A^\mathrm{car}(E\calH)^0$ is the unique ground state.

\vspace{.1cm}

If the infimum of the positive spectrum of $H$ is an eigenvalue 
$\lambda$ with eigenprojection $E^\lambda$, then an extremal ground state of 
$(A^\mathrm{car}(E\calH)^0, \beta)$ is either the restriction of 
$\omega_{E}$ to $A^\mathrm{car}(E\calH)^0$ or 
the quasifree state $\omega_\nu$ constructed from the basis projection 
$E - P_\nu + \Gamma P_\nu \Gamma$, 
where $E^\lambda \nu = \nu$ and $P_\nu (v) = \langle \tfrac{\nu}{\|\nu\|} ,v\rangle \tfrac{\nu}{\|\nu\|}$. 
The representations of the states $\{\omega_\nu\}_{\nu \in \mathrm{Ran}(E^\lambda)}$ 
are all equivalent and disjoint from the 
restriction of $\omega_{E}$ to $A^\mathrm{car}(E\calH)^0$.
\end{theo}


\subsection{The index map on canonical transformations}

This section uses an index map for canonical transformations on infinite systems to assign a 
topological phase label to quasifree ground states and BdG Hamiltonians. This index was previously studied by 
Araki~\cite{Araki}, Araki--Evans \cite{ArakiEvans} and Carey--O'Brien~\cite{COB}. A similar exposition to ours 
can be found in~\cite{Reyes-Lega}.

\vspace{0.2cm}

For the sake of concreteness, let us fix a countable set $\Lambda$ 
and the Nambu space $\calH_\ph = \ell^2(\Lambda) \otimes \mathbb{C}^2$ with 
particle-hole involution $\Gamma = \mathcal{C}(\one \otimes \sigma_1)$. The results below can 
readily be adapted to the case of an arbitrary separable Hilbert space with real structure.

\vspace{0.2cm}

Let $E$ be a basis projection on $\calH_\ph$ and $J = \ii (2E-\one)$ a skew-adjoint 
unitary such that $\Gamma J \Gamma = J$. In particular, $J$ is well-defined on the 
real subspace $\calH^{\Gamma}_\mathbb{R} = \{v \in\calH_\ph \,:\, \Gamma v = v\}$. 
If $\tilde{E}$ is another basis projection giving rise to another $\tilde{J}$,  there is a unitary $W\in \mathcal{U}(\calH_\ph)$, 
$\Gamma W \Gamma = W$ such that $\tilde{J} = WJW^*$, see Lemma~\ref{lem:On_transitive_on_J}. 
One obtains a Bogoliubov transformation $\mathbf{U}_W$ on $\mathcal{F}(\ell^2(\Lambda))$ 
and the two representations 
$\pi_E$ and $\pi_{\tilde{E}}$ of $A^\mathrm{car}_\Lambda$ are equivalent if and only 
if $[W, J] \in \mathcal{L}^2(\calH_\ph)$, the ideal of Hilbert-Schmidt operators~\cite{ShaleStinespring,NNS}.

\begin{lemma}[\cite{COB}]
Let $E$ be a basis projection and $J = \ii (2E-\one)$ a complex structure on $\calH_{\mathbb{R}}^{\Gamma}$. 
Define 
$$
  \calU_J(\calH_\ph, \Gamma) \;=\; \big\{W \in \calU(\calH_{\ph})\,:\,  \Gamma W \Gamma = W ,\,\,  
   [J,W] \in \calL^2(\calH_{\ph}) \big\}\; .
$$
\begin{enumerate}
  \item[{\rm (i)}] If $W\in \calU_J(\calH_\ph,\Gamma)$, then $\frac{1}{2}(J + WJW^*)$ is Fredholm.
  \item[{\rm (ii)}] The Banach Lie group $\calU_J(\calH_\ph,\Gamma)$ has the same homotopy type 
  as the group $\varinjlim \calO_{2n}/\calU_n$.
  In particular, $\pi_0(\calU_J(\calH_\ph,\Gamma)) \cong \ZM_2$.
\end{enumerate}
\end{lemma}

Given $W \in \calU_J(\calH_\ph,\Gamma)$, $\| J - WJW^* \|_\mathcal{Q} = 0$ and so 
we can apply the continuous 
index map from Proposition \ref{prop:Z_2_unitary_map}. 

\begin{proposi}[\cite{COB, DSBW, BCLR}] \label{prop:bog_transform_Z2_index}
For $W \in \calU_J(\calH_\ph,\Gamma)$ the $\mathbb{Z}_2$-index of 
Proposition \ref{prop:Z_2_unitary_map},
$$
   j_J(W) \;=\; \mathrm{Ind}_2(J, WJW^*) \;=\;  (-1)^{ \frac{1}{2}\,  \mathrm{dim}\, \Ker(J + WJW^*)  } \; ,
$$
induces an isomorphism of $\pi_0(\calU_J(\calH_\ph,\Gamma))$ to $\ZM_2$.
\end{proposi}

Note that $j_J(W)=j_J(W^*)$ and that $j_{VJV^*}(VWV^*)=j_J(W)$ 
for any canonical transformation $V=\Gamma V\Gamma$.
The index map from Proposition~\ref{prop:bog_transform_Z2_index} 
requires a choice of complex structure $J$, which is equivalent 
to a choice of  basis projection on $\calH_\ph$. 
By imposing stronger conditions on the unitaries, one can 
remove the necessity of making a choice of complex structure.

\begin{proposi}
Let $W \in \mathcal{U}(\calH_\ph)$ satisfy $\Gamma W \Gamma = W$. 
Then $W \in \calU_J(\calH_\ph,\Gamma)$ 
for \emph{any} complex structure 
$J = \ii (2E-\one)$  if and only if 
$W+\one$ or $W-\one$ is Hilbert-Schmidt. In this particular situation,
$j_J(W)$ is independent of $J$.
\end{proposi}

\noindent {\bf Proof.} 
The equivalence is shown in \cite[Theorem 8]{Araki}. For the second claim, let $J'=VJV^*$ be 
another complex structure. Then $j_{J'}(W)=j_J(V^*WV)$ and $s\in[0,1]\mapsto (V^s)^*WV^s$ is a 
path in $\mathcal{U}_J(\calH_\ph,\Gamma)$ along which the index does not change
by Proposition~\ref{prop:bog_transform_Z2_index}, so that $j_J(V^*WV)=j_J(W)$.
\hfill $\Box$

\begin{remark}
{\rm 
For $W \in \calU_J(\calH_\ph,\Gamma)$, 
one can consider the path of skew-adjoint Fredholm operators 
$$
   [0,1] \ni t \mapsto J_t \;=\; (1-t) J + t WJ W^* \; ,  \qquad t \in [0,1] \; .
$$
Then 
$$
    j_J(W) \;=\; (-1)^{ \frac{1}{2}\, \mathrm{dim}\, \Ker( J + WJW^*)  } 
    \;=\; {\mathrm{Sf}_2(t\in [0,1] \mapsto (1-t)J + t WJ W^* )} 
$$
by the definition of $\mathbb{Z}_2$-valued spectral flow.
\hfill $\diamond$
}
\end{remark}

\begin{example} 
{\rm 
Let us consider the case of $J =\ii \sigma_3$.
Then any $W \in \calU_{\ii \sigma_3}(\calH_\ph, \Gamma)$ has the form 
$$
  W \;=\; \begin{pmatrix} u & v \\ \ol{v} & \ol{u} \end{pmatrix}\; , \qquad 
  v \in \calL^2(\ell^2(\Lambda))\; , \qquad 
  u\, \text{ Fredholm.}
$$
In this case, the expression for the index map 
$j_{i\sigma_3} :\calU_{\ii \sigma_3}(\calH_\ph,\Gamma) \to \mathbb{Z}_2$ can be written more simply. Namely, 
\begin{equation} \label{eq:inf_index_def}
   j_{i\sigma_3} (W) \;=\; (-1)^{\mathrm{dim}\, \Ker(u) }\; .
\end{equation}
For a finite lattice $\Lambda$,  
any unitary $W = \Gamma W \Gamma \in \calU(\calH_\ph)$ will be in the group 
$\calU_{\ii \sigma_3}(\calH_\ph, \Gamma)$. In this case,  
$j_{i\sigma_3}(W) = \mathrm{sgn}\,\mathrm{det}(W)$, so the index map in Equation \eqref{eq:inf_index_def} 
provides a generalisation of Kitaev's index from Section \ref{sec:Z_2_index_def} to infinite chains.

\vspace{0.1cm}

Suppose that $\Lambda$ is countably infinite and  
let $P \in \calB(\ell^2(\Lambda))$ be a finite rank projection. Define 
$$
   W_P \;=\; \begin{pmatrix} 1-P & P \\ P & 1-P \end{pmatrix}\; .
$$
It is immediate that $W_P \in \calU_{\ii \sigma_3}(\calH_\ph,\Gamma)$ and, furthermore, 
$j_{i\sigma_3}(W_P) = (-1)^{\dim(P)}$.
\hfill $\diamond$
}
\end{example}

\begin{remark} 
{\rm
As the previous example shows, one can construct canonical transformations on 
$\calH_\ph$ that are non-trivial for \emph{any} countable lattice $\Lambda$. In particular, 
taking $\Lambda = \mathbb{Z}^\nu$ for any $\nu \geq 1$, we obtain non-trivial 
indices in any lattice dimension. 
In contrast, the strong topological phase associated to free-fermionic Hamiltonians with 
even particle-hole symmetry is non-trivial only in certain dimensions~\cite{GSB}. Hence, the above index 
map is \emph{distinct} from the strong topological phase.

\vspace{0.1cm}

We can conclude from this discussion that the index map on Bogoliubov transformations 
is in general a coarser invariant for topological superconductors as it is unable to 
distinguish dimension in infinite systems. This result is not so surprising since, while the index 
has a $K$-theoretic interpretation, it does not arise as a pairing with 
a Dirac element as is the case for strong topological phases~\cite{GSB}.
\hfill $\diamond$
}
\end{remark}

\subsection{A \texorpdfstring{$\mathbb{Z}_2$}--index on pairs of BdG Hamiltonians}
\label{sec-IndBdGPairs}

Next the  index map $j_J:\calU_J(\calH_\ph, \Gamma) \to \mathbb{Z}_2$ is used to 
write an explicit $\mathbb{Z}_2$-index between a pair of quasifree dynamics  
with gapped BdG Hamiltonians. The definition works for BdG Hamiltonians over an arbitrary countable
set $\Lambda$ and is thus not restricted to dimension $1$. As before, our constructions 
readily extend to an arbitrary complex Hilbert space $\calH$ with real structure $\Gamma$.

\begin{defini}
Let $H_k$, $k=0,1$, be a pair of gapped BdG Hamiltonians 
on $\calH_\ph$  coming from quasifree dynamics on $A^\mathrm{car}_\mathrm{sd}(\calH_\ph,\Gamma)$  
and satisfying $0 \notin \sigma( H_k )$. Suppose that the positive energy spectral projections
$E_k = \chi_{(0,\infty)}(H_k)$  are such that  
$E_0 - E_1$ is a Hilbert-Schmidt operator. Then index of the pair of gapped BdG Hamiltonians is 
defined by
$$
  j(H_0, H_1) \;=\; (-1)^{ \frac{1}{2}\, \mathrm{dim}\, \Ker(  J_0 + J_1  )} 
    \;=\;    (-1)^{ \mathrm{dim} \, ( E_0 \wedge (1- E_1 )  ) } \; ,
$$
where $J_k = \ii  H_k | H_k|^{-1}$.
\end{defini}

Let us note that for $j(H_0,H_1)$ to be defined, the ground 
states $\omega_{E_0}$ and $\omega_{E_1}$ for $A^\mathrm{car}_\Lambda$ are 
unitarily equivalent by Theorem~\ref{thm:Araki_quasifree}(iii). 
The index $j(H_0,H_1)$ is a re-writing of the index on canonical transformations. 
More precisely, because the orthogonal group acts transitively on the space of complex structures by Lemma~\ref{lem:On_transitive_on_J}, 
there exists a $W \in \calU_{J_0}(\calH_\ph,\Gamma)$ such that $J_1 = W J_0 W^*$, and then 
$j(H_0,H_1) = j_{J_0}(W)$. The index also coincides with the index 
from \cite{ArakiEvans} which is reproduced in Equation (6.10.9) of \cite{EvansKawahigashi}.

\vspace{0.2cm}

The index map is a homomorphism by Proposition \ref{prop:bog_transform_Z2_index}; 
so if $j(H_0,H_1)$ and 
$j(H_1,H_2)$ are well-defined, then 
$$
   j(H_0,H_2)  \;=\; j(H_0,H_1) \; j(H_1,H_2) \; .
$$

By Theorem~\ref{theo:even_GS_condition}, the $\mathbb{Z}_2$-index 
encodes whether the restriction of the states $\omega_{E_k}$ 
to the even subalgebra $(A^\mathrm{car}_\Lambda)^0$ give rise to equivalent 
representations.

\subsection{Connections to \texorpdfstring{$\mathbb{Z}_2$}--valued spectral flow}
\label{sec:Z_2_index_as_sf}

Let $\beta$ be a quasifree dynamics with BdG Hamiltonian $H$ such that $0 \notin \sigma_\mathrm{ess}(H)$. 
Then 
$\ii  H$ defines a skew-adjoint Fredholm operator 
on the real Hilbert space $\calH_{\mathbb{R}}^{\Gamma}$. Therefore, Fredholm paths $t\in[0,1]\mapsto \ii H(t)$ of BdG Hamiltonians 
give paths of skew-adjoint Fredholm operators on $\calH_{\mathbb{R}}^{\Gamma}$. For paths 
with invertible (gapped) 
endpoints,  then 
one can 
consider $\mathrm{Sf}_2(t\in[0,1]\mapsto \ii H(t))$.

\vspace{0.2cm}

We now prove an infinite-dimensional analogue of Proposition \ref{prop:sf_as_Z2_obstruction}.
\begin{proposi}
\label{prop-sf_as_Z2_obstruction2}
Let $H_0$ and $H_1$ be invertible BdG Hamiltonians on 
$\calH_\ph$ with $j(H_0, H_1)$  well-defined. 
Then for any continuous path of self-adjoint Fredholm operators 
$H_t$ connecting $H_0$ and $H_1$, 
$$
j(H_0,H_1)
\;=\;
\mathrm{Sf}_2(t\in[0,1]\mapsto  \ii H_t )
\;.
$$
\end{proposi}
\noindent {\bf Proof.} 
Let $J_0= iH_0|H_0|^{-1}$ and $J_1 = iH_1 |H_1|^{-1}$. 
As $\| J_0 - J_1 \|_\mathcal{Q} =0$, 
one can take the trivial partition of $[0,1]$ in the definition of the $\mathbb{Z}_2$-spectral flow, 
and so 
$$
  \mathrm{Sf}_2(t\in[0,1]\mapsto  \ii H_t )  \;=\;  
  (-1)^{ \frac{1}{2} \mathrm{dim}\,\Ker (  J_0 +  J_1 )}  \; =\; j(H_0,H_1) \; ,
$$
completing the proof. 
\hfill $\Box$

\vspace{0.2cm}

There is also an infinite-dimensional analogue of Proposition \ref{prop:Z2_flow_and_degeneracy}.
\begin{proposi} \label{prop:quasifree_sf_gap_close}
Let $H_0$ and $H_1$ be invertible BdG Hamiltonians on 
$\calH_\ph$ with $j(H_0,H_1)$ well-defined.  
If $j(H_0,H_1)=-1$, then 
for any continuous path of self-adjoint and particle-hole symmetric Fredholm operators $H(t)$
connecting $H_0$ and $H_1$,  
there is some $t_0 \in (0,1)$ such that $H(t_0)$ has a double degenerate kernel.
\end{proposi}
\noindent {\bf Proof.} 
The assumptions ensure 
that  $\mathrm{Sf}_2(t\in[0,1]\mapsto  \ii H(t))$ is well-defined and non-trivial. 
Therefore 
there is at least one $t_0 \in (0,1)$ such that $\Ker( \ii H(t_0)) = \Ker( H(t_0))$ 
is even-dimensional.
\hfill $\Box$

\vspace{0.2cm}

Propositions \ref{prop:quasifree_sf_gap_close}  shows 
that the  index on pairs of BdG Hamiltonians precisely encodes the topological obstruction for two BdG 
Hamiltonians to be in the same topological phase. 
Let us now consider the relationship between the  $\mathbb{Z}_2$-index, the 
$\mathbb{Z}_2$-valued spectral flow and gapped ground states on the CAR algebra.

\begin{proposi} \label{prop:quasifree_GNS_gap_close}
Let $H_0$ and $H_1$ be invertible BdG Hamiltonians on 
$\calH_\ph$ that give gapped ground states $\omega_{E_0}$ and 
$\omega_{E_1}$ on $A^\mathrm{car}_\Lambda$. Let $H(t)$ be any continuous 
path of self-adjoint and particle-hole symmetric Fredholm operators 
connecting $H_0$ and $H_1$. 
Suppose $j(H_0,H_1) = -1$. Then there exists a $t_0 <1$ such that the 
path $[0,t_0) \ni t\mapsto \omega_{E_t}$ of ground states of the quasifree dynamics 
generated by $H(t)$ as in Proposition \ref{prop:gappedBdG_gappedGS} will not 
be uniformly gapped.
\end{proposi}
\noindent {\bf Proof.} 
By Proposition \ref{prop:quasifree_sf_gap_close}  there is a smallest $t_0 \in(0,1)$ 
such that $0 \in \sigma(H(t_0))$. For all $t \in [0,t_0)$, one has has $0 \notin\sigma(H(t))$. 
Then we obtain a path of ground states $[0,t_0) \ni t \mapsto \omega_{E_t}$ 
with $E_{t} = \chi_{(0,\infty)}(H(t))$ by Proposition \ref{prop:gappedBdG_gappedGS}.
For every $t\in [0,t_0)$,  the GNS space is 
$$
   \mathfrak{h}_{E_t} \;\cong \; \bigoplus_{n=0}^\infty \bigwedge\nolimits^{\! n} E_{t}\, \calH_\ph \; , 
   \qquad \bigwedge\nolimits^{\! 0} E_{t}\, \calH_\ph  \;=\; \mathbb{C}\, \Omega_{E_t} \; ,
$$
and the GNS Hamiltonian $h_{\omega_t}$ is the second quantisation 
of $H(t)$ restricted to anti-symmetric tensors on $E_t\, \calH_{\ph}$. 
As the spectral gap of $H(t)$ above $0$ 
goes to $0$ as $t \to t_0$, so too will the spectral gap of $h_{\omega_t}$. 
Thus for any $\gamma>0$, one has $\sigma(h_{\omega_t})\cap(0,\gamma)\neq \emptyset$ for 
any $t_0 - t$ sufficiently small. 
\hfill $\Box$

\vspace{0.2cm}

Let us now elaborate on the example of the Kitaev chain on $\ZM$ studied in Section~\ref{sec-QuasifreeDynBdG}
to produce an example of a non-trivial spectral flow, again given by a flux insertion as in the case of the closed 
finite chain studied in Sections~\ref{sec-KitaevClosedChain} and \ref{sec:closed_interacting_chain}.


\begin{example}[Flux insertion in infinite Kitaev chain] \label{ex:Kitaev_inf_flux}
{\rm 
The Hamiltonian will be a local perturbation of \eqref{eq-KitaevMuw1}. Let us first focus 
on the topological phase and thus set $\mu=0$, and for sake of simplicity $w=-1$. 
The local perturbation is then given by the flux insertion as in Proposition~\ref{prop:Maj_flux_flow}, 
but between site $0$ and $1$:
\begin{align*}
\mathbf{H}_{[a,b]}^\mathrm{Kit}(\alpha) 
\;=\; 
&
\sum_{j=a}^{b-1} \delta_{j\not=0}\big( \fraka_j^* \fraka_{j+1} + \fraka_{j+1}^*\fraka_j 
+\ii \fraka_{j}\fraka_{j+1} -\ii \fraka_{j+1}^* \fraka_j^* \big)  
\\
&
\;\;\;+\; 
\big(e^{\ii \alpha}\fraka_0^* \fraka_1 + e^{-\ii \alpha}\fraka_1^* \fraka_0 
 +\ii e^{{-\ii \alpha }}\fraka_0 \fraka_1 -\ii e^{{\ii \alpha}} \fraka_1^*\fraka_0^*
 \big)
\; . 
\end{align*}
Let us note that inserting a half-flux is implemented by an automorphism of $A^\mathrm{car}_\ZM$ 
$$
\gamma_-( \fraka_j) \;=\; \begin{cases}  \fraka_j \, , & j \geq 1\, , \\  -\fraka_j\, , & j \leq 0\;, \end{cases}
$$
namely one has
$$
\mathbf{H}_{[a,b]}^\mathrm{Kit}(\pi) 
\;=\;\gamma_-\big(\mathbf{H}_{[a,b]}^\mathrm{Kit}(0)\big) 
\;.
$$
The BdG Hamiltonian is now given by $H^\mathrm{Kit}_\ZM(\alpha)=S_\alpha+S_\alpha^*$
where the translations with inserted flux are
$$
S_\alpha\;=\;
S\otimes\frac{1}{2}\begin{pmatrix} 1 & \ii \\ \ii & -1 \end{pmatrix}
\,+\,\nu_1(\nu_0)^*\otimes
\frac{1}{2}
\begin{pmatrix}
e^{-\ii\alpha}-1 & \ii(e^{\ii\alpha}-1)
\\
\ii(e^{-\ii\alpha}-1) & -(e^{\ii\alpha}-1)
\end{pmatrix}
\;.
$$
with $\nu_n$ the partial isometry onto the site $n\in \mathbb{Z}$.
Note that $H^\mathrm{Kit}_\ZM(\alpha)$ is a finite rank perturbation of 
\eqref{eq-KitaevMuw2}, which is gapped. Hence the  $\mathbb{Z}_2$-valued spectral flow
of the path $\alpha\in[0,\pi]\mapsto  \ii H^\mathrm{Kit}_\ZM(\alpha)$ is well-defined. It has
been shown by an explicit calculation in~\cite[Section 10]{CPS} that it is equal to $-1$. By Proposition~\ref{prop-sf_as_Z2_obstruction2}
and homotopy invariance of $\mathrm{Sf}_2$,
one hence has $j(H^\mathrm{Kit}_\ZM(0),H^\mathrm{Kit}_\ZM(\pi))=-1$.

\vspace{.1cm}

Now let us consider the topologically trivial phase of the Kitaev chain, namely set $\mu=1$ and $w=\Delta=0$.
As the Hamiltonian has no kinetic part now, the flux insertion does not change the Hamiltonian, that is, $H^\mathrm{Kit}_\ZM(\alpha)=H^\mathrm{Kit}_\ZM(0)$. In particular, $j(H^\mathrm{Kit}_\ZM(0),H^\mathrm{Kit}_\ZM(\pi))=1$.

\vspace{.1cm}

Hence the flux insertion is a test of the topologically non-trivial nature of the ground state. 
In Section~\ref{sec:general_Z2}, it is shown how this concept extends to 
systems which are not quasifree.
}
\hfill $\diamond$
\end{example}

\begin{remark} \label{rem:QFGS_path_with_closing}
{\rm
This remark provides further understanding of the GNS-representation spaces along a flux insertion.
Let $(\mathcal{H}, \Gamma)$ be a complex Hilbert space with real structure and consider a 
norm-continuous path 
of BdG Hamiltonians $H(s)$ such that $0 \notin \sigma(H(s))$ for all $s \in [0,1]\setminus \{s_0\}$. 
At the point $s_0\in (0,1)$ let us assume that the $0$-energy eigenspace of $H(s_0)$ is finite 
dimensional. Hence one has a continuous path of Fredholm BdG Hamiltonians with a 
gap-closing point at $s_0$. We now consider the family of $\mathbb{R}$-actions on 
$A^\mathrm{car}_\mathrm{sd}(\mathcal{H},\Gamma)$ given by 
$$
   \alpha_{s,t}( \frakc(v) ) \;=\; \frakc\big( e^{i t H(s)} v \big) \; , \qquad t \in \mathbb{R}\; , \quad s \in [0,1]\; .
$$
Example \ref{ex:Kitaev_inf_flux} is a special case of the above setting.

\vspace{0.1cm}

Applying Proposition \ref{prop:gappedBdG_gappedGS}, outside of the point $s_0$, the dynamics 
$\alpha_s$ has a unique pure ground state $\omega_s$ constructed by the basis projection 
$E_s = \chi_{(0,\infty)}(H(s))$ with the GNS space 
$\mathfrak{h}_{\omega_s} = \bigoplus_n \bigwedge^n E_s \mathcal{H}$.

\vspace{0.1cm}

At the crossing point $s_0$, 
let $E_0 = \chi_{\{0\}}(H(s_0))$ and $E_+ = \chi_{(0,\infty)}(H(s_0))$. Then 
one can decompose the CAR algebra 
$A^\mathrm{car}_\mathrm{sd}(\mathcal{H}, \Gamma) \simeq A^\mathrm{car}(E_0\mathcal{H}) \,\hat\otimes\, A^\mathrm{car}(E_+ \mathcal{H})$ with 
$A^\mathrm{car}(E_0\mathcal{H})$ finite-dimensional. Given an arbitrary state $\omega_0$ 
on $A^\mathrm{car}(E_0 \mathcal{H})$, then by~\cite[Proposition 6.37]{EvansKawahigashi} 
$$
   \omega( a_0 a_1) \;=\; \omega_0(a_0) \, \omega_{E_+}( a_1 ) \; ,  \qquad 
   a_0 \in A^\mathrm{car}(E_0 \mathcal{H})\; , \quad a_1 \in A^\mathrm{car}(E_+ \mathcal{H})
$$
will be a ground state of the dynamics $\alpha_{s_0}$. In particular, by the tensor 
product structure, the GNS triple of this ground state is given by 
$$
  ( \pi_{\omega_{s_0}}, \mathfrak{h}_{\omega_{s_0}}, \Omega_{\omega_{s_0}} )  \; \cong \; 
  \big( \pi_{0} \hat\otimes \one_{\mathfrak{h}_{E_+}} + \one_{\mathfrak{h}_0}   
  \hat\otimes \pi_{E_+}, \, \mathfrak{h}_0 \hat\otimes \mathfrak{h}_{E_+}, \, 
  \Omega_0 \hat\otimes \Omega_{E_+} \big)\,,
$$
with $(\pi_0, \mathfrak{h}_0, \Omega_0)$ the (unique) GNS triple of the 
finite dimensional algebra $A^\mathrm{car}(E_0\mathcal{H})$ and state $\omega_0$. 
In particular, as $\mathfrak{h}_0$ is finite-dimensional, there is some $N$ such 
that $\mathfrak{h}_{\omega_{s_0}} \cong \mathbb{C}^{N} \hat\otimes \,\mathfrak{h}_{E_+}$.
}
\hfill $\diamond$
\end{remark}

The relative $\mathbb{Z}_2$-index provides a topological obstruction 
for a pair of quasifree ground states to be connected such that the corresponding infinite GNS 
Hamiltonian retains a spectral gap above $0$. This closely aligns with the heuristic 
physical picture of a (relative) topological or SPT phase of parity-symmetric 
gapped ground states in the fermionic setting. The next task is to consider 
ground states that are not quasifree.


\section{A \texorpdfstring{$\mathbb{Z}_2$}--index for pure gapped ground states}  \label{sec:general_Z2}

In this section, we define a candidate $\mathbb{Z}_2$-phase label for one-dimensional 
ground states that are not necessarily quasifree. 
The constructions rely heavily on the Jordan--Wigner transform and, as such, are 
restricted to the one-dimensional lattice $\mathbb{Z}$.

\vspace{0.1cm}

The interactions are assumed to be even (parity-preserving), finite range and with the 
property that for $X\subset \mathbb{Z}$ finite 
\begin{equation}  \label{eq:interaction_bound_assumption}
  \sup_{j\in \mathbb{Z}} \sum_{{X\ni j}} \frac{\| \Phi(X)\|}{|X|}  \;<\; \infty  \;.
\end{equation}
Note that Equation \eqref{eq:interaction_bound_assumption} is satisfied for any 
finite range  Hamiltonian with uniformly bounded $\Phi$,  {\it e.g.} a translation invariant finite range Hamiltonian.
All states on $A^\mathrm{car}_\mathbb{Z}$ considered here are assumed to be 
parity invariant, $\omega \circ \Theta = \omega$ for $\Theta$. 
This ensures the existence of a self-adjoint unitary $\Sigma$ on $\mathfrak{h}_\omega$ such 
that $\Sigma \Omega_\omega = \Omega_\omega$ and a decomposition 
$$
  \mathfrak{h}_\omega \;=\; \mathfrak{h}_\omega^0 \oplus \mathfrak{h}_\omega^1 \; , \qquad 
  \mathfrak{h}_\omega^i \;=\; \frac{1}{2}(1+(-1)^i \Sigma)\mathfrak{h}_\omega \;=\; 
      \overline{\pi_\omega((A^\mathrm{car}_\mathbb{Z})^i) \Omega_\omega } \; .
$$
Interactions satisfying the bound \eqref{eq:interaction_bound_assumption} also 
satisfy a Lieb--Robinson bound and so the automorphism 
$\beta: \mathbb{R} \to \mathrm{Aut}(A^\mathrm{car}_\mathbb{Z})$ given by 
$$
  \beta_t( a) \;=\;   \lim_{N\to\infty} e^{\ii t \mathbf{H}_N} a e^{-\ii t\mathbf{H}_N} \; , \qquad 
  \mathbf{H}_N = \sum_{X \subset [-N,N]\cap \mathbb{Z}} \Phi(X) \; 
$$
exists for any $t\in \mathbb{R}$ \cite[Theorem 3.5]{NSY18}. 
In this section, ground states on $A^\mathrm{car}_\mathbb{Z}$ will always be with respect to this dynamics.

\subsection{The Jordan--Wigner transform} \label{subsec:JordanWigner}
In order to apply techniques from spin-chains to fermionic systems, one needs to clearly understand 
the way to pass between the two in the infinite volume limit. This will be established by  
the Jordan--Wigner transform, so we now restrict to the one-dimensional lattice 
$\Lambda=\mathbb{Z}$.
The basic references here are~\cite[Example 6.2.14]{BR2} and~\cite[Chapter 6.5]{EvansKawahigashi}. 

\vspace{0.1cm}

For one-dimensional fermionic interactions that are even, there are three $C^*$-algebras of interest in the infinite volume limit: the fermion algebra 
$A_\mathbb{Z}^\mathrm{car}= \varinjlim A^\mathrm{car}_{[-a,b]\cap \mathbb{Z}}$, the Pauli algebra 
$A^{P}_{\ZM}= \bigotimes_\mathbb{Z} M_2(\mathbb{C})$ given by the $C^*$-algebraic closure 
of the tensor algebra generated by the spin matrices at each site, 
and a crossed product algebra $\widehat{A}_\mathbb{Z} = A_\mathbb{Z}^\mathrm{car} \rtimes_{\gamma_-} \mathbb{Z}_2$, 
where the (outer)  action of $\mathbb{Z}_2$ is 
\begin{equation} \label{eq:gamma_-_defn}
    \gamma_- ( \fraka_j) \;=\; \begin{cases}  \fraka_j \, , & j \geq 1\, , \\  - \fraka_j\, , & j \leq 0 \end{cases}\; .
\end{equation}
One can abstractly characterise $\widehat{A}_\mathbb{Z}$ as the $C^*$-algebra generated by 
$A_\mathbb{Z}^\mathrm{car}$ and the self-adjoint unitary $T$ such that $Ta = \gamma_{-}(a)T$ for 
any $a \in A^\mathrm{car}_\mathbb{Z}$. The grading $\Theta$ of $A^\mathrm{car}_\mathbb{Z}$ extends 
to a grading on $\widehat{A}_\mathbb{Z}$ by defining $\Theta(T) = T$. 

\vspace{0.1cm}

There is a $\ast$-embedding of the Pauli algebra $A^{P}_{\ZM}$ in $\widehat{A}_\mathbb{Z}$ by the 
map
\begin{align*}
  &\sigma_j^x \; \mapsto \; TS_j(\fraka_j+\fraka_j^*) \; ,   &&\sigma_j^y \; \mapsto \; \ii  \,TS_j (\fraka_j - \fraka_j^*) \; ,
  &&\sigma_j^z \;\mapsto \; 2\fraka_j^* \fraka_j - \one \; , 
\end{align*}
where 
$$
   S_j \;=\; \begin{cases} \prod_{i=1}^{j-1} \sigma_i^z \, ,  &  j  \,\geq \, 1\, ,  \\  \one \, ,  & j \,=\, 1 \\  
         \prod_{i=j}^0 \sigma_i^z \, ,  &  j \,\leq\, 0 \end{cases} \; .
$$
Thus, both $A_\mathbb{Z}^\mathrm{car}$ and the Pauli algebra $A^{P}_{\ZM}$
can be embedded within a larger algebra $\widehat{A}_\mathbb{Z}$. 

\vspace{0.1cm}

To better compare $A^\mathrm{car}_\mathbb{Z}$ and $A^{P}_{\ZM}$ embedded within $\widehat{A}_\mathbb{Z}$, 
let us give the Pauli algebra a grading, where at each site $j \in \mathbb{Z}$, $\sigma_j^z$ is even 
and $\sigma_j^x, \,\sigma_j^y$ are odd. This gives a decomposition $A^{P}_{\ZM} = (A^{P}_{\ZM})^0 \oplus (A^{P}_{\ZM})^1$ and 
ensures that the embedding $A^{P}_{\ZM} \hookrightarrow  \widehat{A}_\mathbb{Z}$ is graded.  
Using the decomposition of $\widehat{A}_\mathbb{Z}$, 
$$
  \widehat{A}_\mathbb{Z} \;\cong \;  (\widehat{A}_\mathbb{Z})^0 \,\oplus \, (\widehat{A}_\mathbb{Z})^1  
   \;\cong \; \big( ( A^\mathrm{car}_\mathbb{Z})^0 \,\oplus \, T(A^\mathrm{car}_\mathbb{Z})^0 \big)
    \,\oplus\, \big( (A^\mathrm{car}_\mathbb{Z})^1 \,\oplus\, T(A^\mathrm{car}_\mathbb{Z})^1 \big) \; ,
$$
one then has the following equivalences of algebras and vector spaces respectively,
\begin{align*}
  (A^{P}_{\ZM})^0 \;\cong\;  (A^\mathrm{car}_\mathbb{Z})^0 \; ,  \qquad (A^{P}_{\ZM})^1 \;\cong\; T (A^\mathrm{car}_\mathbb{Z})^1 \; .
\end{align*}
Lastly, let us note that, for half-infinite systems where $\Lambda = \mathbb{N}$,
the automorphism $\gamma_-$ on $A^\mathrm{car}_\mathbb{N}$ is the identity automorphism 
and one can naturally identify
$\widehat{A}_\mathbb{N} \cong A^\mathrm{car}_\mathbb{N} \cong A^P_\mathbb{N}$ as 
graded algebras, where $A^P_\mathbb{N} = \bigotimes_\mathbb{N} M_2(\mathbb{C})$.

\subsection*{States under the Jordan--Wigner transform}

Having analyzed the connections between $A^\mathrm{car}_\mathbb{Z}$ and $A^{P}_{\ZM}$, let 
us now discuss links between states on these algebras. 
Any $\Theta$-invariant state $\omega$ on $A^\mathrm{car}_\mathbb{Z}$ has a  restriction 
$\omega |_{(A^\mathrm{car}_\mathbb{Z})^0}$. If $\omega$ is pure, then this restriction is 
pure as well~\cite[Lemma 6.23]{EvansKawahigashi}.  
One can extend $\omega$ to a state $\hat\omega$ on $\widehat{A}_\mathbb{Z}$ by 
setting $\hat{\omega}(a_0 + Ta_1 ) = \omega(a_0)$ where $a_0,a_1\in {A}^\mathrm{car}_\mathbb{Z}$. 
This provides a state $\omega^P$ on the Pauli 
algebra $A^{P}_{\ZM} \subset \widehat{A}_\mathbb{Z}$ as the restriction of $\hat{\omega}$. 
Because $(A^\mathrm{car}_\mathbb{Z})^0 \cong (A^{P}_{\ZM})^0$, the state 
$\omega^P |_{(A^{P}_{\ZM})^0}$ of $(A^{P}_{\ZM})^0$ is pure if $\omega$ is so, but 
$\omega^P$ itself need not be pure.

\begin{theo}[\cite{EvansKawahigashi}, Theorem 6.25] \label{thm:car_even_state_equivalence}
Let $\omega$ be a pure $\Theta$-invariant state on $A^\mathrm{car}_\mathbb{Z}$. Then 
$\omega^P$, the restriction of $\hat{\omega}$ to $A^{P}_{\ZM}$, is not pure if and only if the following two conditions hold:
\begin{enumerate}
\item[{\rm (i)}] $\omega$ and $\omega \circ \gamma_-$ are equivalent states on $A^\mathrm{car}_\mathbb{Z}$,
\item[{\rm (ii)}]  $\omega |_{(A^\mathrm{car}_\mathbb{Z})^0}$ and $\omega |_{(A^\mathrm{car}_\mathbb{Z})^0} \circ \gamma_-$ 
 are not equivalent states on $(A^\mathrm{car}_\mathbb{Z})^0$.
\end{enumerate}
If $\omega^P$ is not pure, then it is a mixture of $2$ inequivalent pure states.
\end{theo}

Let us now specialise Theorem~\ref{thm:car_even_state_equivalence} to a quasifree pure $\Theta$-invariant state. 
Let $E$ be a basis projection on $\Hh_\ph=\ell^2(\ZM)\otimes\CM^2$.
Then the quasifree state $\omega_{E}$ on $A^\mathrm{car}_\ZM$ is pure and $\Theta$-invariant. 
To know if $\omega^P_{E}$ is pure or not, 
by Theorem \ref{thm:car_even_state_equivalence},  we need to compare the states 
$\omega_{E}$ and $\omega_{E} \circ \gamma_-$ on  $A^\mathrm{car}_\mathbb{Z}$ and $(A^\mathrm{car}_\mathbb{Z})^0$  with 
the $\mathbb{Z}_2$-action $\gamma_-$   from Equation \eqref{eq:gamma_-_defn}.
For this purpose it is useful to introduce the operator 
\begin{equation} \label{eq:theta_operator_defn}
   \theta_- \;:\; \ell^2(\mathbb{Z}) \;\to\; \ell^2(\mathbb{Z}) \; , \qquad 
   \theta_- e_j \;=\; \begin{cases} e_j \, ,  & j \, \geq\, 1 \, , \\ - e_j \, ,  &  j \, \leq \, 0 \, \end{cases} \; 
\end{equation}
with $\{e_j\}_{j\in\mathbb{Z}}$ the canonical basis of $\ell^2(\mathbb{Z})$. We also denote by 
 $\theta_-$ the diagonal extension $\theta_- \otimes \one_{\mathbb{C}^2}$ 
to $\calH_\ph$. Then $\theta_- E \theta_-$ is a basis projection and 
$$
   \omega_{\theta_- E \theta_-}(a)  \;=\; \omega_{E}\circ\gamma_- (a) \; , \qquad a \in A^\mathrm{car}_\mathbb{Z} \; .
$$
By Theorem~\ref{theo:even_GS_condition}, the restrictions of $\omega_{E}$ and 
$\omega_{E} \circ \gamma_-$ give equivalent representations of 
$(A^\mathrm{car}_\mathbb{Z})^0$  if and only if 
$E - \theta_- E \theta_-$ is Hilbert-Schmidt and 
 $\mathrm{dim}\big(\theta_- E \theta_- \wedge (1- E) \big)$ is even. 
On the other hand, by  the last 
item of Theorem~\ref{thm:Araki_quasifree}, $E - \theta_- E \theta_-$ is Hilbert-Schmidt if and only if
$\omega_E$ and $\omega_E\circ\gamma_-$ are equivalent. Therefore one concludes from Theorem~\ref{thm:car_even_state_equivalence}: 

\begin{coro} \label{coro:car_even_state_equivalence}
Let $E$ be a basis projection and $\omega_{E}$ be the corresponding pure, $\Theta$-invariant and 
quasi\-free state on $A^\mathrm{car}_\mathbb{Z}$. If $\omega_E$ is equivalent to $\omega_E\circ\gamma_-$, 
then for $J=i(2E-1)$: 
$$
\omega^P_{E}\;\;\mbox{\rm pure}
\quad
\Longleftrightarrow
\quad
\mathrm{dim}\big( \theta_- E \theta_- \wedge (1- E) \big)\;\;\mbox{\rm even}
\quad
\Longleftrightarrow
\quad
j_J(\theta_- )\,=\,1
\;.
$$
with $j_J$ the index map on canonical transformations from Proposition~\ref{prop:bog_transform_Z2_index}.
\end{coro}

\subsection{Ground states of the \texorpdfstring{$XY$}--Hamiltonian}

This section gives a detailed review of results on the ground states of 
the $XY$-Hamiltonian on the lattice $\mathbb{Z}$, based on the work of 
Araki and Matsui~\cite{ArakiMatsui85} which is also described in detail in
\cite[Chapter 6-7]{EvansKawahigashi}. The $XY$-Hamiltonian reduces 
to the Kitaev chain and quantum Ising model for special values of the input parameters, and the 
exposition motivates how we deal with more general fermionic chains in 
Section \ref{subsec:split_prop} and \ref{subsec:general_Z2}. 
While the $XY$-Hamiltonian is typically defined on the Pauli algebra $A^{P}_{\ZM}$, we will work on 
the larger algebra $\widehat{A}_\mathbb{Z}$, where one can pass between fermionic and 
spin-chain descriptions without issue. 

\vspace{0.2cm}

The Hamiltonian, written using the fermion operators, is defined on the local region 
$[a,b]\cap \mathbb{Z}$ as 
\begin{equation}  \label{eq:fermionic_XY}
  \mathbf{H}_{[a,b]}^{XY} \;=\; \sum_{j=a}^{b-1} \big[ -(\fraka_j^* \fraka_{j+1} + \fraka_{j+1}^*\fraka_j)  
  +\rho( \fraka_j\fraka_{j+1} + \fraka_{j+1}^* \fraka_{j}^*) \big] \;+ \;\mu \sum_{j=a}^b (\fraka_j^* \fraka_j -\tfrac{1}{2} \one) \; .
\end{equation}
with $\rho, \,\mu \in \mathbb{R}$. Note that we use a different scaling of the parameters 
to~\cite{ArakiMatsui85} in order to better align with the rest of the paper. 
The Hamiltonian $\mathbf{H}_{[a,b]}^{XY}$ conserves parity and can be 
written in terms of the Pauli operators:
\begin{equation} \label{eq:spin_XY}
\mathbf{H}_{[a,b]}^{XY} 
\;=\;  
\sum_{j=a}^{b-1} \big[ (1+\rho)\sigma_j^x \sigma_{j+1}^x 
    + (1-\rho)\sigma_j^y \sigma_{j+1}^y  \big] \,+ \,\mu \sum_{j=a}^b \sigma_j^z \; . 
\end{equation}
Comparing with \eqref{eq:Kitaev_model} shows that $\mathbf{H}_{[a,b]}^{XY}$ with $\rho=1$   recovers the 
Kitaev with $w=\Delta=1$. For the parameters $(\mu,\rho) = (0,\pm 1)$, the $XY$-Hamiltonian reduces to the 
quantum Ising chain.

\vspace{0.2cm}

The $XY$-Hamiltonian gives the BdG Hamiltonian on $\calH_\ph = \ell^2(\mathbb{Z})\otimes \mathbb{C}^2$, 
$$
   H_\mathbb{Z}^{XY} \;=\; -2\begin{pmatrix}  S + S^* - \mu & \rho(S-S^*) \\ -\rho(S-S^*) & -(S+S^*-\mu) \end{pmatrix} \; , 
$$
with $S$ the bilateral shift operator. 
One can check using the Fourier transform that 
for $\mu=0$ and $\rho\neq 0, \pm 1$, $\sigma(H_\mathbb{Z}^{XY} ) = [-2,-2\rho]\cup [2\rho,2]$ 
or $[-2\rho,2]\cup [2,2\rho]$ with constant multiplicity $4$. 
If $(\mu,\rho) =(0,1)$, then 
$\sigma(H_\mathbb{Z}^{XY}) = \{\pm 4\}$. We also note 
that if $(\mu,\rho)\neq(0, \pm 1)$, then the point spectrum $\sigma_p(H_\mathbb{Z}^{XY}) = \emptyset$~\cite{ArakiMatsui85}. 
In particular, for $(\mu,\rho)$ such that $0 \notin \sigma(H_\mathbb{Z}^{XY})$, 
Proposition \ref{prop:gappedBdG_gappedGS} applies and  says that for 
$E = \chi_{(0,\infty)}(H_\mathbb{Z}^{XY})$, $\omega_{E}$ is the unique 
ground state on $A^\mathrm{car}_\mathbb{Z}$, the representation 
$\pi_{E}$ is irreducible and the infinite GNS Hamiltonian is also gapped.

\vspace{0.2cm}

Let us also consider 
the ground states on the even subalgebra $(A^\mathrm{car}_\mathbb{Z})^0$, where 
Theorem \ref{thm:even_CAR_GS} applies. Specifically, in the case of 
$(\mu,\rho)\neq (0,\pm 1)$, the restriction of $\omega_{E}$ to $(A^\mathrm{car}_\mathbb{Z})^0$ 
is the unique ground state. If $(\mu,\rho)=(0,\pm 1)$, then $\sigma(H_\mathbb{Z}^{XY}) =
\{\pm 4\}$ and each eigenvalue has infinite multiplicity. If 
$\{\nu_j\}_{j\in\mathbb{Z}}$ are mutually orthogonal eigenvectors of 
$+4$, they each give basis projections $E - P_{\nu_j} + \Gamma P_{\nu_j} \Gamma$ 
with $P_{\nu_j} (v) = \langle \tfrac{\nu_j}{\|\nu_j\|} ,v\rangle \tfrac{\nu_j}{\|\nu_j\|}$. Therefore an arbitrary ground state of 
$(A^\mathrm{car}_\mathbb{Z})^0$  is a convex combination of 
the restrictions of $\omega_{E}$ and $\omega_{\nu_j}$. 
The GNS representations associated to $\omega_{\nu_j}$ are 
all equivalent. Hence, if ground states are counted up to equivalence of GNS 
representations, then $\mathbf{H}^{XY}$ has two distinct ground states 
for $(\mu,\rho)=(0,\pm 1)$. 

\vspace{0.2cm}

We have so far shown that the number of ground states of the 
even subalgebra $(A^\mathrm{car}_\mathbb{Z})^0$ in the 
infinite volume limit depends on the parameters $(\mu,\rho)$ in the $XY$-Hamiltonian. 
In particular, the case $(\mu,\rho) = (0, 1)$ which has $2$ distinct ground states coincides with the 
infinite Kitaev chain with $w=\Delta=1$.
However, at the level of ground states of $(A^\mathrm{car}_\mathbb{Z})^0$ in the region 
$(\mu,\rho)\neq (0,\pm 1)$, we currently cannot distinguish between what is considered 
the trivial region, $|\mu|\geq \tfrac{1}{2}$ or $\rho=0$ and $|\mu|<\frac{1}{2}$, with the non-trivial region 
$|\mu|<\tfrac{1}{2}$ and $\rho\neq 0$. These regions can be distinguished by 
looking at ground states of the Pauli algebra 
$A^{P}_{\ZM}$.

\vspace{0.2cm}

Suppose $0 \notin\sigma(H^{XY}_\mathbb{Z})$ and let $\omega$ be the pure ground 
state of the $XY$-chain on $A^\mathrm{car}_\mathbb{Z}$. 
As previously explained, one obtains a state $\omega^P$ on $A^{P}_{\ZM}$ by 
extending $\omega$ to $\widehat{A}_\mathbb{Z}$ and then restricting to $A^{P}_{\ZM}$. 
Theorem \ref{thm:XY_soln} below analyses the purity of $\omega^P$ based on  
Corollary~\ref{coro:car_even_state_equivalence} and the following:

\begin{proposi}[\cite{ArakiMatsui85}, Lemma 4.5] \label{prop:graded_XY_states}
Recall that $(\mu,\rho)$ are the parameters  in $\mathbf{H}^{XY}$.
\begin{enumerate}
  \item[{\rm (i)}] If either $|\mu|=\tfrac{1}{2}$ or $|\mu|<\tfrac{1}{2}$ and $\rho=0$, then 
  $E - \theta_- E \theta_-$ is not Hilbert-Schmidt.
  \item[{\rm (ii)}] If either $|\mu|>\tfrac{1}{2}$ or $(\mu,\rho) = (0,\pm 1)$, then 
  $E - \theta_- E \theta_-$ is Hilbert-Schmidt and 
$\mathrm{dim}\big( \theta_- E \theta_- \wedge (1- E) \big)$ is even.
  \item[{\rm (iii)}] If $|\mu|<\tfrac{1}{2}$ and $\rho \neq 0$, then $E - \theta_- E \theta_-$ is Hilbert-Schmidt and 
$\mathrm{dim}\big( \theta_- E \theta_- \wedge (1- E) \big)$ is odd.
\end{enumerate}
\end{proposi}

\begin{theo}[\cite{BR2}, Example 6.2.56; \cite{ArakiMatsui85}, Theorem 1] \label{thm:XY_soln}
The number of extremal (and thus pure) 
ground states of the 
$XY$-Hamiltonian on the Pauli algebra $A^{P}_{\ZM}$ is as follows
\begin{enumerate}
  \item[{\rm (i)}] $1$ if $|\mu|\geq \tfrac{1}{2}$ or if $|\mu|<\tfrac{1}{2}$ and $\rho=0$,
  \item[{\rm (ii)}] $2$ if $|\mu|< \tfrac{1}{2}$, $\rho\neq 0$ and $(\mu,\rho)\neq (0,\pm 1)$. 
  The grading automorphism $\Theta$ on $A^{P}_{\ZM}$ maps between these ground states.
  \item[{\rm (iii)}] $\infty$ if $(\mu,\rho) = (0, \pm 1)$.
\end{enumerate}
\end{theo}

In the quantum Ising region $(\mu,\rho) =(0,\pm 1)$, there are $4$ ground states up to unitary equivalence.
Namely, for $\nu_j$ 
any $+4$-eigenvector of $H_\mathbb{Z}^{XY}$, 
the states $\omega_{E}$ and $\omega_{\nu_i}$ 
 both split into a 
sum of two extremal ground states $\omega_{E}^j$ and $\omega_{\nu_i}^j$, 
$j\in \{0,1\}$ such that $\omega_{E}^0 \circ \Theta =  \omega_{E}^1$ 
and $\omega_{\nu_j}^0 \circ \Theta =  \omega_{\nu_j}^1$ with $\Theta$ the grading on $A^{P}_{\ZM}$.

\vspace{0.1cm}

To summarise our discussion, one obtains a richer characterisation of the ground states of the 
infinite $XY$-chain by considering   both $A^\mathrm{car}_\mathbb{Z}$ and 
the Pauli algebra $A^{P}_{\ZM}$ (or, equivalently, studying the states 
$\omega$ and $\omega\circ \gamma_-$ restricted to the even subalgebra  $(A^\mathrm{car}_\mathbb{Z})^0$).

\subsection{The split property} \label{subsec:split_prop}

The split property has its roots 
in algebraic quantum field theory~\cite{DL84} but was adapted to 
fermion and spin chains by Matsui~\cite{Matsui01, Matsui13}. 
More recently, the application of the split property to the analytic 
approach to SPT phases has been developed by 
Ogata {\it et al.}~\cite{OgataTRS, OgataReflection, OgataTasaki}. A long range 
version of~\cite{OgataTRS} is given by Moon~\cite{Moon19}.
Given a subset $\Lambda \subset \mathbb{Z}$  
with complement $\Lambda^c = \mathbb{Z}\setminus \Lambda$  and 
a $\Theta$-invariant state $\omega$, one introduces the product state of the restrictions by 
$$
  \omega_\Lambda \otimes_F \omega_{\Lambda^c}(A_1 A_2) \;=\; 
  \omega_\Lambda(A_1) \,\omega_{\Lambda^c}(A_2) \; , \qquad 
  A_1 \in A^\mathrm{car}_\Lambda \; , \; A_2 \in A^\mathrm{car}_{\Lambda^c} \; , \; 
  A_1 A_2 \in A^\mathrm{car}_\mathbb{Z} \; . 
$$
To briefly indicate why $\omega_\Lambda \otimes_F \omega_{\Lambda^c}$ is a state, 
first note that for $A_1 \in A^\mathrm{car}_\Lambda$, $A_2 \in A^\mathrm{car}_{\Lambda^c}$ 
and $A_1 A_2 \in A^\mathrm{car}_\mathbb{Z}$, one always has that 
$A_2^* A_1^* A_1 A_2 = A_1^* A_1 A_2^* A_2$ and so 
$$
  \omega_\Lambda \otimes_F \omega_{\Lambda^c}\big((A_1 A_2)^* A_1 A_2\big) 
  \;=\;  \omega_\Lambda \otimes_F \omega_{\Lambda^c}(A_1^* A_1 A_2^* A_2) 
  \;=\; \omega_\Lambda(A_1^* A_1) \, \omega_{\Lambda^c}(A_2^*A_2) \; , 
$$
which will be positive as $\omega_\Lambda$ and $\omega_{\Lambda^c}$ are states.

\vspace{0.2cm}

We will mainly use $\Lambda=\NM$ and then denote 
$\omega_R= \omega_\NM$ and $\omega_L=\omega_{\NM^c}$. 
These are states on $A^\mathrm{car}_L = A^\mathrm{car}_{(-\infty,0]\cap\mathbb{Z}}$ and 
$A^\mathrm{car}_R= A^\mathrm{car}_{[1,\infty)\cap\mathbb{Z}}=A^\mathrm{car}_\mathbb{N}$.
Recall that $2$ states $\omega_0,\,\omega_1$ on a $C^*$-algebra $A$ are quasiequivalent if 
there is an isomorphism $\rho: \pi_{\omega_0}(A)'' \to \pi_{\omega_1}(A)''$ with 
$\rho\circ \pi_{\omega_0}(a) = \pi_{\omega_1}(a)$ for all $a \in A$~\cite[Section 2.4.4]{BR1}. 
Pure states are either disjoint or unitarily equivalent~\cite{DixC}, 
so  if two pure states are quasiequivalent they are necessarily unitarily equivalent.

\begin{defini}
A $\Theta$-invariant state $\omega$ on $A^\mathrm{car}_\mathbb{Z}$ satisfies the split property 
if $\omega$ is quasiequivalent to $\omega_L \otimes_F \omega_R$.
\end{defini}

The following proposition is stated in~\cite[page 6]{Matsui13} without proof. We provide a 
proof based on~\cite[Proposition 2.2]{Matsui01}.

\begin{proposi} \label{prop:PureSplit_Type1}
Let $\omega$ be a pure $\Theta$-invariant state on $A^\mathrm{car}_\mathbb{Z}$. 
Then $\omega$ satisfies the split property if and only if 
$\pi_\omega(A^\mathrm{car}_L)''$ and $\pi_\omega(A^\mathrm{car}_R )''$ are 
type I von Neumann algebras.
\end{proposi}
\noindent {\bf Proof.} 
Suppose that $\omega$ is pure, $\Theta$-invariant and quasiequivalent to $\omega_L\otimes_F \omega_R$. 
Then the restrictions $\omega_L$ and $\omega_R$ are also $\Theta$-invariant. Therefore 
$\omega_L \otimes_F \omega_R (a_L a_R) = \omega_L(a_L^0) \omega_R(a_R^0)$ as any 
odd part of $a_L$ and $a_R$ will vanish. Therefore there is an ungraded tensor decomposition 
$\omega \sim_{qe} \omega_L \otimes \omega_R$ and because $\omega$ is pure, it is type I. 
Therefore $\omega_L$ and $\omega_R$ must also be type I as a non type I tensor product cannot 
be type I.

\vspace{.1cm}

Now suppose that $\pi_\omega(A^\mathrm{car}_R )''$ is type I. 
Let $\mathfrak{h}_R = \overline{\pi_\omega(A^\mathrm{car}_R)\Omega_\omega}$ and $\pi_R(A^\mathrm{car}_R)$ the restriction of $\pi_\omega(A^\mathrm{car}_R)$ 
to $\mathfrak{h}_R$. 
Because $\pi_\omega(A^\mathrm{car}_R )''$ is a type I factor, the center of
$\pi_\omega(A^\mathrm{car}_R)''$ is trivial and 
any subrepresentation of $\pi_\omega(A^\mathrm{car}_R)$ is 
quasiequivalent to $\pi_\omega(A^\mathrm{car}_R)$ itself.
This implies that 
$\pi_\omega(A^\mathrm{car}_R)$ and $\pi_R(A^\mathrm{car}_R)$ are 
quasiequivalent and, hence, $\pi_R(A^\mathrm{car}_R)''$ is a type I factor.

\vspace{.1cm}

Next, recall~\cite[Chapter V, Theorem 1.31]{Takesaki1}, where given a 
type I factor $M$ on a separable Hilbert space $\calH_0$ with commutant $M'$, there are 
separable Hilbert spaces $\calH_1$ and $\calH_2$ with a unitary 
$W: \calH_0 \to \calH_1 \otimes \calH_2$ such that 
$$
  W M W^{-1} \;=\; \calB(\calH_1) \otimes \one_{\calH_2} \; , \qquad 
  W M' W^{-1} \;=\; \one_{\calH_1} \otimes \calB( \calH_2) \; .
$$
Using this result, the state $\omega$ of $A^\mathrm{car}_\mathbb{Z}$ is 
equivalent to a state $\phi_L \otimes \phi_R$. Because $\omega$ is 
$\Theta$-invariant, so are $\phi_L$ and $\phi_R$, and so 
$\phi_L$ and $\phi_R$ are quasiequivalent to $\omega_L$ and $\omega_R$ 
respectively. Hence $\omega$ is quasiequivalent to $\omega_L \otimes_F \omega_R$.
\hfill $\Box$

\vspace{.2cm}

For spin systems, a factor state on the left and right chains 
imply a locality property of $\omega$ away from 
the boundary, \cite[Corollary 2.6.11]{BR1} or \cite[Proposition~2.1]{Matsui01}.  

\vspace{.2cm}

There are many one-dimensional models whose ground states do not satisfy the 
split property. For example, adapting the results in~\cite[Section 16]{Wassermann} 
to our setting,  the ground state of the $XY$-Hamiltonian from Equation \eqref{eq:fermionic_XY} 
with parameters $(\mu,\rho) = (0,0)$ generates a type III${}_1$-representation. 
However, there is an important connection between gapped ground states 
and the split property in one-dimensional systems. 

\begin{theo}[Corollary 1.9 in \cite{Matsui13}] \label{thm:gapped_implies_split_property}
Let $\mathbf{H}$ be a one-dimensional $\Theta$-invariant finite range Hamiltonian 
\begin{equation} \label{eq:gapped_fr_Ham}
  \mathbf{H} \;=\; \sum_{j\in \mathbb{Z}} \Phi_j \; , \qquad 
  \Phi_j \in A^\mathrm{car}_{[j-r,j+r]\cap\mathbb{Z}} \; , \qquad 
  \Theta(\Phi_j) \;=\; \Phi_j \; , \qquad 
  \| \Phi_j \| \;\leq\; C \; 
\end{equation}
and satisfying the bound \eqref{eq:interaction_bound_assumption}. 
If $\omega$ is a gapped ground state of $H$, then 
$\pi_\omega(A^\mathrm{car}_L)''$ and $\pi_\omega(A^\mathrm{car}_R )''$ are 
type I von Neumann algebras. In particular, if $\omega$ is pure, then 
it satisfies the split property. 
\end{theo}

The relationship between the split property and gapped ground states is a 
one-dimensional phenomena and the proof of Theorem \ref{thm:gapped_implies_split_property} 
relies on the Jordan--Wigner transform and the area law for the decay of 
entanglement entropy in spin chains.
Results in higher dimensional spin systems have been considered using a 
weaker notion of the split property,  see~\cite{Cha18}.

\subsection{The \texorpdfstring{$\mathbb{Z}_2$}--phase label} \label{subsec:general_Z2}

The next aim is to distinguish different gapped ground states of fermionic Hamiltonians, ideally via 
a topological phase label. To this end, we again utilize the following decomposition obtained from the 
Jordan--Wigner transform, see Section~\ref{subsec:JordanWigner}:
\begin{align}
\label{eq-AhatDecomp}
 &\widehat{A}_\mathbb{Z} \;=\; A^\mathrm{car}_\mathbb{Z} \rtimes_{\gamma_-}\mathbb{Z}_2 
   \;\cong \; A^\mathrm{car}_\mathbb{Z} \,\oplus\, T \,A^\mathrm{car}_\mathbb{Z} \; , 
 &&A^{P}_{\ZM} \;\cong\; (A^{P}_{\ZM})^0 \oplus (A^{P}_{\ZM})^1 \;\cong\; (A^\mathrm{car}_\mathbb{Z} )^0 \,\oplus\, T(A^\mathrm{car}_\mathbb{Z} )^1 \; .
\end{align}
Here $\gamma_-$ is the $\mathbb{Z}_2$-action from Equation \eqref{eq:gamma_-_defn}.
One can extend any state $\omega$ on 
$A^\mathrm{car}_\mathbb{Z}$ to a state on $\widehat{A}_\mathbb{Z}$ and then 
restrict to a state $\omega^P$ on $A^{P}_{\ZM}$.
If one starts with a $\Theta$-invariant and pure state  on $A^\mathrm{car}_\mathbb{Z}$, 
by Theorem \ref{thm:car_even_state_equivalence} the purity of $\omega^P$ depends on 
the representations of $\omega$ and $\omega\circ\gamma_-$ on $A^\mathrm{car}_\mathbb{Z}$ 
and $(A^\mathrm{car}_\mathbb{Z})^0$. In the quasifree case, this obstruction can be 
expressed in terms of a Hilbert-Schmidt condition and a $\mathbb{Z}_2$-index on canonical transformations. 
Let us now consider 
this question for more general states. The next results do not need $\omega$ to be a ground state.

\begin{lemma} \label{lem:split_gives_quasequiv}
Let $\omega$ be a $\Theta$-invariant  state on $A^\mathrm{car}_\mathbb{Z}$ that 
satisfies the split property. Then $\omega$ is quasiequiva\-lent to $\omega\circ \gamma_-$.
\end{lemma}
\noindent {\bf Proof.} 
If $\omega$ is $\Theta$-invariant then so are the restrictions $\omega_L$ and $\omega_R$ 
to the subalgebras $A^\mathrm{car}_L$ and $A^\mathrm{car}_R$. Furthermore, we observe 
that 
\begin{align*}
   &\gamma_- \big|_{A^\mathrm{car}_{L}} \;=\; \Theta \big|_{A^\mathrm{car}_{L}} \; , 
   &&\gamma_- \big|_{A^\mathrm{car}_{R}} \;=\; \mathrm{Id}_{A^\mathrm{car}_R} \; ,
\end{align*}
and so 
$$
  \omega_{L} \otimes_F \omega_{R} (\gamma_-(a_L a_R)) \;=\; \omega_L(\gamma_-(a_L))\, \omega_R(\gamma_-(a_R)) \;=\; 
  \omega_L(\Theta(a_L))\, \omega_R(a_R) \;=\;  \omega_{L} (a_L)\, \omega_R(a_R) \; .
$$
That is, $\omega_L \otimes_F \omega_R \circ \gamma_- = \omega_L \otimes_F \omega_R$. 
Therefore by Corollary 2.3.17 of \cite{BR1}, there is a unitary $W \in \mathcal{B}(\mathfrak{h}_{\omega_L \otimes_F \omega_R})$ such that 
$W\Omega_{\omega_L \otimes_F \omega_R} = \Omega_{\omega_L \otimes_F \omega_R}$ and 
$W \pi_{\omega_L \otimes_F \omega_R}(a) W^* = \pi_{\omega_L \otimes_F \omega_R}(\gamma_-(a))$.

\vspace{0.2cm}

Because $\omega_L\otimes_F \omega_R$ is quasiequivalent to $\omega$, there is an 
isomorphism 
$\varphi: \pi_{\omega_L\otimes_F \omega_R}(A^\mathrm{car}_\mathbb{Z})'' \to \pi_\omega(A^\mathrm{car}_\mathbb{Z})''$ 
such that $\varphi(\pi_{\omega_L\otimes_F \omega_R}(a)) = \pi_\omega(a)$ for all $a \in A^\mathrm{car}_\mathbb{Z}$. 
Let us now consider the map $\varphi\circ \mathrm{Ad}_W$ which has the property that 
$$
  \varphi( W \pi_{\omega_L\otimes_F \omega_R}(a) W^*) \;=\; \varphi( \pi_{\omega_L\otimes_F\omega_R}(\gamma_-(a))) \;=\; 
  \pi_\omega( \gamma_-(a)) \;=\; \pi_{\omega\circ\gamma_-}(a)  \; , \qquad a \in A^\mathrm{car}_\mathbb{Z} \; .
$$
Hence $\varphi \circ \mathrm{Ad}_W$ gives 
an isomorphism 
$\pi_{\omega_L\otimes_F\omega_R}(A^\mathrm{car}_\mathbb{Z})'' \cong \pi_{\omega\circ\gamma_-}(A^\mathrm{car}_\mathbb{Z})''$
that implements a quasiequivalence between $\omega_L\otimes_F \omega_R$ and $\omega\circ \gamma_-$.
 Because quasiequivalence is transitive, $\omega$ is quasiequivalent to $\omega \circ \gamma_-$.
\hfill $\Box$

\vspace{0.2cm}

Let us now assume that $\omega$ is pure and $\Theta$-invariant. 
In particular, $\pi_\omega(A^\mathrm{car}_\mathbb{Z})'' = \mathcal{B}(\mathfrak{h}_\omega)$ 
and the GNS space is graded by a self-adjoint 
unitary $\Sigma$. 
If, moreover, $\omega$ is 
equivalent to $\omega \circ \gamma_-$, 
there exists a  unitary $V \in \mathcal{B}(\mathfrak{h}_\omega)$ 
such that $\pi_\omega(\gamma_-(a)) = V \pi_\omega(a)V^*$.  It turns out that this 
unitary can be either even or odd.

\begin{proposi} \label{prop:equiv_unitary_graded}
Let $\omega$ be a pure $\Theta$-invariant state on $A^\mathrm{car}_\mathbb{Z}$ equivalent 
to $\omega \circ \gamma_-$. 
\begin{enumerate}
  \item[{\rm (i)}] The states 
$\omega |_{(A^\mathrm{car}_\mathbb{Z})^0}$ and $\omega |_{(A^\mathrm{car}_\mathbb{Z})^0} \circ \gamma_-$
 are equivalent (that is,  $\omega^P$ is pure)
if and only if there is a self-adjoint unitary $V_0 \in \pi_\omega((A^\mathrm{car}_\mathbb{Z})^0)''$ 
such that $\pi_\omega(\gamma_-(a)) = V_0 \pi_\omega(a)V_0^*$ for all $a \in A^\mathrm{car}_\mathbb{Z}$.
 \item[{\rm (ii)}] If $\omega |_{(A^\mathrm{car}_\mathbb{Z})^0}$ and $\omega |_{(A^\mathrm{car}_\mathbb{Z})^0} \circ \gamma_-$
 are not equivalent (that is,  $\omega^P$ is not pure), then 
there exists a unitary $V_1 \in \pi_\omega((A^\mathrm{car}_\mathbb{Z})^1)''$ 
such that $\pi_\omega(\gamma_-(a)) = V_1 \pi_\omega(a)V_1^*$ for all $a \in A^\mathrm{car}_\mathbb{Z}$. 
Furthermore, $\omega^P$ is a mixture of two inequivalent pure states.
\end{enumerate}
\end{proposi}

We note that there is a large overlap between the above proposition and~\cite[Proposition 6.3]{Matsui13}. 

\vspace{0.2cm}

\noindent {\bf Proof.} 
(i) Given the state $\omega$, one can identify the GNS space 
$\mathfrak{h}_{\omega |_{(A^\mathrm{car}_\mathbb{Z})^0}}$ of its restriction to the even algebra with 
$\mathfrak{h}_\omega^0 = \overline{\pi_\omega((A^\mathrm{car}_\mathbb{Z})^0)\Omega_\omega} \cong \frac{1}{2}(1+{\Sigma})\mathfrak{h}_\omega$. Because 
$\omega$ is $\Theta$-invariant and pure, $\omega |_{(A^\mathrm{car}_\mathbb{Z})^0}$ 
is pure~\cite[Lemma 6.23]{EvansKawahigashi}. In particular, 
the states $\omega |_{(A^\mathrm{car}_\mathbb{Z})^0}$ and 
$\omega |_{(A^\mathrm{car}_\mathbb{Z})^0} \circ \gamma_-$ on $(A^\mathrm{car}_\mathbb{Z})^0$ will 
be equivalent if and only if there is a self-adjoint unitary 
$V=V_0\in \pi_\omega((A^\mathrm{car}_\mathbb{Z})^0)''$ implementing 
$\gamma_-$ on $\mathfrak{h}_\omega^0$, \textit{i.e.}  ${\Sigma} V {\Sigma} = V$.

\vspace{0.2cm}

For part (ii), let us fix some $j\in\mathbb{N}$ and set $Z_j = \fraka_j + \fraka_j^*$ 
which is an odd self-adjoint unitary in $A^\mathrm{car}_\mathbb{Z}$. 
By~\cite[Lemma 6.27]{EvansKawahigashi} (applied with $U=Z_j$ and $\beta=\gamma_-$), 
the pure state $\omega |_{(A^\mathrm{car}_\mathbb{Z})^0}$ on $(A^\mathrm{car}_\mathbb{Z})^0$ is equivalent to 
$\omega |_{(A^\mathrm{car}_\mathbb{Z})^0} \circ \gamma_- \circ \mathrm{Ad}_{Z_j}$.
Therefore there is some $\tilde{W} \in \pi_\omega((A^\mathrm{car}_\mathbb{Z})^0)''$ such that 
$\mathrm{Ad}_{\tilde{W}}$ implements $\gamma_- \circ \mathrm{Ad}_{Z_j}$ on 
$\mathfrak{h}_\omega^0\cong \frac{1}{2}(1+{\Sigma})\mathfrak{h}_\omega$. Because 
$(\gamma_- \circ \mathrm{Ad}_{Z_j})^2 = \mathrm{Id}$, for an appropriate phase we 
can take $W = e^{i\phi}\tilde{W}$ self-adjoint with $\mathrm{Ad}_W$ implementing 
$\gamma_-\circ \mathrm{Ad}_{Z_j}$ on the GNS space. 
We then compute that
\begin{align*}
    \pi_\omega(Z_j)W \pi_\omega(a) W \pi_\omega(Z_j) \;&=\;  \pi_\omega (\mathrm{Ad}_{Z_j} \circ \gamma_- \circ \mathrm{Ad}_{Z_j}(a) ) 
    \;=\; \pi_\omega( \gamma_-(a) ) \; , \quad a \in (A^\mathrm{car}_\mathbb{Z})^0 \; .
\end{align*}
Once again, because $\gamma_-^2 =\mathrm{Id}$, the operator $\pi_\omega(Z_j)W$ is 
self-adjoint up to a phase. In particular, $\pi_\omega(Z_j)W = e^{i\nu} W \pi_\omega(Z_j)$ 
for some $\nu$.

\vspace{0.2cm}

We now consider odd elements, where we compute that, for $a_1 \in (A^\mathrm{car}_\mathbb{Z})^1$,
\begin{align} \label{eq:odd_conjugation_gamma-}
   \pi_\omega(Z_j)W \pi_\omega(a_1) W \pi_\omega(Z_j) \;&=\; 
   e^{i\nu} W \pi_\omega( Z_j a_1) W \pi_\omega(Z_j) 
   \;=\; e^{i\nu} \pi_\omega( \gamma_- \circ \mathrm{Ad}_{Z_j} (Z_j a_1) ) \pi_\omega(Z_j) \nonumber \\
   &=\; e^{i\nu} \pi_\omega( \gamma_-(a_1) Z_j) \pi_\omega(Z_j) 
   \;=\; e^{i\nu} \pi_\omega(\gamma_-(a_1)) \; , 
\end{align}
where we have used that $Z_j a_1$ is even and our results on even elements. 
Because Equation \eqref{eq:odd_conjugation_gamma-} is true for all odd elements, 
we have that 
\begin{equation} \label{eq:odd_conjugation_Zj}
   \pi_\omega(Z_j)W \pi_\omega( Z_j) W \pi_\omega(Z_j) \;=\; 
  e^{i\eta}  \pi_\omega( \gamma_-(Z_j) ) \;=\;  e^{i\eta} \pi_\omega( Z_j) \; .
\end{equation}
Because the left-hand side of Equation \eqref{eq:odd_conjugation_Zj} is self-adjoint, 
so must be the right-hand side, which implies that $e^{i\eta} = \pm 1$. 
If $e^{i\eta} = 1$ we are done and can take the unitary 
$V_1 = \pi_\omega(Z_j) W \in \pi_\omega((A^\mathrm{car}_\mathbb{Z})^1)''$. 
If $e^{i\eta} = -1$, then instead we consider $\pi_\omega(Z_j) W \Sigma$, where 
for any $a \in A^\mathrm{car}_\mathbb{Z}$ with homogeneous grading $|a| \in \{0,1\}$,
\begin{align*}
  \pi_\omega(Z_j) W \Sigma \pi_\omega( a) \Sigma W \pi_\omega(Z_j) 
  \;&=\; (-1)^{|a|} \pi_\omega(Z_j) W \pi_\omega(a) W \pi_\omega(Z_j)  \\
  \;&=\; (-1)^{|a|} (-1)^{|a|} \pi_\omega( \gamma_-(a) ) \;=\; \pi_\omega( \gamma_-(a) ) \; .
\end{align*}
Thus $V_1 =\pi_\omega(Z_j) W \Sigma \in \pi_\omega((A^\mathrm{car}_\mathbb{Z})^1)''$ gives 
the required result. 
The last statement is Theorem \ref{thm:car_even_state_equivalence}.
\hfill $\Box$

\vspace{0.2cm}

For completeness, let us now construct the corresponding states $\omega^P$ on $A^{P}_{\ZM}$  in the  two settings of Proposition \ref{prop:equiv_unitary_graded}. If
for $i=0$ or $i=1$ there is an element $V_i \in \pi_\omega((A^\mathrm{car}_\mathbb{Z})^i)''$
such that $\pi_\omega(\gamma_-(a)) = V_i \pi_\omega(a)V_i^*$, then 
recalling the decomposition~\eqref{eq-AhatDecomp} of  $A^{P}_{\ZM}$, one can 
define a representation $\pi:A^{P}_{\ZM} \to \mathcal{B}(\mathfrak{h}_\omega)$ by
$$
  \pi( a_0 + T a_1 ) \;=\; \pi_\omega(a_0) + V_i \,\pi_\omega(a_1)\; , \qquad 
  a_j  \in  (A^\mathrm{car}_\mathbb{Z})^j \; .
$$
We then set
$$
   \omega^P( Q) \;=\; \langle \Omega_\omega, \, \pi(Q) \Omega_\omega \rangle_{\mathfrak{h}_\omega}  
   \;=\;  \langle \Omega_\omega, \, \pi_\omega(a_0) \Omega_\omega \rangle_{\mathfrak{h}_\omega}  
   +  \langle \Omega_\omega, \, V_i \pi_\omega(a_1) \Omega_\omega \rangle_{\mathfrak{h}_\omega} \;, \quad 
   Q \;=\; a_0+Ta_1 \in  A^{P}_{\ZM} \; .
$$
For the even unitary $V_0$,  the second term in $\omega^P( Q)$ will vanish as 
$\Omega_\omega$ is even and $V_0 \pi_\omega(a_1) \Omega_\omega$ odd. 
By~\cite[Proposition 6.3 (ii)]{Matsui13}, $\omega^P$ is the unique $\Theta$-invariant 
pure state on $A^{P}_{\ZM}$ coming from the state $\omega$ on $A^\mathrm{car}_\mathbb{Z}$. 
If the unitary $V_1$ is odd, then the second term does not vanish and $\omega^P$ is 
a sum of two states.

\vspace{.2cm}

Let us now define a $\ZM_2$-phase label for a class of pure $\Theta$-invariant 
states on $A^\mathrm{car}_\mathbb{Z}$ that are not necessarily quasifree. 
The definition distinguishes the two cases considered in Proposition~\ref{prop:equiv_unitary_graded}.
Recall that $\Sigma$ is the implementation of the parity $\Theta$ in the GNS representation.

\begin{defini}
\label{def-GenZ2}
Let $\omega$ be a pure $\Theta$-invariant state on $A^\mathrm{car}_\mathbb{Z}$ 
that is equivalent to $\omega \circ \gamma_-$.   
Further let $V \in \pi_\omega((A^\mathrm{car}_\mathbb{Z})^i)''$ be a  
unitary such that  $\pi_\omega(\gamma_-(a)) = V \pi_\omega(a)V^*$ for all 
$a \in A^\mathrm{car}_\mathbb{Z}$. Then a $\ZM_2$-phase label of $\omega$ 
is assigned by $ j(\omega) = (-1)^{i}\in\ZM_2$ with $i=0,1$  as above, 
namely ${\Sigma} V {\Sigma} = (-1)^i V$. 
\end{defini}

Let us make some first comments on this definition. First, we note that 
any $V$ implementing $\gamma_-$ on $\mathfrak{h}_\omega$ has indeed homogeneous parity 
by Proposition \ref{prop:equiv_unitary_graded}. Such a unitary $V$ is determined up to 
unitary equivalence and, because $\pi_\omega$ is irreducible, 
any other operator $U VU^*$ implementing $\gamma_-$ is the same as $V$ up to a 
complex scalar of modulus one. Hence the parity of all unitaries implementing 
$\gamma_-$ is constant and thus the phase-label is well-defined. 
Moreover, Lemma \ref{lem:split_gives_quasequiv} implies that the $\mathbb{Z}_2$-phase 
label is well-defined for pure and $\Theta$-invariant  states that satisfy the split property. In particular, 
the $\mathbb{Z}_2$-phase label is defined 
for any pure gapped ground state of a Hamiltonian for the form considered in Theorem~\ref{thm:gapped_implies_split_property}.
Moreover, for quasifree states the $\ZM_2$-phase label is linked to a $\ZM_2$-valued Fredholm index.

\begin{proposi} \label{prop:index_link}
Let $E$ be a basis projection and $\omega_E$ the corresponding pure, $\Theta$-invariant  and 
quasifree state on $A^\mathrm{car}_\mathbb{Z}$. 
If  $\omega$ is equivalent to $\omega \circ \gamma_-$, then for $J=\ii (2E-1)$ 
and $\theta_-$ the diagonal extension of \eqref{eq:theta_operator_defn},
$$
j(\omega_E)\;=\;j_J(\theta_-)
\;.
$$
\end{proposi}

\noindent {\bf Proof.}
By Theorem~\ref{thm:car_even_state_equivalence}, $\omega_E^P$ is pure in case (i) of 
Proposition~\ref{prop:equiv_unitary_graded} and not pure in case (ii). 
These cases correspond to $j(\omega_E)=1$ and $j(\omega_E)=-1$ respectively. 
Therefore Corollary~\ref{coro:car_even_state_equivalence} implies the claim.
\hfill $\Box$

\vspace{0.2cm}

Recalling Example \ref{ex:Kitaev_inf_flux} in the quasifree setting, the automorphism 
$\gamma_-$ can be implemented by inserting a local half-flux through a Hamiltonian. 
Because the index $j(\omega)$ is a comparison between the state $\omega$ and 
the `half-flux-inserted state' $\omega \circ \gamma_-$, if $j(\omega) = -1$, this indicates 
that a flux insertion  induces a change in the ground state. In the quasifree 
setting, such a change of the ground state is detected by the $\mathbb{Z}_2$-valued spectral 
flow.

\vspace{0.2cm}

We now consider some basic stability properties of the phase label. The following is a simple 
application of standard properties of the  GNS representation of pure states.

\begin{proposi} \label{prop:weak_automorphic_equivalence}
Let $\omega_0$ and $\omega_1$ be pure $\Theta$-invariant states on $A^\mathrm{car}_\mathbb{Z}$ 
equivalent to $\omega_0 \circ \gamma_-$ and $\omega_1 \circ \gamma_-$ respectively. Suppose that there is an automorphism 
$\eta \in \mathrm{Aut}(A^\mathrm{car}_\mathbb{Z})$ commuting with 
$\Theta$ and $\gamma_-$ and such that $\omega_1 = \omega_0 \circ \eta$. Then $j(\omega_0) = j(\omega_1)$.
\end{proposi}

The hypothesis that $\eta$ commutes  
with $\Theta$ and $\gamma_-$ is quite strong, though it is sufficient to assume that 
$\eta$ commutes with $\Theta$ and leaves $A^\mathrm{car}_L$ and $A^\mathrm{car}_R$ invariant. 
Proposition~\ref{prop:weak_automorphic_equivalence} combined with the following remark shows that
the $\ZM_2$-phase label is perturbatively stable, for example, when weak interactions are added to a quasifree system.

\begin{remark} \label{rk:C1_aut_equiv}
{\rm 
Examples of such automorphisms $\eta$ of $A^\mathrm{car}_\mathbb{Z}$ that satisfy the 
hypothesis of Proposition \ref{prop:weak_automorphic_equivalence} can be constructed using 
the quasilocal structure of $A^\mathrm{car}_\mathbb{Z}$ and the quasiadiabatic evolution 
(also called the spectral flow) of uniformly gapped $C^1$-interactions~\cite{NSY18}. In particular, 
let us consider a path of local Hamiltonians for all $X \subset \mathbb{Z}$ finite, where 
$$
  \mathbf{H}_X (s) \;=\;  \mathbf{H}_X + \Phi_X(s)  \; 
$$
and the  path satisfies several assumptions.
 First, 
the ground state gap of $\mathbf{H}_X (s)$ is required to be uniformly bounded for all $s\in [0,1]$. 
Furthermore, $\Phi_X(s) \in \mathcal{B}_F$ for all $s\in [0,1]$ and $X \in \mathcal{P}_0(\mathbb{Z})$, where  
$\mathcal{B}_F$ is the space of strongly $C^1$-interactions satisfying~\cite[Assumption 6.12]{NSY18} 
with the additional property that $\Theta( \Phi_X(s)) = \Phi_X(s)$ \emph{and} $\gamma_-( \Phi_X(s)) = \Phi_X(s)$ 
for all $s \in [0,1]$. If these assumptions are satisfied, then the results in~\cite[Section 6-7]{NSY18} 
(adapted to the fermionic case, where the property $\Theta( \Phi_X(s)) = \Phi_X(s)$ 
is crucial)
guarantee the existence of an automorphism $\eta_s^\Phi$ in the infinite-volume limit 
that maps between the ground states on $A^\mathrm{car}_\mathbb{Z}$ with the 
property that $\Theta \circ \eta^\Phi_s = \eta^\Phi_s \circ \Theta$ and 
$\gamma_- \circ \eta^\Phi_s = \eta^\Phi_s \circ \gamma_-$ for all $s\in [0,1]$. 

\vspace{0.1cm}

To summarise, if $j(\omega)$ is well-defined and comes from the thermodynamic 
limit of a finite-volume Hamiltonian $ \mathbf{H}_X (0)$ with gapped ground state, then 
$j(\omega \circ \eta^\Phi_s) = j(\omega)$ for all $s \in [0,1]$.
While this result shows an important stability property of the  $\mathbb{Z}_2$-phase label, 
the assumption that $\gamma_-( \Phi_X(s)) = \Phi_X(s)$ is somewhat artificial. 
Given a $\Theta$-invariant interaction $\Phi$, one can consider 
$\tilde{\Phi} = \frac{1}{2}\big( \Phi + \gamma_-(\Phi) \big)$ which is $\gamma_-$-invariant, 
but it is 
interesting to investigate to what degree the $\gamma_-$-invariant assumption can be lessened. 
One may be able to use a construction similar to~\cite{OgataTRS} in order to work with 
paths of interactions that need not be $\gamma_-$-invariant.
\hfill $\diamond$
}
\end{remark}

\begin{proposi} \label{prop:cts_unitaries_same_index}
Let $\omega_0$ be a pure and $\Theta$-invariant state on $A^\mathrm{car}_\mathbb{Z}$ 
that is equivalent to $\omega_0 \circ \gamma_-$. Suppose that 
there is a path of  states $\{\omega_s\}_{s\in [0,1]}$ with an
associated family of Hilbert spaces $\{\mathfrak{h}_{\omega_s}\}_{s\in [0,1]}$, as well as 
unitaries $\{U_s \}_{s\in[0,1]}$ such that
$U_s : \mathfrak{h}_{\omega_0} \rightarrow \mathfrak{h}_{\omega_s}$. 
Then, $j(\omega_s) = j(\omega_0)$ for all $s \in [0,1]$.
\end{proposi}
\noindent {\bf Proof.} 
Given such a path of unitaries,  for any $A_s \in \mathcal{B}(\mathfrak{h}_{\omega_s})$ there is an 
operator $A_0 \in \mathcal{B}(\mathfrak{h}_{\omega_0})$ such that 
$A_s = U_s A_0 U_s^*$. We can therefore define a representation 
$\pi_{\omega_s} = \mathrm{Ad}_{U_s}\circ \pi_{\omega_0}$. Because $\pi_{\omega_0}$ is 
irreducible, so is $\pi_{\omega_s}$. Furthermore, 
for $V_s = U_s V_0 U_s^*$, $\Sigma_s = U_s \Sigma_0 U_s^*$ one has
\begin{align*}
  &V_s \pi_{\omega_s}(a) V_s \;=\; \pi_{\omega_s}(\gamma_-(a)) \; , 
  &&\Sigma_s \pi_{\omega_s}(a) \Sigma_s \;=\; \pi_{\omega_s}(\Theta(a)) \; ,
\end{align*}
so that
$$
  \Sigma_s V_s \Sigma_s \;=\; U_s \Sigma_0 V_0 \Sigma_0 U_s^* \;=\; (-1)^{|V_0|} V_s \; .
$$
Thus for all $s\in [0,1]$, $j(\omega_s)$ is well-defined with $j(\omega_s) = j(\omega_0)$.
\hfill $\Box$

\vspace{0.2cm}

Results from~\cite{NSY18} guarantee that our $\mathbb{Z}_2$-index is stable under strongly $C^1$-paths
of interactions that are $\Theta$-symmetric, $\gamma_-$-symmetric and satisfy~\cite[Assumption 6.12]{NSY18}. 
In particular, if two pure gapped ground states $\omega_0$ and $\omega_1$ have different indices, 
$j(\omega_0)= - j(\omega_1)$, these ground states cannot be connected by such a path. Similarly, 
by Proposition \ref{prop:cts_unitaries_same_index} there cannot be family of unitaries  
of unitaries connecting $\mathfrak{h}_{\omega_0}$ and $\mathfrak{h}_{\omega_1}$.

\vspace{0.2cm}

Let us now state a stability result of the $\ZM_2$-phase label in the quasifree setting.

\begin{proposi} \label{prop:Fred_path_index_invariant}
Let $(\calH, \Gamma)$ be a complex Hilbert space with real structure. Let 
$H_0$ and $H_1$ be gapped BdG Hamiltonians on $\calH$ with quasifree 
ground states $\omega_{E_0}$ and $\omega_{E_1}$ such that 
$j(\omega_{E_0})$ and $j(\omega_{E_1})$  are well-defined. Suppose that 
$H_0$ and $H_1$ can be connected by a norm-continuous path of self-adjoint Fredholm 
operators $[0,1]\ni t \mapsto H_t$ such that $\Gamma H_t \Gamma = -H_t$ for 
all $t\in [0,1]$. Then $j(\omega_{E_0}) = j(\omega_{E_1})$.
\end{proposi}
\noindent {\bf Proof.} 
By the assumptions on the path $H_t$, the $\mathbb{Z}_2$-valued spectral flow 
$\mathrm{Sf}_2( iH_t)$ is well-defined. In particular, there is a partition 
$0= t_0 < t_1 < \ldots < t_n =1$ such that 
$\mathrm{Ind}_{2}( J_{t_j}, J_{t_{j+1}}) = (-1)^{\frac{1}{2} \Ker( J_{t_j} + J_{t_{j+1}} )}$ 
is well-defined for $J_{t_j} = i H_{t_j}|H_{t_j}|^{-1}$ (with an arbitrary complex 
structure on $\Ker(H_{t_j})$ if needed). Now Proposition \ref{prop:index_link} 
implies that
$$
   j(\omega_{E_0}) \;=\; j_{J_{t_0}} (\theta_-) \;=\; \mathrm{Ind}_2( J_{t_0}, \theta_- J_{t_0} \theta_- ) \; 
$$
with $\theta_-$ the diagonal extension of \eqref{eq:theta_operator_defn}. 
Recalling the concatenation and invariance properties of $\mathrm{Ind}_2$, in particular 
$$
\mathrm{Ind}_2(J_{t_j}, J_{t_{j+1}}) \;=\; \mathrm{Ind}_2(J_{t_{j+1}}, J_{t_{j}}) \;=\; 
\mathrm{Ind}_2( V J_{t_j} V^*, V J_{t_{j+1}}V^*)
$$
for any unitary $V$ with $\Gamma V \Gamma = V$,
we compute 
\begin{align*}
   j(\omega_{E_0}) \;&=\; \mathrm{Ind}_2( J_{t_0}, \theta_- J_{t_0} \theta_- ) \\
    &=\;   \mathrm{Ind}_2(J_{t_0}, J_{t_1}) \cdots  \mathrm{Ind}_2( J_{t_{n-1}}, J_{t_n} ) 
      \,  \mathrm{Ind}_2( J_{t_n}, \theta_- J_{t_n} \theta_- ) \,  
      \mathrm{Ind}_2(\theta_- J_{t_n} \theta_-, \theta_- J_{t_{n-1}} \theta_- )  \\
      &\qquad \times \cdots   \mathrm{Ind}_2(\theta_- J_{t_1} \theta_- , \theta_- J_{t_0} \theta_- ) \\
    &=\;  \mathrm{Ind}_2( J_{t_n}, \theta_- J_{t_n} \theta_- ) 
\\
& =\; j(\omega_{E_1} ) \; 
\end{align*}
as all other terms cancel.
\hfill $\Box$

\vspace{0.2cm}

Proposition \ref{prop:Fred_path_index_invariant}, in comparison with 
Proposition \ref{prop:cts_unitaries_same_index}, shows that in special cases 
we can take paths of ground states such that the GNS spaces are not unitarily equivalent, 
but where the $\mathbb{Z}_2$-phase label remains constant. Furthermore, 
recalling Proposition \ref{prop:quasifree_GNS_gap_close}, if the path $iH_t$ from 
Proposition \ref{prop:Fred_path_index_invariant} has a non-trivial $\mathbb{Z}_2$-valued 
spectral flow, 
then the spectral gap of the GNS Hamiltonians will close. Therefore, we see that 
in special cases the index $j(\omega)$ is invariant on  paths that can close the 
ground state gap.


\subsection{Changes in the \texorpdfstring{$\mathbb{Z}_2$}--phase label}

In Section \ref{subsec:general_Z2} we introduced a $\mathbb{Z}_2$-phase label for 
a class of pure and $\Theta$-invariant states on $A^\mathrm{car}_\mathbb{Z}$ and 
showed some basic stability properties of this label. In this 
section, we wish to consider to consider paths of ground states that are capable of 
accommodating a change in the $\mathbb{Z}_2$-phase label. The following example 
from the quasifree setting gives some motivation.

\begin{example}
{\rm 
Recall the example of the non-interacting but infinite Kitaev chain 
$\mathbf{H}^\mathrm{Kit}(\mu, w)$ from Example \ref{ex:inf_Kitaev_chain}. 
Using our results on flux insertion from Example \ref{ex:Kitaev_inf_flux} or 
alternatively using Proposition \ref{prop:graded_XY_states}, in the region 
$w=0$ and $|\mu|> \frac{1}{2}$, then the unique quasifree ground state 
$\omega_{E}$ is such that $j(\omega_{E}) = 1$. 
If $\mu = 0$ and $w \neq 0$, then $j(\omega_{E}) = -1$. 

\vspace{0.1cm}

Recall that the BdG Hamiltonians $H^\mathrm{Kit}_\mathbb{Z}(\mu, 0)$ 
and $H^\mathrm{Kit}_\mathbb{Z}(0, w)$ can be related by the unitary 
$$
  W \;=\;  \frac{\ii }{2} \begin{pmatrix} (\one +S) & \ii(\one -S) \\ \ii (\one -S) & -(\one +S) \end{pmatrix}\; , \qquad W^*W \;=\; WW^* \;=\; \one  \; , 
  \qquad \Gamma W \Gamma \;=\; W \; ,
$$
but where $W$ does \emph{not} give rise to a unitary operator between 
GNS spaces. Thus the two systems can be connected, but in a way where 
singularities emerge.
}
\hfill $\diamond$
\end{example}

This motivates the following definition.

\begin{defini} \label{def:GS_path}
Let $A$ be a unital $C^*$-algebra.
Two ground states $(\omega_0,\beta_0)$ and $(\omega_1,\beta_1)$ 
on $A$ are said to be connected by  path of ground states if 
there is a family of $\mathbb{R}$-actions $\{\beta_s\}_{s\in[0,1]}$ 
and states $\{\omega_s\}_{s\in[0,1]}$ on $A$ such that 
\begin{enumerate}
  \item[{\rm (i)}]  For all $s\in [0,1]$, $\omega_s$ is a ground state  for  $\beta_s$.
  \item[{\rm (ii)}] There is at most a finite set $S_C = \{s_1,\ldots, s_N\} \subset (0,1)$ 
  such that:
  \begin{enumerate}
    \item[{\rm (a)}]  for all $a \in A$, the map 
    $[0,1] \setminus S_C \ni s \mapsto \| \pi_{\omega_s}(a)\| \in [0,\infty)$  is continuous;
    \item[{\rm (b)}] if $h_{\omega_s}$ is the generator of the dynamics $\beta_s$ on 
    $\mathfrak{h}_{\omega_s}$, the map $[0,1]\setminus S_C \ni s \mapsto \|(z - h_{\omega_s})^{-1} \|$ 
    is continuous for all  $z\in \mathbb{C}\setminus \mathbb{R}$.
  \end{enumerate}
\end{enumerate}
If $\tau$ is an automorphism of $A$, then the family of states $\{\omega_s\}_{s\in[0,1]}$ is said to be 
$\tau$-invariant if $\omega_s \circ \tau = \omega_s$ for all $s\in [0,1]$. 
\end{defini}

Let us further comment on this definition. 
For the case $A = A^\mathrm{car}_\mathbb{Z}$, a strongly continuous family 
of actions $\beta_s:  \mathbb{R} \to \mathrm{Aut}(A^\mathrm{car}_\mathbb{Z})$ 
and ground  states satisfying part  (i) of Definition \ref{def:GS_path} can be obtained by 
using the quasilocal structure of $A^\mathrm{car}_\mathbb{Z}$ and results from (amongst others)~\cite{NSY18}. 
Condition (ii) is stronger, but allows us to study paths of operators over the 
different GNS spaces. Indeed, for all $a\in A^\mathrm{car}_\mathbb{Z}$, 
the map $s \mapsto \pi_{\omega_s}(a)$ defines a  continuous section of 
a $C^*$-bundle $p:B \to [0,1]\setminus S_C$ with fibres 
$p^{-1}(s) \cong \pi_{\omega_s}(A^\mathrm{car}_\mathbb{Z})$, 
\textit{cf.}~\cite[Appendix C]{Williams}. 
 By~\cite[Theorem 2]{BeckusBel}, 
condition (ii)(b) is equivalent to the spectral edges of 
$\sigma(h_{\omega_s})$ being continuous in $s$ outside 
the finite points $S_C = \{s_1,\ldots,s_N\}$. The set $S_C$ can be thought of as the points 
where the spectral gaps of $h_{\omega_s}$ close. At best, one expects a fractional H\"{o}lder continuity 
of the spectral edges when a gap closes and condition (ii) requires that such gap closings 
happen at most finitely many times. See~\cite{BeckusBel} for more details on the 
continuity of spectral edges at gap closing points.

\begin{example}
{\rm
Consider a path of local and parity-symmetric Hamiltonians, 
$$
  \mathbf{H}(s) \;=\; \sum_{X \subset \mathbb{Z}\;{\rm finite}}  \Phi(X, s) \; , \quad   s \in [0,1] \; , 
$$
where the interactions $s \mapsto \Phi(X,s)$ are sufficiently smooth and local so 
that the interaction satisfies a Lieb--Robinson bound for all $s \in [0,1]$. 
Therefore by~\cite[Theorem 3.5]{NSY18}, one obtains a dynamics 
$$
\alpha_{s,t} \;=\; \lim_{X \to \mathbb{Z}}  \mathrm{Ad}_{e^{it \mathbf{H}_X(s)}} , \;  \quad s \in [0,1] \; , 
\quad t \in \mathbb{R} \; .
$$
We also 
require the Hamiltonians at the end points, $\mathbf{H}(0)$ and $\mathbf{H}(1)$, to be such that the 
weak $\ast$-limit of the finite-volume ground states gives a unique ground state for the 
dynamics on $A^\mathrm{car}_\mathbb{Z}$. Similarly, the weak $\ast$-limit of the 
finite volume ground states for all $s\in [0,1]$ will give a $\Theta$-invariant path of ground states 
$\{\omega_s\}_{s\in [0,1]}$ of  $A^\mathrm{car}_\mathbb{Z}$.
If the end points of the path of ground states satisfy the split property, \textit{e.g.}  
$\mathbf{H}(0)$ and $\mathbf{H}(1)$ are gapped interactions satisfying the 
conditions of Theorem \ref{thm:gapped_implies_split_property}, then the $\mathbb{Z}_2$-phase 
label $j(\omega_0)$ and $j(\omega_1)$ can be defined. Thus, if 
$j(\omega_0) \neq j(\omega_1)$ the path of finite-volume Hamiltonians 
$\mathbf{H}(s)$ and corresponding path of ground states $\{\omega_s\}_{s\in [0,1]}$ 
can potentially model this $\mathbb{Z}_2$-phase label change.
\hfill $\diamond$
}
\end{example}

\begin{lemma} \label{lem:symm_breaking}
Let $\omega_0$ and $\omega_1$ be $\Theta$-invariant ground states on $A^\mathrm{car}_\mathbb{Z}$ 
and suppose that $j(\omega_0)$ and $j(\omega_1)$ are well-defined with 
$j(\omega_0) \neq j(\omega_1)$. 
Then $\omega_0$ and $\omega_1$ cannot be connected by a {$\Theta$-invariant} path of 
pure ground states satisfying the split property and without discontinuities.
\end{lemma}
\noindent {\bf Proof.} 
Let us suppose the contrary, so there is a family $\{\omega_s\}_{s\in[0,1]}$ connecting 
$\omega_0$ and $\omega_1$ with each $\omega_s$ a $\Theta$-invariant 
pure ground state satisfying the 
split property.  
By Lemma \ref{lem:split_gives_quasequiv}, $\omega_s$ is 
equivalent to $\omega_s \circ \gamma_-$ for all $s\in[0,1]$. Let  $V_s$ and $\Sigma_s$ be 
the unitaries implementing $\gamma_-$ and $\Theta$ respectively  on $\mathfrak{h}_{\omega_s}$. 
By the continuity of the map $s \mapsto \pi_{\omega_s}(\gamma_-(a)) = V_s \pi_{\omega_s}(a)V_s^*$ 
for all $a \in A^\mathrm{car}_\mathbb{Z}$, 
the map $s \mapsto V_s$ is also continuous. By the same argument, $s\mapsto \Sigma_s$ is 
 continuous and, furthermore, $\Sigma_s \Omega_s = \Omega_s$ for all $s\in[0,1]$.
By Proposition \ref{prop:equiv_unitary_graded}, $V_s$ has homogeneous parity for 
all $s\in[0,1]$, namely
$\Sigma_s V_s \Sigma_s = (-1)^{|V_s|}V_s$ with $|V_s| \in\{0,1\}$ being the parity. 
In particular, ${\Sigma}_s V_s \Omega_s = (-1)^{|V_s|} V_s \Omega_s$. 
By the hypothesis, one also has 
${\Sigma}_0 V_0 \Omega_0 = \sigma V_0 \Omega_0$ and ${\Sigma}_1 V_1 \Omega_1 = -\sigma V_1 \Omega_1$ 
for a sign $\sigma$. Thus there is at least one point $s_0$ with a neighbourhood $U\subset (0,1)$ 
such that $\Sigma_s V_s$ is a self-adjoint (resp. skew-adjoint) unitary 
for $s<s_0$ and $\Sigma_s V_s$ is a skew-adjoint (resp. self-adjoint) unitary for $s>s_0$. 
But such a change would violate the continuity of the section $\Sigma_s V_s$.
\hfill $\Box$

\begin{theo} \label{thm:GS_gap_close_general}
Let $\omega_0$ and $\omega_1$ be pure $\Theta$-invariant and gapped ground states on $A^\mathrm{car}_\mathbb{Z}$ 
(in particular, $j(\omega_0)$ and $j(\omega_1)$ are well-defined). 
Suppose that $j(\omega_0)\neq j(\omega_1)$.
Let $\{\omega_s\}_{s\in[0,1]}$ be a $\Theta$-invariant path of ground states connecting $\omega_0$ 
and $\omega_1$. Then there is at least one $s_0 \in (0,1)$ such that $\omega_{s_0}$ cannot 
come from the ground state of a $\Theta$-invariant and gapped interaction of the form \eqref{eq:gapped_fr_Ham}. 

\vspace{.1cm}

If the path of ground states is constructed from a uniformly bounded 
path of interactions $\Phi(s)$ satisfying  \eqref{eq:gapped_fr_Ham} pointwise, 
then the spectral gap of the infinite 
GNS Hamiltonian $h_{\omega_s}$ above $0$ will close along the path.
\end{theo}
\noindent {\bf Proof.} 
By Lemma \ref{lem:symm_breaking}, there is a $s_0\in (0,1)$ such that 
either $\omega_{s_0}$ is not pure or $\omega_{s_0}$ is not split (or both). 

\vspace{.1cm}

If $\omega_{s_0}$ is pure and not split, then $\pi_{\omega_{s_0}}(A^\mathrm{car}_R)''$ 
is not a type I factor. By the contrapositive of 
Theorem \ref{thm:gapped_implies_split_property}, $\omega_{s_0}$ cannot come 
from the ground state of a gapped, finite-range and parity-symmetric fermionic interaction. 
If the path of ground sates is constructed from a uniformly bounded 
path of interactions $\Phi(s)$ satisfying  \eqref{eq:gapped_fr_Ham} pointwise, 
then only the gap hypothesis of Theorem \ref{thm:gapped_implies_split_property} fails. 
At the endpoints, $h_{\omega_0}$ and $h_{\omega_1}$ 
have a spectral gap above $0$.
Because 
the spectral edges of the infinite GNS Hamiltonian are continuous outside a 
gap closing point~\cite[Theorem 2]{BeckusBel}, 
 the spectral gap above $0$ of $h_{\omega_s}$ must therefore close as $s\to s_0$.

\vspace{.1cm}

If $\omega_{s_0}$ is not pure, then there is a decomposition 
$\omega_{s_0} = c_a \omega_a + c_b \omega_b$. Consider then the 
GNS representations of $\omega_a$ and $\omega_b$ with 
cyclic vectors $\Omega_{\omega_a}$ and $\Omega_{\omega_b}$ which can 
be embedded within $\mathfrak{h}_{\omega_{s_0}}$. Because 
$\omega_{s_0}$ is a ground-state, 
both $\Omega_{\omega_a}$ and $\Omega_{\omega_b}$ are $0$-energy 
eigenvectors of the GNS Hamiltonian $h_{\omega_{s_0}}$. As the 
state is not pure, these eigenvectors are distinct and the spectrum is 
degenerate at $0$. Because the endpoints $h_{\omega_0}$ and $h_{\omega_1}$ 
have non-degenerate $0$-energy spectrum with a non-zero spectral gap,  
 the continuity of the spectral edges outside gap closing points implies that 
 for any $\gamma>0$ one can find a sufficiently small $\varepsilon$ such 
 that $\sigma(h_{\omega_{s_0-\varepsilon}})\cap (0,\gamma)$ is non-empty.
\hfill $\Box$

\subsection{Concluding remarks}

We have defined a $\mathbb{Z}_2$-index for 
one-dimensional many-body fermionic gapped ground states. 
While some basic properties of this index have been studied,
let us list some additional questions that we 
hope to investigate further in future work.

\begin{enumerate}
  \item As already stated, Propositions \ref{prop:weak_automorphic_equivalence}, \ref{prop:cts_unitaries_same_index} 
  and Remark \ref{rk:C1_aut_equiv} 
  have shown  stability properties of the  $\mathbb{Z}_2$-index, though the assumptions 
  are quite strong. A more systematic treatment similar to recent studies of $\mathbb{Z}_2$-indices 
  of ground states of spin chains satisfying the split property with time-reversal or reflection 
  symmetry~\cite{ Moon19, OgataTRS, OgataReflection} will 
  hopefully give more optimised results. 
  
  Similarly, the definition of a  path of gapped ground states is quite rigid and 
  a result similar to Theorem \ref{thm:GS_gap_close_general} may hold for a weaker notion of 
 a path of ground states.

  \item If one takes a half-infinite lattice $\mathbb{N}$, then $\gamma_- = \mathrm{Id}$ and the  
  phase label is trivial. Hence, a different method to define the $\mathbb{Z}_2$-phase 
  label is required in half-infinite chains. For one-dimensional spin systems, the left and right degeneracy of edge 
  ground states in half-infinite chains is a complete invariant of the $C^1$-classification of 
  frustration-free and 
  translation invariant interactions~\cite{OgataIII}. One can similarly investigate such a 
  characterisation in fermionic systems. Furthermore, if a connection between 
 edge states in half-infinite systems with the   $\mathbb{Z}_2$-phase label 
  for $\mathbb{Z}$-lattices can be established, this would give an interesting bulk-boundary correspondence in 
  the interacting setting.
  
  \item For the case of quasifree ground states on the full discrete line, the insertion of a flux 
  quanta leads to a non-trivial $\mathbb{Z}_2$-valued spectral flow if the ground state is topologically non-trivial. 
  The $\ZM_2$-phase label $j(\omega)$ extends this probing of the state $\omega$ to a wider class of ground states,
  even though only the ``half-flux added'' state is used and not the flux insertion itself.
  This flux insertion 
  was studied numerically in an interacting finite chain  in \cite{TwistedKitaev} and the same behavior 
  of level crossing was found for the many-body states. For an infinite chain, the flux insertion
  implemented as in Section~\ref{sec:closed_chain_flux} leads to a path of Hamiltonians and dynamics that fits into
  the framework of Definition~\ref{def:GS_path}, but much more is actually expected to hold, see 
  Remark~\ref{rem:QFGS_path_with_closing}. 
  To show a Fredholm-like property for flux insertion for an interaction chain is an interesting open
  problem.  If so, one could introduce  a  $\mathbb{Z}_2$-spectral 
  flow of the infinite GNS Hamiltonian. A more systematic study of such a $\mathbb{Z}_2$-flow 
  would give a more clear picture of an index-theoretic interpretation of the $\mathbb{Z}_2$-phase label.
  Such a viewpoint offers possible future directions for the studies of 
  phase labels and invariants of interacting systems using 
  flux insertions and higher-dimensional analogues.

  \item We have considered an operator algebraic formulation of gapped one-dimensional 
  fermionic ground states associated with parity conserving Hamiltonians. A natural extension is to consider fermionic 
  SPT phases for other symmetries and group actions. It was shown by 
  Ogata~\cite[Appendix B]{OgataTRS} and more recently in~\cite{OgataSPT} that 
  for $G$-symmetric ground states of spin chains with the split property, there is a 
  projective representation of $G$ on a GNS space whose cohomology class is 
  invariant under the quasiadiabatic evolution of gapped 
  symmetric Hamiltonians. We would expect a similar result to hold in the fermionic case 
  that takes into account the parity symmetry. This has already been studied for fermionic 
  matrix product states~\cite{BWH, FK, KTY18, TurzilloYou}.
\end{enumerate}

\vspace{.5cm}

\noindent {\bf Acknowledgements:}  This work is supported by the DFG and 
the World Premier International Research Center Initiative (WPI), MEXT, Japan. CB is 
supported by a JSPS Grant-in-Aid for Early-Career Scientists (No. 19K14548) and   thanks 
Nick Bultinck, Hosho Katsura and Yoshiko Ogata for helpful discussions. 
Both authors thank the Erwin Schr\"{o}dinger Institute program Bivariant K-Theory in Geometry and 
Physics and HSB the Institut-Mittag-Leffler for hospitality and support during the production of this work. 
Finally, we thank the referee for a careful reading and whose comments have 
greatly improved the manuscript.


\end{document}